%% file: main.tex
\documentclass[12pt]{report}

\usepackage[a4paper,margin=1in]{geometry}
\usepackage[utf8]{inputenc}
\usepackage{graphicx}
\usepackage{amsmath,amssymb}
\usepackage{booktabs}
\usepackage{xcolor}
\usepackage{xurl}
\usepackage{hyperref}
\usepackage{caption}
\usepackage{subcaption}
\usepackage{float}
\usepackage{tikz}
\usetikzlibrary{positioning, arrows.meta, calc, fit, backgrounds, decorations.pathreplacing}
\input{figures/architectures}

\hypersetup{colorlinks=true, linkcolor=blue!60!black, citecolor=blue!60!black, urlcolor=blue!60!black}

\begin{document}

\begin{titlepage}
\centering

\vspace*{1cm}

\includegraphics[height=2.5cm]{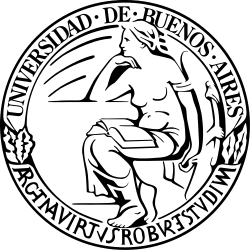}
\hspace{2cm}
\includegraphics[height=2.5cm]{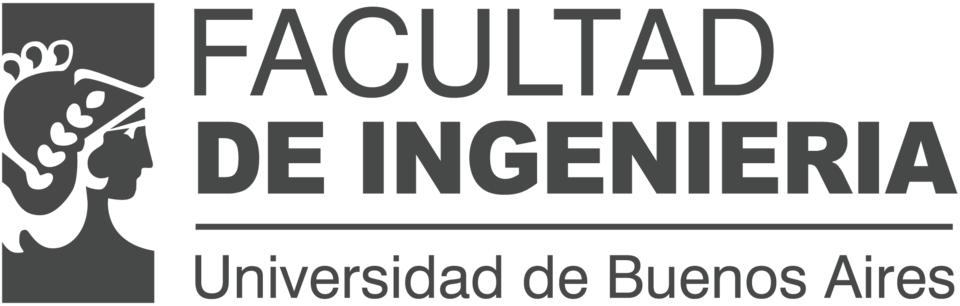}

\vspace{1.5cm}

{\Large Universidad de Buenos Aires \par}
{\large Facultad de Ingenier\'ia \par}

\vspace{2cm}

{\LARGE \bfseries Volatility Surface Reconstruction \\ Using Deep Learning \\ Under No-Arbitrage Constraints \par}

\vspace{2cm}

{\Large Pablo Rodriguez Manzi \par}
\vspace{0.5cm}
{\large Electronic Engineering Thesis \par}

\vspace{2cm}

{\large Supervisor: Ing.\ Ricardo Veiga \par}
\vspace{0.3cm}
{\large Co-supervisor: Ing.\ Alberto Dams \par}

\vfill

{\large 2026 \par}

\end{titlepage}

\chapter*{}
\vspace*{0.1\textheight}
\begin{flushright}
\itshape
A mi abuela, la persona m\'as inteligente que conoc\'i.

\vspace{0.5cm}

Esta tesis tambi\'en es suya.

\vspace{1cm}

To my grandmother, the most intelligent person I ever knew.

\vspace{0.5cm}

This thesis is also hers.
\end{flushright}

\chapter*{Acknowledgements}
\addcontentsline{toc}{chapter}{Acknowledgements}

This thesis is the result of a long journey that would not have been possible without the help and support of many people who stood by me throughout these years.

I would first like to thank my thesis supervisors, Ing. Ricardo Veiga and Ing. Alberto Dams, not only for their guidance throughout the project, but above all for being the ones who contacted me and encouraged me to complete this work when even I no longer believed it was possible.

To the public university, and to all the professors of this faculty who have always shown vocation, effort, and a strong commitment to educating engineers in service of this country.

To my friends, classmates, colleagues, and all the people who have been by my side throughout these years and who have been essential in this journey.

Finally, I would like to thank my family. To my parents and my sister, for showing me the way and for always believing in me, even in the moments when I stopped doing so myself. And to my grandmother, who taught me many of my values and who could not see this result. I take comfort in knowing that time is merely an illusion.

\chapter*{Abstract}
\addcontentsline{toc}{chapter}{Abstract}

Implied volatility surfaces are fundamental to the pricing of financial derivatives and to risk management. In practice, however, they are often incomplete due to low liquidity in certain strike and maturity regions. In this thesis, the problem of reconstructing a complete volatility surface from partial observations is formulated as a two-dimensional signal reconstruction problem, and a systematic comparison is conducted between six deep learning architectures (MLP, CNN, U-Net, Transformer, and two variants of variational autoencoders) and the industry-standard Stochastic Volatility Inspired (SVI) parameterization.

All neural models are calibrated to have a similar number of parameters in order to isolate the effect of the structural properties of each architecture. An encoder-decoder Transformer architecture with coordinate-based Fourier positional encoding is proposed. The results show that the Transformer and the U-Net achieve the highest reconstruction accuracy under standard conditions, but the Transformer stands out for its robustness when data are missing in specific regions of the surface and under extreme sparsity scenarios.

Calendar spread and butterfly convexity no-arbitrage conditions are formulated as differentiable penalty terms in the loss function, and the relationship between accuracy and arbitrage is analyzed for different penalty levels. On synthetic data, one of the central findings is that, for convolutional architectures, these constraints act as a form of regularization that simultaneously improves reconstruction accuracy and reduces arbitrage violations, rather than creating a trade-off between the two objectives.

All models are validated on 3,900 real SPY option surfaces from 2008 to 2025. The transfer learning analysis shows that pretraining on synthetic data improves the accuracy of the Transformer, but not that of the convolutional models, while consistently reducing arbitrage violation rates across all architectures. This suggests that structural properties learned from synthetic data transfer effectively to real markets.

The source code is available at \url{https://github.com/pablisho/vol-surface-reconstruction}.

\vfill

\tableofcontents
\listoffigures
\listoftables

\chapter{Introduction}
\label{ch:intro}
\input{chapters/01_introduction}

\chapter{Financial and Mathematical Background}
\label{ch:finance}
\input{chapters/02_finance_background}

\chapter{Deep Learning Background}
\label{ch:dl}
\input{chapters/03_dl_background}

\chapter{Data and Surface Representation}
\label{ch:data}
\input{chapters/04_data}

\chapter{Methodology}
\label{ch:methodology}
\input{chapters/05_methodology}

\chapter{Experiments and Results}
\label{ch:results}
\input{chapters/06_results}

\chapter{No-Arbitrage Analysis}
\label{ch:arbitrage}
\input{chapters/07_arbitrage}

\chapter{Conclusion}
\label{ch:conclusion}
\input{chapters/08_conclusion}

\bibliographystyle{plain}
\bibliography{references}

\end{document}

%% file: figures/architectures.tex
\definecolor{inputcolor}{HTML}{4ECDC4}
\definecolor{hiddencolor}{HTML}{45B7D1}
\definecolor{outputcolor}{HTML}{F7DC6F}
\definecolor{convcolor}{HTML}{5DADE2}
\definecolor{poolcolor}{HTML}{E74C3C}
\definecolor{skipcolor}{HTML}{2ECC71}
\definecolor{bottleneckcolor}{HTML}{9B59B6}
\definecolor{encodercolor}{HTML}{3498DB}
\definecolor{decodercolor}{HTML}{E67E22}
\definecolor{latentcolor}{HTML}{1ABC9C}
\definecolor{attentioncolor}{HTML}{FF6B6B}
\definecolor{ffncolor}{HTML}{A29BFE}
\definecolor{poscolor}{HTML}{FDCB6E}

\newcommand{\mlpdiagram}{%
\begin{tikzpicture}[
    node distance=1.2cm,
    layer/.style={rectangle, draw, minimum width=1.8cm, minimum height=0.7cm, font=\small, rounded corners=2pt},
    arrow/.style={-{Stealth[length=2mm]}, thick},
    label/.style={font=\scriptsize, text=gray}
]
    \node[layer, fill=inputcolor!30] (input) {Input};
    \node[label, right=0.3cm of input] {$(B, 2, 8, 25)$};

    \node[layer, fill=inputcolor!15, below=0.6cm of input] (flat) {Flatten};
    \node[label, right=0.3cm of flat] {$(B, 400)$};

    \node[layer, fill=hiddencolor!30, below=0.6cm of flat] (h1) {Linear + ReLU};
    \node[label, right=0.3cm of h1] {$(B, 256)$};

    \node[layer, fill=hiddencolor!30, below=0.6cm of h1] (h2) {Linear + ReLU};
    \node[label, right=0.3cm of h2] {$(B, 256)$};

    \node[layer, fill=hiddencolor!30, below=0.6cm of h2] (h3) {Linear + ReLU};
    \node[label, right=0.3cm of h3] {$(B, 256)$};

    \node[layer, fill=outputcolor!50, below=0.6cm of h3] (out) {Linear};
    \node[label, right=0.3cm of out] {$(B, 200)$};

    \node[layer, fill=outputcolor!30, below=0.6cm of out] (reshape) {Reshape};
    \node[label, right=0.3cm of reshape] {$(B, 1, 8, 25)$};

    \foreach \a/\b in {input/flat, flat/h1, h1/h2, h2/h3, h3/out, out/reshape}
        \draw[arrow] (\a) -- (\b);
\end{tikzpicture}%
}

\newcommand{\cnndiagram}{%
\begin{tikzpicture}[
    node distance=0.8cm,
    conv/.style={rectangle, draw, minimum width=3.5cm, minimum height=0.7cm, font=\small, rounded corners=2pt, fill=convcolor!25},
    io/.style={rectangle, draw, minimum width=3.5cm, minimum height=0.7cm, font=\small, rounded corners=2pt},
    arrow/.style={-{Stealth[length=2mm]}, thick},
    label/.style={font=\scriptsize, text=gray}
]
    \node[io, fill=inputcolor!30] (input) {Input $(2, 8, 25)$};

    \node[conv, below=0.5cm of input] (c1) {Conv2d $3{\times}3$ + ReLU};
    \node[label, right=0.2cm of c1] {$(104, 8, 25)$};

    \node[conv, below=0.5cm of c1] (c2) {Conv2d $3{\times}3$ + ReLU};
    \node[label, right=0.2cm of c2] {$(104, 8, 25)$};

    \node[conv, below=0.5cm of c2] (c3) {Conv2d $3{\times}3$ + ReLU};
    \node[label, right=0.2cm of c3] {$(104, 8, 25)$};

    \node[conv, below=0.5cm of c3] (c4) {Conv2d $3{\times}3$ + ReLU};
    \node[label, right=0.2cm of c4] {$(104, 8, 25)$};

    \node[io, fill=outputcolor!40, below=0.5cm of c4] (c5) {Conv2d $3{\times}3$};
    \node[label, right=0.2cm of c5] {$(1, 8, 25)$};

    \foreach \a/\b in {input/c1, c1/c2, c2/c3, c3/c4, c4/c5}
        \draw[arrow] (\a) -- (\b);

    \draw[decorate, decoration={brace, amplitude=5pt, mirror}, thick, gray]
        ([xshift=-0.3cm]c1.south west) -- ([xshift=-0.3cm]c4.south west)
        node[midway, left=0.3cm, font=\scriptsize, text=gray, align=center] {$11{\times}11$\\receptive\\field};
\end{tikzpicture}%
}

\newcommand{\unetdiagram}{%
\begin{tikzpicture}[
    node distance=0.6cm and 1.5cm,
    block/.style={rectangle, draw, minimum width=2.2cm, minimum height=0.6cm, font=\scriptsize, rounded corners=2pt},
    op/.style={circle, draw, inner sep=1pt, font=\scriptsize},
    arrow/.style={-{Stealth[length=2mm]}, thick},
    skip/.style={-{Stealth[length=2mm]}, thick, dashed, color=skipcolor!70!black},
    label/.style={font=\tiny, text=gray}
]
    \node[block, fill=encodercolor!20] (e1) {ConvBlock$(2, 24)$};
    \node[label, right=0.1cm of e1] {$(24, 8, 25)$};

    \node[block, fill=poolcolor!20, below=0.4cm of e1] (p1) {MaxPool $2{\times}2$};
    \node[label, right=0.1cm of p1] {$(24, 4, 12)$};

    \node[block, fill=encodercolor!20, below=0.4cm of p1] (e2) {ConvBlock$(24, 48)$};
    \node[label, right=0.1cm of e2] {$(48, 4, 12)$};

    \node[block, fill=poolcolor!20, below=0.4cm of e2] (p2) {MaxPool $2{\times}2$};
    \node[label, right=0.1cm of p2] {$(48, 2, 6)$};

    \node[block, fill=bottleneckcolor!25, below=0.4cm of p2] (bn) {ConvBlock$(48, 96)$};
    \node[label, right=0.1cm of bn] {$(96, 2, 6)$};

    \node[block, fill=decodercolor!20, below=0.4cm of bn] (u1) {Upsample + Concat};
    \node[label, right=0.1cm of u1] {$(144, 4, 12)$};

    \node[block, fill=decodercolor!20, below=0.4cm of u1] (d1) {ConvBlock$(144, 48)$};
    \node[label, right=0.1cm of d1] {$(48, 4, 12)$};

    \node[block, fill=decodercolor!20, below=0.4cm of d1] (u2) {Upsample + Concat};
    \node[label, right=0.1cm of u2] {$(72, 8, 25)$};

    \node[block, fill=decodercolor!20, below=0.4cm of u2] (d2) {ConvBlock$(72, 24)$};
    \node[label, right=0.1cm of d2] {$(24, 8, 25)$};

    \node[block, fill=outputcolor!40, below=0.4cm of d2] (out) {Conv $1{\times}1$};
    \node[label, right=0.1cm of out] {$(1, 8, 25)$};

    \foreach \a/\b in {e1/p1, p1/e2, e2/p2, p2/bn, bn/u1, u1/d1, d1/u2, u2/d2, d2/out}
        \draw[arrow] (\a) -- (\b);

    \draw[skip] (e2.west) -- +(-0.8,0) |- (u1.west)
        node[pos=0.25, left, font=\tiny, color=skipcolor!70!black] {skip};
    \draw[skip] (e1.west) -- +(-1.2,0) |- (u2.west)
        node[pos=0.15, left, font=\tiny, color=skipcolor!70!black] {skip};
\end{tikzpicture}%
}

\newcommand{\transformerdiagram}{%
\begin{tikzpicture}[
    node distance=0.5cm and 2.5cm,
    block/.style={rectangle, draw, minimum width=2.8cm, minimum height=0.6cm, font=\scriptsize, rounded corners=2pt},
    smallblock/.style={rectangle, draw, minimum width=2.4cm, minimum height=0.5cm, font=\tiny, rounded corners=2pt},
    arrow/.style={-{Stealth[length=2mm]}, thick},
    crossarrow/.style={-{Stealth[length=2mm]}, thick, color=attentioncolor!70!black},
    label/.style={font=\tiny, text=gray},
    grouplabel/.style={font=\small\bfseries, text=gray}
]
    \node[block, fill=poscolor!30] (coords) {Grid coords $(\tau, m)$};
    \node[block, fill=poscolor!20, below=0.4cm of coords] (fourier) {Fourier encoding $\gamma$};
    \node[label, below=0.1cm of fourier] {$L{=}8 \to 34$ features};

    \draw[arrow] (coords) -- (fourier);

    \node[block, fill=encodercolor!20, right=2.5cm of coords] (enc_in) {Observed IVs + $\gamma(\tau,m)$};
    \node[label, right=0.1cm of enc_in] {$(N_{\text{obs}}, 35)$};

    \node[block, fill=encodercolor!15, below=0.4cm of enc_in] (enc_proj) {Linear proj.};
    \node[label, right=0.1cm of enc_proj] {$(N_{\text{obs}}, 64)$};

    \node[smallblock, fill=attentioncolor!15, below=0.35cm of enc_proj] (enc_sa1) {Self-Attention (4 heads)};
    \node[smallblock, fill=ffncolor!15, below=0.25cm of enc_sa1] (enc_ff1) {FFN $(64 \to 256 \to 64)$};

    \node[label, left=0.1cm of enc_sa1] {$\times 3$};

    \begin{scope}[on background layer]
        \node[fit=(enc_sa1)(enc_ff1), draw=encodercolor, dashed, rounded corners=3pt, inner sep=3pt] (enc_layer) {};
    \end{scope}

    \node[block, fill=encodercolor!30, below=0.5cm of enc_ff1] (memory) {Memory};
    \node[label, right=0.1cm of memory] {$(N_{\text{obs}}, 64)$};

    \draw[arrow] (enc_in) -- (enc_proj);
    \draw[arrow] (enc_proj) -- (enc_sa1);
    \draw[arrow] (enc_sa1) -- (enc_ff1);
    \draw[arrow] (enc_ff1) -- (memory);
    \draw[arrow] (fourier.east) -- ++(0.5, 0) |- (enc_in.west);

    \node[block, fill=decodercolor!20, right=2.5cm of enc_in] (dec_in) {All IVs (0 if missing) + $\gamma(\tau,m)$};
    \node[label, right=0.1cm of dec_in] {$(200, 35)$};

    \node[block, fill=decodercolor!15, below=0.4cm of dec_in] (dec_proj) {Linear proj.};
    \node[label, right=0.1cm of dec_proj] {$(200, 64)$};

    \node[smallblock, fill=attentioncolor!15, below=0.35cm of dec_proj] (dec_sa) {Self-Attention (4 heads)};
    \node[smallblock, fill=attentioncolor!30, below=0.25cm of dec_sa] (dec_ca) {Cross-Attention};
    \node[smallblock, fill=ffncolor!15, below=0.25cm of dec_ca] (dec_ff) {FFN $(64 \to 256 \to 64)$};

    \node[label, left=0.1cm of dec_sa] {$\times 2$};

    \begin{scope}[on background layer]
        \node[fit=(dec_sa)(dec_ca)(dec_ff), draw=decodercolor, dashed, rounded corners=3pt, inner sep=3pt] (dec_layer) {};
    \end{scope}

    \node[block, fill=outputcolor!40, below=0.5cm of dec_ff] (out_proj) {Linear $(64 \to 1)$};
    \node[label, right=0.1cm of out_proj] {$(200, 1)$};

    \node[block, fill=outputcolor!30, below=0.4cm of out_proj] (reshape) {Reshape};
    \node[label, right=0.1cm of reshape] {$(1, 8, 25)$};

    \draw[arrow] (dec_in) -- (dec_proj);
    \draw[arrow] (dec_proj) -- (dec_sa);
    \draw[arrow] (dec_sa) -- (dec_ca);
    \draw[arrow] (dec_ca) -- (dec_ff);
    \draw[arrow] (dec_ff) -- (out_proj);
    \draw[arrow] (out_proj) -- (reshape);

    \draw[crossarrow] (memory.east) -- ++(0.8, 0) |- (dec_ca.west)
        node[pos=0.3, above, font=\tiny, color=attentioncolor!70!black] {cross-attn};

    \node[grouplabel, above=0.2cm of enc_in] {Encoder};
    \node[grouplabel, above=0.2cm of dec_in] {Decoder};
\end{tikzpicture}%
}

\newcommand{\vaediagram}{%
\begin{tikzpicture}[
    node distance=0.5cm and 1.5cm,
    block/.style={rectangle, draw, minimum width=2.2cm, minimum height=0.55cm, font=\scriptsize, rounded corners=2pt},
    arrow/.style={-{Stealth[length=2mm]}, thick},
    label/.style={font=\tiny, text=gray},
    title/.style={font=\small\bfseries}
]
    \node[title] (fctitle) {FC VAE};

    \node[block, fill=inputcolor!30, below=0.3cm of fctitle] (fc_in) {Flatten};
    \node[label, right=0.1cm of fc_in] {$400$};

    \node[block, fill=encodercolor!20, below=0.35cm of fc_in] (fc_e1) {Linear+ELU};
    \node[label, right=0.1cm of fc_e1] {$288$};

    \node[block, fill=encodercolor!20, below=0.35cm of fc_e1] (fc_e2) {Linear+ELU};
    \node[label, right=0.1cm of fc_e2] {$144$};

    \node[block, fill=encodercolor!20, below=0.35cm of fc_e2] (fc_e3) {Linear+ELU};
    \node[label, right=0.1cm of fc_e3] {$72$};

    \node[block, fill=latentcolor!30, below=0.35cm of fc_e3] (fc_mu) {$\mu$, $\log\sigma^2$};
    \node[label, right=0.1cm of fc_mu] {$d_z{=}32$};

    \node[block, fill=latentcolor!50, below=0.35cm of fc_mu] (fc_z) {$\mathbf{z} = \mu + \sigma \odot \epsilon$};

    \node[block, fill=decodercolor!20, below=0.35cm of fc_z] (fc_d1) {Linear+ELU};
    \node[label, right=0.1cm of fc_d1] {$72$};

    \node[block, fill=decodercolor!20, below=0.35cm of fc_d1] (fc_d2) {Linear+ELU};
    \node[label, right=0.1cm of fc_d2] {$144$};

    \node[block, fill=decodercolor!20, below=0.35cm of fc_d2] (fc_d3) {Linear+ELU};
    \node[label, right=0.1cm of fc_d3] {$288$};

    \node[block, fill=outputcolor!40, below=0.35cm of fc_d3] (fc_out) {Linear + Reshape};
    \node[label, right=0.1cm of fc_out] {$(1,8,25)$};

    \foreach \a/\b in {fc_in/fc_e1, fc_e1/fc_e2, fc_e2/fc_e3, fc_e3/fc_mu, fc_mu/fc_z, fc_z/fc_d1, fc_d1/fc_d2, fc_d2/fc_d3, fc_d3/fc_out}
        \draw[arrow] (\a) -- (\b);

    \node[title, right=4cm of fctitle] (cvtitle) {Conv VAE};

    \node[block, fill=inputcolor!30, below=0.3cm of cvtitle] (cv_in) {Input};
    \node[label, right=0.1cm of cv_in] {$(2,8,25)$};

    \node[block, fill=encodercolor!20, below=0.35cm of cv_in] (cv_e1) {Conv+ReLU};
    \node[label, right=0.1cm of cv_e1] {$(32,8,25)$};

    \node[block, fill=encodercolor!20, below=0.35cm of cv_e1] (cv_e2) {Conv+ReLU, s=2};
    \node[label, right=0.1cm of cv_e2] {$(64,4,13)$};

    \node[block, fill=encodercolor!20, below=0.35cm of cv_e2] (cv_e3) {Conv+ReLU, s=2};
    \node[label, right=0.1cm of cv_e3] {$(128,2,7)$};

    \node[block, fill=latentcolor!30, below=0.35cm of cv_e3] (cv_mu) {Flatten $\to$ $\mu$, $\log\sigma^2$};
    \node[label, right=0.1cm of cv_mu] {$d_z{=}16$};

    \node[block, fill=latentcolor!50, below=0.35cm of cv_mu] (cv_z) {$\mathbf{z} = \mu + \sigma \odot \epsilon$};

    \node[block, fill=decodercolor!20, below=0.35cm of cv_z] (cv_d0) {Linear + Reshape};
    \node[label, right=0.1cm of cv_d0] {$(128,2,7)$};

    \node[block, fill=decodercolor!20, below=0.35cm of cv_d0] (cv_d1) {Upsample + Conv};
    \node[label, right=0.1cm of cv_d1] {$(64,4,13)$};

    \node[block, fill=decodercolor!20, below=0.35cm of cv_d1] (cv_d2) {Upsample + Conv};
    \node[label, right=0.1cm of cv_d2] {$(32,8,25)$};

    \node[block, fill=outputcolor!40, below=0.35cm of cv_d2] (cv_out) {Conv $3{\times}3$};
    \node[label, right=0.1cm of cv_out] {$(1,8,25)$};

    \foreach \a/\b in {cv_in/cv_e1, cv_e1/cv_e2, cv_e2/cv_e3, cv_e3/cv_mu, cv_mu/cv_z, cv_z/cv_d0, cv_d0/cv_d1, cv_d1/cv_d2, cv_d2/cv_out}
        \draw[arrow] (\a) -- (\b);

    \node[block, fill=attentioncolor!15, below=0.5cm of fc_out, minimum width=5cm] (opt) {Inference: optimize $\mathbf{z}$ (200 Adam steps on observed MSE)};
    \draw[arrow, dashed] (fc_out.south) -- ++(0,-0.2) -| (opt.north);

\end{tikzpicture}%
}

%% file: chapters/01_introduction.tex

This chapter presents the motivation for this work: volatility surfaces are essential for the pricing of financial derivatives and risk management, but in practice they are often incomplete. The central problem is formulated as the reconstruction of a two-dimensional signal from partial observations. Previous approaches are then reviewed, and the contributions of this thesis are positioned relative to existing work. Finally, the structure of the remaining chapters is described.

\section{Motivation}
Volatility surfaces are central to the pricing of financial derivatives and to risk management. They arise from the observation that implied volatilities vary across strike and maturity, a phenomenon that contradicts the constant-volatility assumption of the Black-Scholes model \cite{black1973pricing, merton1973theory}. In the financial industry, significant resources are devoted to constructing up-to-date volatility surfaces, since accurate pricing depends on having a complete surface for a given asset at each point in time.

In practice, volatility surfaces observed in the market are both noisy and incomplete, particularly at the extremes of the surface (strikes far from the current price) and at short maturities, where liquidity is low. However, for a volatility surface to be useful, it must be complete, smooth, and free of arbitrage; that is, it must satisfy certain structural constraints imposed by the absence of riskless profit opportunities.

In this thesis, we formulate the problem of obtaining a complete volatility surface from incomplete observations as the reconstruction of a two-dimensional signal from partial inputs. This formulation is analogous to image inpainting, where missing regions of an image are filled by exploiting spatial structure, and to magnetic resonance image reconstruction, where complete images are recovered from undersampled measurements by leveraging anatomical regularity.

\begin{figure}[ht]
    \centering
    \includegraphics[width=0.7\textwidth]{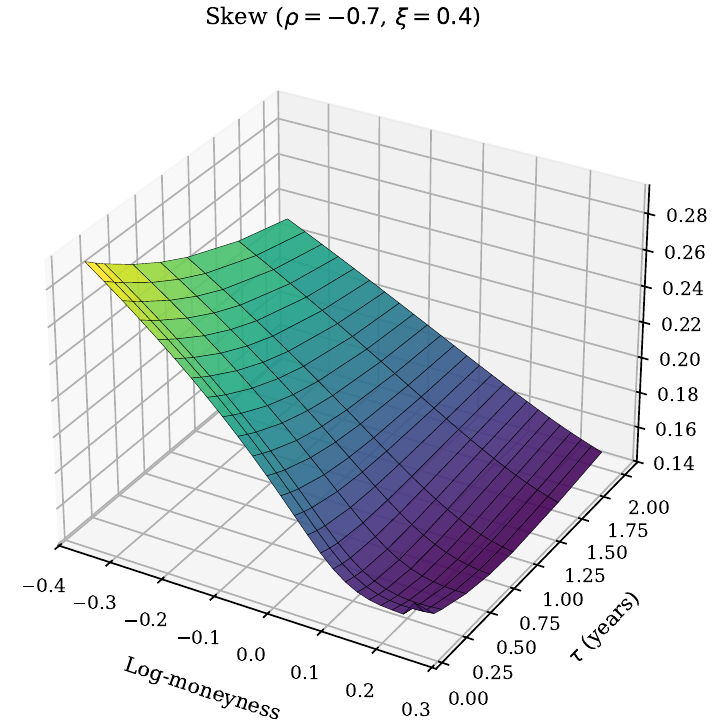}
    \caption[Heston implied volatility surface]{Example implied volatility surface generated with the Heston stochastic volatility model. The surface exhibits the characteristic smirk across strikes and a term structure across maturities.}
    \label{fig:vol-surface-intro}
\end{figure}

The industry has traditionally relied on parametric methods to construct volatility surfaces, with the Stochastic Volatility Inspired (SVI) parameterization \cite{gatheral2014svi} being the most widely adopted. More recently, machine learning techniques have been explored for this and related problems. Parametric methods can impose arbitrage-free surfaces by construction, but at the cost of limited reconstruction accuracy, particularly when the observed data are sparse. Machine learning methods have shown better reconstruction accuracy, but may introduce undesired arbitrage violations. In addition, the field of \emph{deep learning} has advanced rapidly, with new architectures and training procedures offering opportunities that have not yet been systematically evaluated for this task.

Despite these advances, no prior work has systematically compared \emph{deep learning} architectures for volatility surface reconstruction under controlled conditions, nor studied the effect of incorporating no-arbitrage constraints directly into the training objective. In particular, the Transformer architecture \cite{vaswani2017attention}, which has driven state-of-the-art results in natural language processing and \emph{computer vision}, has not been applied to this problem. This thesis seeks to address these gaps.

\section{Problem Statement}
The central problem addressed in this thesis is the reconstruction of a complete implied volatility surface from a partial observation. Formally, the input consists of a two-channel tensor: an $8 \times 25$ grid of implied volatility values (organized into 8 maturities and 25 strikes in log-moneyness), together with a binary mask indicating which points were observed. The output is a complete $8 \times 25$ surface of implied volatility values.

\begin{figure}[ht]
    \centering
    \includegraphics[width=0.9\textwidth]{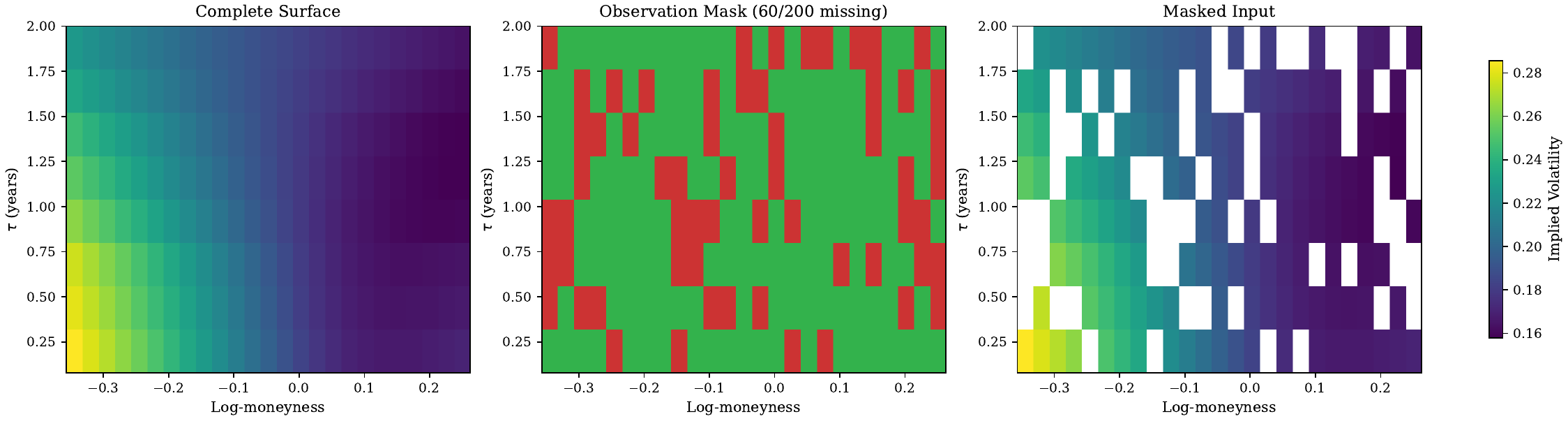}
    \caption[Illustration of the reconstruction task]{Illustration of the reconstruction task. Left: a complete volatility surface. Right: the same surface with 30\% of the points masked (missing). The objective is to recover the complete surface from the partial observation.}
    \label{fig:masking-intro}
\end{figure}

The main objective is to minimize the reconstruction error on the missing points, measured by the root mean squared error (RMSE). Beyond accuracy, we also investigate whether the reconstructed surfaces can satisfy no-arbitrage conditions: total implied variance must be non-decreasing in maturity (calendar spread condition), and implied variance must be convex in log-moneyness for each maturity (butterfly condition). All architectures are first evaluated in terms of reconstruction accuracy (Chapter~\ref{ch:results}), and the effect of incorporating these constraints as differentiable penalties in the loss function is then studied (Chapter~\ref{ch:arbitrage}).

This formulation establishes a direct connection with masked prediction tasks in other domains. In \emph{computer vision}, masked autoencoders \cite{he2022mae} reconstruct images from randomly masked patches by learning spatial structure. In tabular data, similar approaches have been applied to impute missing values \cite{du2024remasker}. Volatility surface reconstruction lies between these two settings: the data are arranged on a regular 2D grid (as in an image), but they also possess domain-specific structural constraints that generic inpainting methods do not impose.

\section{Related Work}
\label{sec:related-work}

\subsection{Parametric Volatility Surface Fitting}
The most widely used approach for constructing volatility surfaces is the Stochastic Volatility Inspired (SVI) parameterization introduced by Gatheral and Jacquier \cite{gatheral2014svi}, which fits a five-parameter function to each maturity slice independently. The SVI formula expresses total implied variance as a function of log-moneyness using five parameters that control the level, slope, curvature, and skew of each volatility smile. Extensions such as the Surface SVI (SSVI) parameterization impose consistency across slices by restricting the parameters to vary smoothly across maturities, reducing the risk of calendar spread arbitrage.

These methods are interpretable and, when properly constrained, can guarantee arbitrage-free surfaces analytically. However, they operate on each maturity slice in isolation: information from one maturity cannot inform the fit at another. This independence becomes a critical limitation when the observed data are sparse, since entire slices may have very few observed points.

\subsection{Neural Networks for Volatility Surfaces}
Several recent works have applied \emph{deep learning} to volatility modeling. Bayer and Stemper \cite{bayer2019deep} used neural networks for fast calibration of stochastic volatility models, training networks to map observed option prices to model parameters. This approach accelerates calibration but does not address surface reconstruction from incomplete data. Horvath et al.\ \cite{horvath2021deep} explored the use of deep neural networks for pricing and calibration of stochastic volatility models, learning continuous representations of implied volatility. Ackerer et al.\ \cite{ackerer2020deep} applied deep learning to smooth option prices directly, using neural networks as flexible interpolators of the pricing function. More recently, Zhang et al.\ \cite{zhang2025volnp} introduced VolNP, an attention-based neural process for implied volatility fitting that uses meta-learning to adapt to new surfaces with few observations.

The most closely related line of work uses variational autoencoders (VAEs) for volatility surface modeling. Bergeron et al.\ \cite{bergeron2022vae} proposed a VAE-based approach in which the model first learns a latent representation of complete volatility surfaces and then reconstructs incomplete surfaces by optimizing the latent vector so that it matches the observed points at inference time. Ning et al.\ \cite{ning2023arbfree} extended this approach to impose arbitrage-free surfaces by incorporating soft constraints into VAE training, demonstrating that the generative framework can produce surfaces consistent with no-arbitrage conditions. Nteumagn\'e et al.\ \cite{nteumagnevaecomplete2025} evaluated VAE-based surface reconstruction, comparing against thin-plate spline interpolation, SABR, SVI, and deterministic autoencoders, and reported good performance even with very few observed data points.

These VAE-based approaches directly address the surface reconstruction task, but they focus exclusively on the VAE architecture. There is no prior work that presents a controlled comparison across multiple families of \emph{deep learning} architectures for this task, nor one that evaluates non-generative architectures such as CNNs, U-Nets, or Transformers under the same conditions.

\subsection{Positioning Relative to Prior Work}

Compared with the most closely related prior work, this thesis differs in three key respects. First, while existing studies evaluate a single architecture family (typically VAEs \cite{bergeron2022vae, ning2023arbfree, nteumagnevaecomplete2025} or attention-based neural processes \cite{zhang2025volnp}), we perform a controlled comparison across six neural architectures and SVI, all matched in approximate parameter count, isolating the effect of architectural design rather than model capacity. Second, we integrate differentiable no-arbitrage penalties directly into the loss function and systematically study the trade-off between accuracy and arbitrage through Pareto analysis, instead of checking compliance after the fact or relying on architectural guarantees. Third, we evaluate robustness under both random and structured missingness patterns across a broad range of missing-data percentages (10\%--90\%), providing a more complete picture of practical performance than evaluations under a single condition.

\subsection{Image Inpainting and Masked Prediction}
The problem of reconstructing missing regions from partial observations has a long history in \emph{computer vision}. Classical inpainting methods such as the work of Bertalm\'io et al.\ \cite{bertalmio2000image} propagate information from the boundary of known regions into missing areas using partial differential equations. More recently, \emph{deep learning}-based approaches have achieved remarkable results: convolutional networks learn to fill large missing regions by exploiting learned prior knowledge about natural images.

A particularly relevant development is the Masked Autoencoder (MAE) framework of He et al.\ \cite{he2022mae}, which trains Vision Transformers \cite{vaswani2017attention} by randomly masking image patches and reconstructing the missing content. This self-supervised approach learns rich visual representations by solving a reconstruction task that is structurally similar to volatility surface completion. Du et al.\ \cite{du2024remasker} extended this masked autoencoding paradigm to tabular data with ReMasker, demonstrating that the approach generalizes beyond images to structured data with missing values.

However, these methods are designed for generic data and do not incorporate domain-specific structural constraints. Volatility surfaces require not only visual plausibility but also quantitative accuracy and consistency with no-arbitrage conditions, which standard inpainting approaches do not guarantee.

\subsection{Physics-Informed and Constrained Learning}
The idea of incorporating domain knowledge as differentiable penalties in the loss function has been widely studied in numerical simulation and the solution of differential equations. The most prominent example is Physics-Informed Neural Networks (PINNs), introduced by Raissi et al.\ \cite{raissi2019pinn}, which impose partial differential equation constraints by adding residual terms to the loss function. Rather than requiring training data that satisfy the PDE, PINNs learn solutions that approximately satisfy the physics through soft penalties, allowing the network to balance data fidelity against physical consistency.

The same paradigm can be applied to financial constraints: no-arbitrage conditions such as calendar spread and butterfly convexity can be expressed as differentiable functions of the model output and added to the loss function with a tunable penalty strength. This enables a controlled study of the relationship between reconstruction accuracy and compliance with arbitrage conditions, an approach that has not been systematically explored for volatility surface reconstruction.

\subsection{Transfer Learning in Finance}
Transfer from synthetic to real data, often called sim-to-real transfer, is a well-established technique in robotics and \emph{computer vision}, where models are pretrained on large amounts of simulated data before being fine-tuned on smaller real datasets. The simulated environment provides abundant and perfectly labeled data, while fine-tuning closes the gap between simulation and reality.

This approach has been less explored in quantitative finance, despite the availability of realistic simulators such as calibrated stochastic volatility models. The Heston model \cite{heston1993closed}, for example, can generate an unlimited number of synthetic volatility surfaces with known ground-truth values, making it a natural candidate for pretraining. Whether the structural properties learned from these synthetic data transfer effectively to real market surfaces remained an open question.

\section{Contributions}
The main contributions of this thesis are the following:

\begin{enumerate}
    \item We propose an encoder-decoder Transformer architecture with coordinate-based Fourier positional encoding \cite{tancik2020fourier} for volatility surface reconstruction. An ablation study confirms a significant improvement in accuracy (18\%) relative to learnable positional embeddings, validating the design choice. To the best of our knowledge, this is the first application of this architecture to this specific problem.

    \item We conduct a controlled comparative study of six \emph{deep learning} architectures (MLP, CNN, U-Net \cite{ronneberger2015unet}, Transformer \cite{vaswani2017attention}, FC-VAE, Conv-VAE \cite{kingma2014vae}) and the SVI parametric baseline \cite{gatheral2014svi}, all matched at approximately 288,000 parameters. Multi-seed experiments validate the stability of the ranking: the Transformer and U-Net are statistically tied in reconstruction accuracy, with CNN and MLP clearly behind.

    \item We integrate differentiable no-arbitrage penalties (calendar spread and butterfly convexity) into the loss function and perform a full Pareto analysis for different penalty strengths. We find that, for convolutional architectures, the constraints act as regularizers that maintain or improve reconstruction accuracy while substantially reducing arbitrage violations.

    \item We evaluate robustness under both random and structured (wing) missingness patterns, varying the missing-data rate from 10\% to 90\% without retraining. The Transformer achieves $9\times$ lower error than SVI with 90\% missing data and is almost unaffected by structured wing masking (less than 20\% degradation), while convolutional models suffer accuracy losses of 30--47\% when wing information is absent.

    \item We validate all models on real SPY option data comprising 3,900 surfaces between 2008 and 2025, and analyze transfer learning from synthetic to real data. We find that training from scratch outperforms fine-tuning for convolutional models in accuracy, while fine-tuned models exhibit lower arbitrage rates, suggesting that synthetic pretraining teaches smoother surface representations.
\end{enumerate}

\section{Thesis Outline}
The remainder of this thesis is organized as follows.

Chapter~\ref{ch:finance} presents the financial and mathematical background, including the Black-Scholes model, implied volatility, volatility surface construction, the Heston stochastic volatility model, no-arbitrage conditions, and the SVI parameterization.

Chapter~\ref{ch:dl} introduces the foundations of \emph{deep learning}, including the architectures used in this work (MLP, CNN, U-Net, Transformer, VAE), training procedures, and the connection with masked prediction.

Chapter~\ref{ch:data} describes the data and surface representation, including grid discretization, the synthetic dataset generated from the Heston model, the real dataset constructed from SPY options, and the masking strategy used to simulate incomplete observations.

Chapter~\ref{ch:methodology} details the methodology: the formal problem formulation, the seven model architectures and their implementation, the training protocol, the no-arbitrage constraint formulation, and the evaluation metrics.

Chapter~\ref{ch:results} presents the experimental results, including baseline comparisons on synthetic data, robustness to different levels of missing data, regional error analysis, real-data experiments, transfer learning, attention visualization, and computational benchmarks.

Chapter~\ref{ch:arbitrage} presents a dedicated analysis of no-arbitrage constraints, including the impact of constraint strength, lambda sweep experiments, and Pareto frontiers between accuracy and arbitrage compliance.

Chapter~\ref{ch:conclusion} concludes with a summary of contributions, discussion of limitations, and possible directions for future work.

All source code, including the models, experiments, and data generation scripts, is available at \url{https://github.com/pablisho/vol-surface-reconstruction}.

%% file: chapters/02_finance_background.tex

This chapter introduces the financial and mathematical concepts that support volatility surface reconstruction. It begins with the foundations of option pricing, introduces implied volatility and the volatility surface, describes the Heston stochastic volatility model used for synthetic data generation, formalizes the no-arbitrage conditions that constrain valid surfaces, and presents the SVI parameterization used as a benchmark.

\section{Option Pricing Fundamentals}
\label{sec:option-pricing}

\subsection{European Options}

A European option is a financial contract that gives its holder the right, but not the obligation, to buy (call) or sell (put) an underlying asset at a predetermined exercise price (\emph{strike}) $K$ on a specific expiration date $T$. The payoffs at expiration are:
\begin{align}
\text{Call payoff} &= \max(S_T - K, \; 0) \label{eq:call-payoff} \\
\text{Put payoff} &= \max(K - S_T, \; 0) \label{eq:put-payoff}
\end{align}
where $S_T$ is the price of the underlying asset at expiration. The holder exercises the option only when doing so is profitable, limiting the maximum loss to the premium paid for the contract.

\subsection{Risk-Neutral Pricing}

Under the no-arbitrage principle, the fundamental theorem of asset pricing \cite{harrison1979martingales} guarantees the existence of a probability measure $\mathbb{Q}$, called the risk-neutral measure, under which discounted asset prices are martingales; that is, their expected future value equals their current value, with no predictable trend. This measure does not need to be constructed explicitly: it is implicitly defined by the prices observed in the market. Under $\mathbb{Q}$, the price of any financial derivative can be expressed as the discounted expected value of its payoff:
\begin{equation}
V_0 = e^{-r\tau} \, \mathbb{E}^{\mathbb{Q}}\!\left[\text{payoff}(S_T)\right]
\label{eq:risk-neutral}
\end{equation}
where $r$ is the risk-free rate and $\tau = T - t$ is the time to maturity.

\subsection{Put-Call Parity}

European call and put prices on the same underlying, strike, and maturity are linked by put-call parity:
\begin{equation}
C - P = e^{-q\tau} S_0 - e^{-r\tau} K
\label{eq:put-call-parity}
\end{equation}
where $q$ is the continuous dividend yield. This identity follows directly from the no-arbitrage principle: buying a call option and selling a put option with the same strike and maturity replicates a forward contract. Put-call parity implies that any information contained in put prices is also contained in call prices, and vice versa. In practice, \emph{out-of-the-money} (OTM) options are used for each strike (puts for $K < S$, calls for $K \geq S$) because they are more liquid and have tighter bid-ask spreads.

\section{The Black--Scholes--Merton Model}
\label{sec:bsm}

\subsection{Model Assumptions}

The Black--Scholes--Merton (BSM) model \cite{black1973pricing, merton1973theory} derives a closed-form option pricing formula under the following assumptions:
\begin{enumerate}
    \item The underlying asset follows geometric Brownian motion with \emph{constant} volatility $\sigma$:
    \begin{equation}
    dS_t = (r - q) S_t \, dt + \sigma S_t \, dW_t
    \label{eq:gbm}
    \end{equation}
    \item Markets are frictionless (no transaction costs, continuous trading, unlimited short selling).
    \item The risk-free rate $r$ and dividend yield $q$ are constant.
\end{enumerate}

\subsection{The Black--Scholes Formula}

Under these assumptions, the price of a European call option is:
\begin{equation}
C = e^{-q\tau} S_0 \, \Phi(d_1) - e^{-r\tau} K \, \Phi(d_2)
\label{eq:bs-call}
\end{equation}
where $\Phi(\cdot)$ is the standard normal cumulative distribution function and:
\begin{equation}
d_1 = \frac{\log(S_0/K) + (r - q + \sigma^2/2)\tau}{\sigma\sqrt{\tau}}, \qquad d_2 = d_1 - \sigma\sqrt{\tau}
\label{eq:bs-d1d2}
\end{equation}

\subsection{The Black-76 Forward Pricing Variant}

When pricing options on futures or forwards, the underlying is the forward price $F = S_0 e^{(r-q)\tau}$ rather than the spot price. The Black-76 formula \cite{black1976pricing} simplifies the BSM formula by absorbing the drift into the forward:
\begin{equation}
C = e^{-r\tau}\left[F \, \Phi(d_1) - K \, \Phi(d_2)\right]
\label{eq:black76}
\end{equation}
with:
\begin{equation}
d_1 = \frac{\log(F/K) + \sigma^2 \tau / 2}{\sigma\sqrt{\tau}}, \qquad d_2 = d_1 - \sigma\sqrt{\tau}
\end{equation}

This thesis uses the Black-76 formulation because it provides a clearer separation between the forward price level and the volatility structure. The pricing engine implements Equation~\eqref{eq:black76} using the standard normal distribution function computed through the error function ($\Phi(x) = \frac{1}{2}[1 + \text{erf}(x/\sqrt{2})]$), without external dependencies.

\subsection{Limitations}

The main limitation of the BSM model is the assumption of constant volatility. If this assumption were valid, a single volatility parameter $\sigma$ would correctly price all options on the same underlying, regardless of strike and maturity. In practice, implied volatilities vary systematically across strikes and maturities, producing the \emph{volatility smile} and the surface described in Section~\ref{sec:vol-surface}. This discrepancy motivates the entire field of stochastic volatility modeling and, by extension, the reconstruction problem addressed in this thesis.

\section{Implied Volatility}
\label{sec:implied-vol}

\subsection{Definition}

The implied volatility (IV) $\sigma_{\text{IV}}$ of an option is the value of $\sigma$ that, when substituted into the Black-76 formula, reproduces the observed market price:
\begin{equation}
C_{\text{market}} = C_{\text{Black-76}}(F, K, \tau, r, \sigma_{\text{IV}})
\label{eq:iv-def}
\end{equation}

Implied volatility is not directly observable; it must be recovered numerically by inverting the pricing formula. Since $C_{\text{Black-76}}$ is strictly increasing in $\sigma$ (the vega $\partial C / \partial \sigma > 0$ for all $\sigma > 0$), the inversion has a unique solution for any valid option price.

\subsection{Numerical Inversion}

Two methods are implemented for the inversion:

\paragraph{Bisection.} A bracketing method that maintains an interval $[\sigma_l, \sigma_u]$ containing the solution. At each step, the midpoint is evaluated and the interval is halved. Bisection converges unconditionally but slowly (linear convergence). It is used as a fallback to guarantee robustness.

\paragraph{Newton-Raphson.} Using the derivative of price with respect to volatility (known as vega in options terminology), Newton-Raphson converges quadratically:
\begin{equation}
\sigma_{n+1} = \sigma_n - \frac{C_{\text{Black-76}}(\sigma_n) - C_{\text{market}}}{\text{vega}(\sigma_n)}
\label{eq:newton-iv}
\end{equation}
A bisection bracket is maintained as a safeguard: if the Newton-Raphson steps fall outside the interval or fail to converge within 50 iterations, the solver falls back to pure bisection. The convergence tolerance is $10^{-12}$ in both price space and volatility space.

\subsection{The \emph{Volatility Smile}}

If the constant-volatility assumption of the BSM model were valid, implied volatility would be the same for all strikes at a given maturity. Instead, empirical observations consistently show that implied volatility varies with strike, producing a characteristic pattern known as the \emph{volatility smile} (for symmetric shapes) or \emph{skew} (for asymmetric shapes). In equity markets, the typical pattern is a skew: implied volatility increases for low strikes (out-of-the-money puts) and decreases for high strikes (out-of-the-money calls), reflecting the market's expectation of larger downside moves.

\section{The Volatility Surface}
\label{sec:vol-surface}

\subsection{Surface Definition}

The volatility surface is the two-dimensional function $\sigma_{\text{IV}}(K, \tau)$ that maps the exercise price $K$ and time to maturity $\tau$ to implied volatility. It encodes the market's collective view of how the underlying asset's return distribution varies across strikes and time horizons. Rather than using raw exercise prices, it is standard to express the strike axis in \emph{log-moneyness}:
\begin{equation}
m = \log\left(\frac{K}{F}\right)
\label{eq:logm}
\end{equation}
where $F$ is the forward price. Log-moneyness centers the surface at $m = 0$ (at-the-money, ATM), normalizes across different underlying price levels, and produces approximately symmetric smile shapes for stochastic volatility models. In these coordinates, the surface is expressed as $\sigma_{\text{IV}}(m, \tau)$, which is the representation used throughout this thesis.

\subsection{Empirical Properties}

Volatility surfaces in equity markets exhibit three characteristics:

\begin{enumerate}
    \item \textbf{Smile/skew across strike.} For each maturity, implied volatility varies with strike. The shape is typically a skew in equities (higher volatility at low strikes) and a smile in foreign exchange markets (higher volatility at both extremes). The skew reflects the negative correlation between asset returns and volatility changes, known as the \emph{leverage effect} \cite{christie1982stochastic}.
    
    \item \textbf{Term structure.} At-the-money implied volatility varies with maturity. In calm markets, the term structure is upward sloping (longer-dated options have higher IV, reflecting the accumulation of uncertainty). During crises, the term structure inverts when short-term IV rises sharply.
    
    \item \textbf{Smile flattening.} The smile becomes less pronounced at longer maturities. This is consistent with the central limit theorem: as the time horizon grows, the return distribution converges toward a Gaussian, and the constant-volatility assumption of the BSM model becomes a better approximation.
\end{enumerate}

\subsection{Practical Importance}

The volatility surface is central to the pricing of financial derivatives and to risk management:
\begin{itemize}
    \item \textbf{Pricing.} Exotic options (barriers, Asians, among others) depend on the complete volatility surface, not only on a single implied volatility. A complete and arbitrage-free surface is needed to calibrate the pricing models used for these instruments.
    \item \textbf{Hedging.} Delta and vega hedges are computed from the surface. Errors or gaps in the surface propagate directly into hedging errors.
    \item \textbf{Risk management.} Value-at-risk calculations and stress tests require consistent volatility surfaces across all strikes and maturities. Missing or inconsistent points can lead to an underestimation of risk.
\end{itemize}

\section{Stochastic Volatility: The Heston Model}
\label{sec:heston}

The limitations of the constant-volatility assumption of the BSM model motivate stochastic volatility models, which allow volatility to evolve randomly over time. The Heston model \cite{heston1993closed} is the reference model in this class.

\subsection{Model Specification}

Under the Heston model, the asset price $S_t$ and its instantaneous variance $v_t$ follow the coupled stochastic differential equations:
\begin{align}
dS_t &= (r - q) S_t \, dt + \sqrt{v_t} \, S_t \, dW_t^S \label{eq:heston-s} \\
dv_t &= \kappa(\theta - v_t) \, dt + \xi \sqrt{v_t} \, dW_t^v \label{eq:heston-v}
\end{align}
where $dW_t^S$ and $dW_t^v$ are Brownian motions correlated according to $\text{Corr}(dW_t^S, dW_t^v) = \rho \, dt$. The five parameters control different aspects of the surface:

\begin{itemize}
    \item $v_0$ (initial variance): sets the current at-the-money volatility level ($\sigma_{\text{ATM}} \approx \sqrt{v_0}$).
    \item $\kappa$ (mean-reversion speed): controls how quickly variance returns to its long-run level. A larger $\kappa$ produces faster flattening of the term structure.
    \item $\theta$ (long-run variance): the equilibrium variance level. It determines the long-term ATM volatility ($\sigma_{\text{ATM}}^{\infty} \approx \sqrt{\theta}$).
    \item $\xi$ (volatility of volatility): controls the curvature of the smile. A larger $\xi$ produces more pronounced smiles at all maturities.
    \item $\rho$ (spot-volatility correlation): controls the asymmetry of the skew. A negative $\rho$ (typical in equities) produces the characteristic downward skew; $\rho \approx 0$ produces a symmetric smile.
\end{itemize}

\subsection{The Feller Condition}

The variance process $v_t$ must remain strictly positive for the model to be well defined. The Feller condition guarantees this:
\begin{equation}
2\kappa\theta \geq \xi^2
\label{eq:feller}
\end{equation}
When this condition holds, the variance process never reaches zero. It is imposed when generating synthetic data (Section~\ref{sec:synthetic-data}).

\subsection{Semi-Analytical Pricing}

A key advantage of the Heston model is the availability of a semi-analytical pricing formula. The price of a European call is obtained through the Gil-Pelaez Fourier inversion \cite{gilpelaez1951inversion}:
\begin{equation}
C(K, \tau) = e^{-r\tau}\left[F \cdot P_1 - K \cdot P_2\right]
\label{eq:heston-pricing}
\end{equation}
where $P_1$ and $P_2$ are computed from integrals involving the characteristic function of the log price. We use the formulation of Albrecher et al.\ \cite{albrecher2007heston}, which avoids the branch discontinuity known as the ``Little Heston Trap'' that affects naive implementations of the characteristic function.

The characteristic function is evaluated by numerical integration with adaptive quadrature. This makes the Heston model computationally tractable for generating large synthetic datasets: each option price requires a single one-dimensional numerical integral, which can be evaluated to machine precision in microseconds.

\section{No-Arbitrage Conditions}
\label{sec:no-arb-background}

A volatility surface must satisfy certain constraints to be consistent with a valid probability distribution for the underlying asset price. Violations of these constraints create \emph{static arbitrage}: riskless profit opportunities that can be exploited using portfolios of standard (\emph{vanilla}) options without requiring dynamic hedging.

\subsection{Calendar Spread Arbitrage}

The total implied variance $w(m, \tau) = \sigma_{\text{IV}}^2(m, \tau) \cdot \tau$ must be non-decreasing in maturity for each fixed log-moneyness:
\begin{equation}
\frac{\partial w}{\partial \tau}(m, \tau) \geq 0 \quad \text{for all } m
\label{eq:calendar-noarb}
\end{equation}

\textbf{Intuition.} If total variance decreased with maturity for some strike, a trader could sell the shorter-dated option and buy the longer-dated option (a calendar spread), collecting a net premium. The shorter-dated option would expire first, and the remaining longer-dated option would have been acquired at a discount, generating a riskless profit.

\subsection{Butterfly Spread Arbitrage}

For each fixed maturity, the total implied variance must be convex in log-moneyness:
\begin{equation}
\frac{\partial^2 w}{\partial m^2}(m, \tau) \geq 0 \quad \text{for all } \tau
\label{eq:butterfly-noarb}
\end{equation}

\textbf{Intuition.} If the variance curve were locally concave at some strike, a butterfly spread (buying options at the two adjacent strikes and selling two options at the central strike) would have negative cost but non-negative payoff, constituting an arbitrage.

\textbf{Equivalence.} The convexity of total variance in log-moneyness is related (through the result of Breeden and Litzenberger \cite{breeden1978prices}) to the non-negativity of the risk-neutral probability density. A concave region in total variance implies a negative probability density, which has no economic meaning.

\subsection{Discrete Checks on a Grid}

On the discrete $8 \times 25$ grid used in this thesis, these continuous conditions are approximated using finite differences:

\begin{itemize}
    \item \textbf{Calendar}: $w(\tau_{i+1}, m_j) - w(\tau_i, m_j) \geq 0$ for all adjacent maturity pairs and all strikes.
    \item \textbf{Butterfly}: the second-order finite difference of $w$ in the log-moneyness direction must be non-negative at each interior strike (accounting for non-uniform spacing).
\end{itemize}

A numerical tolerance of $10^{-10}$ is applied to avoid counting floating-point noise as violations. The specific finite-difference formulas are presented in Section~\ref{sec:no-arb-constraints}.

\section{Parametric Surface Models: SVI}
\label{sec:svi-background}

The Stochastic Volatility Inspired (SVI) parameterization \cite{gatheral2014svi} provides a parsimonious model for individual implied volatility curves. It is widely used in practice for volatility curve interpolation and serves as the traditional benchmark in this thesis.

\subsection{The SVI Formula}

For a single maturity slice, SVI parameterizes total implied variance as:
\begin{equation}
w(k) = a + b\left[\rho(k - m) + \sqrt{(k - m)^2 + \sigma^2}\right]
\label{eq:svi-raw}
\end{equation}
where $k$ denotes log-moneyness (equivalent to $m$ defined in Equation~\eqref{eq:logm}; $k$ is used here to follow Gatheral's convention for the SVI formula) and the five parameters have the following interpretations \cite{gatheral2011volsurface}:
\begin{itemize}
    \item $a$: overall variance level (vertical shift).
    \item $b$: slope magnitude (controls the slope of the wings).
    \item $\rho$: asymmetry ($\rho < 0$ produces the equity skew; $\rho = 0$ produces a symmetric smile).
    \item $m$: horizontal translation (shifts the center of the curve relative to ATM).
    \item $\sigma$: curvature control (a larger $\sigma$ rounds the minimum of the curve; a smaller $\sigma$ sharpens the kink).
\end{itemize}

The SVI formula is designed to mimic the asymptotic behavior of stochastic volatility models for extreme moneyness values (hence its name): for $|k| \to \infty$, $w(k) \sim a + b(1 + |\rho|)|k|$, producing the linear wing behavior observed in real markets.

\subsection{Strengths and Limitations}

\paragraph{Strengths.} With appropriate constraints on the parameters ($b > 0$, $\sigma > 0$, $|\rho| < 1$), SVI produces smooth and convex curves that naturally satisfy the butterfly no-arbitrage condition. It is interpretable, fast to fit, and requires only five parameters per slice.

\paragraph{Limitations.} SVI fits each maturity slice independently: there is no mechanism for sharing information across maturities. This has two consequences. First, consistency across slices (the calendar spread condition) is not imposed and must be checked separately. Second, when observations are sparse for a given maturity, the fit is underdetermined: five parameters cannot be reliably estimated from two or three data points. This limitation is the fundamental motivation for the neural-network-based approaches explored in this thesis: they process the complete 2D surface jointly, sharing information across both strikes and maturities.

%% file: chapters/03_dl_background.tex

This chapter introduces the deep learning concepts and architectures used in this thesis. We cover the supervised learning framework, the four main architecture families (MLPs, CNNs, U-Nets, Transformers), variational autoencoders, standard training techniques, and the connection between volatility surface reconstruction and the masked prediction paradigm.

\section{Supervised Learning and Loss Functions}
\label{sec:supervised-learning}

\subsection{The Regression Setting}

Supervised learning aims to learn a function $f_\theta: \mathcal{X} \to \mathcal{Y}$ from a dataset of input-output pairs $\{(\mathbf{x}_i, \mathbf{y}_i)\}_{i=1}^N$. In regression (as opposed to classification), the output space $\mathcal{Y} \subseteq \mathbb{R}^d$ is continuous. The function $f_\theta$ is parameterized by a neural network with learnable parameters $\theta$, and training consists of finding $\theta$ that minimizes a loss function $\mathcal{L}$ over the training data.

For volatility surface reconstruction, the input is a partially observed surface with mask ($\mathbf{x} \in \mathbb{R}^{2 \times 8 \times 25}$) and the output is the complete surface ($\mathbf{y} \in \mathbb{R}^{1 \times 8 \times 25}$)---a 200-dimensional regression problem.

\subsection{Mean Squared Error}

The standard loss function for regression is the mean squared error (MSE):
\begin{equation}
\mathcal{L}_{\text{MSE}} = \frac{1}{N} \sum_{i=1}^{N} \| f_\theta(\mathbf{x}_i) - \mathbf{y}_i \|^2
\label{eq:mse-background}
\end{equation}

MSE penalizes large errors quadratically, making it sensitive to outliers but well-suited to problems where prediction accuracy is uniformly important across the output space. Minimizing MSE is equivalent to maximum likelihood estimation under a Gaussian noise model.

\subsection{Data Splits and Generalization}

To assess whether the learned function generalizes beyond the training data, the dataset is partitioned into three disjoint subsets:
\begin{itemize}
    \item \textbf{Training set}: used to compute gradients and update $\theta$.
    \item \textbf{Validation set}: used to monitor generalization during training (e.g., for early stopping and hyperparameter selection). Not used for gradient computation.
    \item \textbf{Test set}: held out entirely until final evaluation. Provides an unbiased estimate of generalization performance.
\end{itemize}

\subsection{Overfitting and Regularization}

A model that fits the training data too closely may fail to generalize, a phenomenon called \emph{overfitting}. The gap between training and validation loss is the primary diagnostic: if training loss continues to decrease while validation loss increases, the model is memorizing training patterns rather than learning generalizable features.

Regularization techniques combat overfitting by constraining the model's capacity:
\begin{itemize}
    \item \textbf{Weight decay} (L2 regularization): adds a penalty $\lambda \|\theta\|^2$ to the loss, discouraging large weights.
    \item \textbf{Dropout} \cite{srivastava2014dropout}: randomly zeroes a fraction of activations during training, preventing co-adaptation of features.
    \item \textbf{Early stopping}: halts training when validation performance stops improving (Section~\ref{sec:early-stopping-bg}).
\end{itemize}

\section{Multilayer Perceptrons}
\label{sec:mlp-background}

\subsection{Architecture}

The multilayer perceptron (MLP) is the simplest neural network architecture. It consists of a sequence of fully connected (dense) layers, where each layer computes:
\begin{equation}
\mathbf{h}^{(\ell)} = \phi\!\left(\mathbf{W}^{(\ell)} \mathbf{h}^{(\ell-1)} + \mathbf{b}^{(\ell)}\right)
\label{eq:mlp-layer}
\end{equation}
where $\mathbf{W}^{(\ell)}$ and $\mathbf{b}^{(\ell)}$ are the learnable weight matrix and bias vector of layer $\ell$, and $\phi$ is a nonlinear activation function. Without the nonlinearity, a stack of linear layers would collapse to a single linear transformation, regardless of depth.

\subsection{Activation Functions}

Common activation functions include:
\begin{itemize}
    \item \textbf{ReLU} (Rectified Linear Unit) \cite{nair2010relu}: $\phi(x) = \max(0, x)$. Simple, computationally efficient, and effective at avoiding the vanishing gradient problem. The default choice for most architectures.
    \item \textbf{GELU} (Gaussian Error Linear Unit) \cite{hendrycks2016gelu}: $\phi(x) = x \cdot \Phi(x)$, where $\Phi$ is the standard normal CDF. A smooth approximation to ReLU, preferred in Transformer architectures for its differentiability at $x = 0$.
    \item \textbf{ELU} (Exponential Linear Unit) \cite{clevert2016elu}: $\phi(x) = x$ for $x > 0$ and $\alpha(e^x - 1)$ for $x \leq 0$. Produces non-zero gradients for negative inputs, sometimes improving convergence.
\end{itemize}

\subsection{Universal Approximation}

The universal approximation theorem \cite{hornik1989approximation} states that an MLP with a single hidden layer of sufficient width can approximate any continuous function on a compact domain to arbitrary accuracy. While theoretically powerful, this result does not guarantee that the approximation is \emph{efficient}---in practice, the required width may be exponentially large. Deep networks (multiple layers of moderate width) often provide much more compact representations than wide, shallow networks.

\subsection{Limitations for Structured Data}

The MLP treats its input as a flat vector with no spatial structure. When applied to grid-structured data like volatility surfaces, the flattening operation discards all information about which values are spatially adjacent. Any spatial patterns must be learned entirely from data, requiring more training examples and larger models compared to architectures that exploit spatial structure by design.

\section{Convolutional Neural Networks}
\label{sec:cnn-background}

Convolutional neural networks (CNNs) \cite{lecun1998gradient} exploit the spatial structure of grid-arranged data through local, translation-equivariant operations.

\subsection{The Convolution Operation}

A 2D convolutional layer applies a set of learnable filters (kernels) to the input feature map. For a single filter $\mathbf{w}$ of size $k \times k$, the output at position $(i, j)$ is:
\begin{equation}
(\mathbf{x} * \mathbf{w})_{ij} = \sum_{p=0}^{k-1} \sum_{q=0}^{k-1} w_{pq} \cdot x_{i+p, \, j+q}
\label{eq:conv2d}
\end{equation}

Each filter produces one output channel (feature map), and a convolutional layer typically applies $n_c$ filters to produce $n_c$ output channels. The key properties are:

\begin{itemize}
    \item \textbf{Local connectivity}: each output depends only on a local $k \times k$ neighborhood, exploiting the assumption that nearby values are more correlated than distant ones.
    \item \textbf{Weight sharing}: the same filter is applied at every spatial position, dramatically reducing the number of parameters compared to a fully connected layer (a $3 \times 3$ filter has 9 parameters regardless of the input size).
    \item \textbf{Translation equivariance}: a shifted input produces an identically shifted output, meaning the network detects features regardless of their position on the grid.
\end{itemize}

\subsection{Receptive Field}

The \emph{receptive field} of a neuron is the region of the input that influences its value. For a stack of $L$ convolutional layers with $k \times k$ kernels, the theoretical receptive field grows to $(1 + L(k-1)) \times (1 + L(k-1))$. For example, five layers of $3 \times 3$ convolutions yield a receptive field of $11 \times 11$. This allows deep CNNs to aggregate information from large spatial regions through successive local operations.

\subsection{Padding and Spatial Preservation}

\emph{Same-padding} (padding the input with $\lfloor k/2 \rfloor$ zeros on each side) ensures that the spatial dimensions are preserved through each layer. This is essential for dense prediction tasks like surface reconstruction, where the output must have the same spatial dimensions as the input. Without padding, each convolutional layer would shrink the spatial dimensions, eventually collapsing the grid to a single point.

\subsection{Pooling and Downsampling}

Pooling layers (e.g., max pooling or average pooling) reduce the spatial dimensions by aggregating local regions. While useful for classification (where spatial precision is not needed), pooling discards spatial information and is generally avoided or carefully managed in dense prediction tasks. Strided convolutions (convolutions with stride $> 1$) provide a learnable alternative to pooling.

\section{The U-Net Architecture}
\label{sec:unet-background}

The U-Net \cite{ronneberger2015unet} is an encoder-decoder architecture with skip connections, originally developed for biomedical image segmentation.

\subsection{Encoder-Decoder Structure}

The \emph{encoder} progressively downsamples the input through convolution and pooling layers, building feature representations at multiple spatial scales. The \emph{decoder} progressively upsamples these representations back to the original resolution using bilinear interpolation or transposed convolutions. This structure captures both local detail (in the early, high-resolution layers) and global context (in the deep, low-resolution bottleneck).

\subsection{Skip Connections}

The defining feature of U-Net is the use of skip connections that concatenate encoder feature maps with the corresponding decoder feature maps at each resolution level. Without skip connections, fine-grained spatial information lost during downsampling would be unrecoverable. The skip connections provide a direct path for high-frequency detail to flow from encoder to decoder, while the bottleneck captures low-frequency, global structure. This combination makes U-Net particularly effective for tasks that require both precise localization and semantic understanding.

\subsection{Application to Dense Prediction}

U-Net and its variants are the dominant architecture for dense prediction tasks where every input pixel requires an output prediction---semantic segmentation, depth estimation, and image inpainting. Volatility surface reconstruction fits naturally into this paradigm: every grid point requires an output (the reconstructed implied volatility), and the model must combine local smoothness (nearby IVs are similar) with global consistency (the surface must be coherent across the full maturity range).

\section{The Transformer Architecture}
\label{sec:transformer-background}

The Transformer \cite{vaswani2017attention} replaces the local operations of CNNs with global attention mechanisms, allowing every position in the sequence to interact directly with every other position.

\subsection{Self-Attention}

The core operation is \emph{scaled dot-product attention}. Given a set of queries $\mathbf{Q}$, keys $\mathbf{K}$, and values $\mathbf{V}$ (all derived from the input via learned linear projections), attention computes a weighted sum of values where the weights are determined by query-key similarity:
\begin{equation}
\text{Attention}(\mathbf{Q}, \mathbf{K}, \mathbf{V}) = \text{softmax}\!\left(\frac{\mathbf{Q}\mathbf{K}^\top}{\sqrt{d_k}}\right) \mathbf{V}
\label{eq:attention}
\end{equation}
where $d_k$ is the dimension of each key (the scaling factor prevents the softmax from saturating in high dimensions).

In \emph{self-attention}, the queries, keys, and values all come from the same input sequence: each position attends to all other positions, with attention weights determined by learned compatibility functions. This gives every position a global receptive field in a single layer---unlike CNNs, which require multiple stacked layers to achieve long-range interactions.

\subsection{Multi-Head Attention}

Rather than computing a single attention function, multi-head attention runs $h$ attention functions in parallel, each with its own learned projections:
\begin{equation}
\text{MHA}(\mathbf{Q}, \mathbf{K}, \mathbf{V}) = \text{Concat}(\text{head}_1, \ldots, \text{head}_h) \, \mathbf{W}^O
\label{eq:mha}
\end{equation}
where $\text{head}_i = \text{Attention}(\mathbf{Q}\mathbf{W}_i^Q, \mathbf{K}\mathbf{W}_i^K, \mathbf{V}\mathbf{W}_i^V)$. Each head can learn different attention patterns---for example, one head might attend to nearby positions while another attends to distant ones.

\subsection{Positional Encoding}

Unlike CNNs, the attention operation is \emph{permutation-equivariant}: it produces the same output regardless of the order of the input sequence. To inject positional information, Transformers add a \emph{positional encoding} to the input embeddings. The original Transformer \cite{vaswani2017attention} uses fixed sinusoidal encodings indexed by sequence position. Alternatives include learned positional embeddings and coordinate-based Fourier features \cite{tancik2020fourier}, the latter of which we use in this thesis (Section~\ref{sec:transformer}).

\subsection{Encoder-Decoder Transformer}

The full Transformer architecture consists of an encoder and a decoder:
\begin{itemize}
    \item \textbf{Encoder}: a stack of self-attention layers that process the input sequence, producing a contextualized representation (memory).
    \item \textbf{Decoder}: a stack of layers that first apply self-attention among the output queries, then \emph{cross-attention} to the encoder memory. Cross-attention allows each decoder position to attend to all encoder positions, enabling information flow from the input to the output.
\end{itemize}

Each layer (both encoder and decoder) also includes a position-wise feed-forward network (two linear layers with a nonlinear activation) and residual connections with layer normalization.

\subsection{Attention Masking}

Attention masks allow selective exclusion of certain positions from the attention computation. By setting mask values to $-\infty$ before the softmax, specific positions receive zero attention weight. This mechanism is used in several ways:
\begin{itemize}
    \item \textbf{Padding masks}: exclude padded positions in variable-length sequences.
    \item \textbf{Causal masks}: prevent positions from attending to future positions (used in autoregressive generation).
    \item \textbf{Observation masks}: in our application, exclude missing grid points from the encoder's self-attention, ensuring that the encoder only processes observed values.
\end{itemize}

\subsection{Attention for Inputs with Missing Data}

Beyond its general applications, the attention mechanism has a natural advantage for processing sparse, irregularly observed data, which motivates its application to the volatility surface reconstruction problem (developed in Section~\ref{sec:transformer}). Unlike CNNs, which apply fixed-size filters regardless of whether the covered positions are observed, attention dynamically routes information from wherever observations exist. When observations are dense, each query attends to many nearby sources; when observations are sparse, the same mechanism adapts by attending to more distant sources. This flexibility is the key architectural advantage of the Transformer architecture for volatility surface reconstruction.

\section{Variational Autoencoders}
\label{sec:vae-background}

Variational autoencoders (VAEs) \cite{kingma2014vae} combine deep learning with probabilistic generative modeling. Unlike the discriminative models above, which learn a direct mapping from input to output, VAEs learn a latent representation of the data distribution.

\subsection{Latent Variable Models}

A VAE posits that the observed data $\mathbf{x}$ is generated from a low-dimensional latent variable $\mathbf{z}$ through a decoder (generator) network:
\begin{equation}
p_\theta(\mathbf{x}) = \int p_\theta(\mathbf{x} | \mathbf{z}) \, p(\mathbf{z}) \, d\mathbf{z}
\end{equation}
where $p(\mathbf{z}) = \mathcal{N}(\mathbf{0}, \mathbf{I})$ is a standard Gaussian prior and $p_\theta(\mathbf{x} | \mathbf{z})$ is the decoder network parameterized by $\theta$. Direct maximization of the log-likelihood $\log p_\theta(\mathbf{x})$ is intractable because it requires integrating over all possible latent codes.

\subsection{The Evidence Lower Bound (ELBO)}

Instead of maximizing the intractable log-likelihood, VAEs maximize a lower bound---the evidence lower bound (ELBO):
\begin{equation}
\log p_\theta(\mathbf{x}) \geq \mathbb{E}_{q_\phi(\mathbf{z}|\mathbf{x})}\!\left[\log p_\theta(\mathbf{x} | \mathbf{z})\right] - D_{\text{KL}}\!\left(q_\phi(\mathbf{z} | \mathbf{x}) \,\|\, p(\mathbf{z})\right)
\label{eq:elbo-background}
\end{equation}
where $q_\phi(\mathbf{z} | \mathbf{x})$ is an encoder network that approximates the intractable posterior $p_\theta(\mathbf{z} | \mathbf{x})$. The first term encourages accurate reconstruction; the second term regularizes the latent space to be close to the prior.

\subsection{The Reparameterization Trick}

Training requires differentiating through the sampling operation $\mathbf{z} \sim q_\phi(\mathbf{z} | \mathbf{x})$. The reparameterization trick expresses the sample as a deterministic function of the encoder output and independent noise:
\begin{equation}
\mathbf{z} = \boldsymbol{\mu}_\phi(\mathbf{x}) + \boldsymbol{\sigma}_\phi(\mathbf{x}) \odot \boldsymbol{\epsilon}, \quad \boldsymbol{\epsilon} \sim \mathcal{N}(\mathbf{0}, \mathbf{I})
\end{equation}
This moves the stochasticity outside the computational graph, enabling standard backpropagation through the encoder.

\subsection{VAEs for Reconstruction}

In the context of surface reconstruction from partial observations, the VAE's latent space provides a learned manifold of plausible complete surfaces. Given a partial observation, the model can search for the latent code $\mathbf{z}^*$ whose decoded surface best matches the observed values---projecting the incomplete data onto the learned manifold. This approach is detailed in Section~\ref{sec:vae}.

\section{Training Techniques}
\label{sec:training-techniques}

\subsection{Stochastic Gradient Descent and Adam}

Neural networks are trained by minimizing the loss function via gradient-based optimization. \emph{Stochastic gradient descent} (SGD) updates parameters using gradients computed on random mini-batches of the training data, reducing computational cost while introducing beneficial noise that can help escape local minima.

The \textbf{Adam} optimizer \cite{kingma2015adam} extends SGD with adaptive per-parameter learning rates based on first and second moment estimates of the gradient:
\begin{align}
\mathbf{m}_t &= \beta_1 \mathbf{m}_{t-1} + (1 - \beta_1) \mathbf{g}_t \label{eq:adam-m} \\
\mathbf{v}_t &= \beta_2 \mathbf{v}_{t-1} + (1 - \beta_2) \mathbf{g}_t^2 \label{eq:adam-v} \\
\theta_{t+1} &= \theta_t - \eta \frac{\hat{\mathbf{m}}_t}{\sqrt{\hat{\mathbf{v}}_t} + \epsilon} \label{eq:adam-update}
\end{align}
where $\mathbf{g}_t$ is the gradient, $\hat{\mathbf{m}}_t$ and $\hat{\mathbf{v}}_t$ are bias-corrected moment estimates, $\eta$ is the learning rate, and $\beta_1 = 0.9$, $\beta_2 = 0.999$ are the default decay rates. Adam is the most widely used optimizer for deep learning and is used for all experiments in this thesis.

\subsection{Learning Rate}

The learning rate $\eta$ is the most critical hyperparameter. Too large, and the optimization oscillates or diverges; too small, and convergence is impractically slow. The optimal learning rate varies across architectures: attention-based models (Transformers) typically require lower learning rates than convolutional or fully connected models, because the attention computation amplifies gradient magnitudes.

Learning rate schedules (e.g., cosine annealing, warm-up) adjust $\eta$ during training. While often beneficial, schedules introduce additional hyperparameters that may interact differently with each architecture, complicating fair comparisons.

\subsection{Early Stopping}
\label{sec:early-stopping-bg}

Early stopping monitors the validation loss during training and halts when it has not improved for a specified number of epochs (the \emph{patience}). The model checkpoint with the best validation loss is retained. This is a form of regularization: it prevents the model from fitting noise in the training data that would degrade generalization. In practice, early stopping is often the most effective and simplest regularization technique.

\subsection{Batch Normalization and Layer Normalization}

\textbf{Batch normalization} \cite{ioffe2015batch} normalizes activations within each mini-batch, stabilizing training and enabling higher learning rates. However, its behavior depends on batch size and differs between training and inference modes.

\textbf{Layer normalization} \cite{ba2016layer} normalizes across the feature dimension within each sample, making it independent of batch size. It is the standard normalization technique in Transformer architectures, where the ``pre-LN'' variant (normalizing before rather than after each sub-layer) has become the default for training stability.

\section{Masked Prediction and Inpainting}
\label{sec:masked-prediction}

\subsection{Masked Language Models}

The masked prediction paradigm was popularized in natural language processing by BERT \cite{devlin2019bert}, which randomly masks tokens in a text sequence and trains a Transformer to predict the masked tokens from context. This self-supervised objective proved remarkably effective at learning transferable representations.

\subsection{Masked Image Modeling}

The concept was extended to computer vision by Masked Autoencoders (MAE) \cite{he2022mae}, which mask random patches of an image and train a Vision Transformer to reconstruct the original. MAE demonstrated that masking 75\% of image patches during training yields powerful visual representations, and that the encoder-decoder Transformer is naturally suited to this task because it can process only the visible patches (reducing computation) and use the decoder to reconstruct all patches.

\subsection{Image Inpainting}

Image inpainting---filling in missing or corrupted regions of an image---is a classical computer vision problem. Modern deep learning approaches use encoder-decoder CNNs or U-Nets that take the corrupted image and a mask as input and produce the completed image. The training objective is the reconstruction error on the missing regions.

\subsection{Volatility Surface Reconstruction as Masked Prediction}

The volatility surface reconstruction task combines elements of all three paradigms:
\begin{itemize}
    \item Like BERT and MAE, we \emph{randomly mask} portions of the input and train the model to reconstruct them. The on-the-fly masking provides data augmentation and prevents memorization.
    \item Like image inpainting, the input is a 2D grid with missing entries, and the model must fill them in while respecting the structure of the data.
    \item Unlike MAE (which is self-supervised), our task is \emph{fully supervised}: we have ground truth surfaces for training. And unlike generic image inpainting, our data has domain-specific structure (smoothness, convexity) that can be exploited through architectural choices and no-arbitrage penalties.
\end{itemize}

This framing positions volatility surface reconstruction at the intersection of financial engineering and representation learning, motivating the diverse set of architectures evaluated in this thesis.

%% file: chapters/04_data.tex

The previous chapters introduced the financial concepts underlying volatility surfaces and the deep learning architectures used to model them. This chapter describes the concrete data representation, the two datasets used throughout this work---one synthetic and one from real equity options markets---and the masking strategy that frames surface reconstruction as a supervised learning problem.

\section{Surface Representation}
\label{sec:surface-representation}

A volatility surface is a two-dimensional function $\sigma(K, \tau)$ mapping strike price $K$ and time to maturity $\tau$ to implied volatility. For computational purposes, we discretize this function on a fixed grid.

\subsection{Grid Discretization}

We represent each surface on a grid of 8 maturities and 25 strikes, yielding 200 grid points per surface. The maturity axis uses the following tenors (in years):
\begin{equation}
\boldsymbol{\tau} = [0.08,\; 0.17,\; 0.25,\; 0.5,\; 0.75,\; 1.0,\; 1.5,\; 2.0]
\label{eq:taus}
\end{equation}
spanning approximately one month to two years. The strike axis uses 25 equally spaced points:
\begin{equation}
\mathbf{K} = \text{linspace}(70, 130, 25) = [70.0,\; 72.5,\; 75.0,\; \ldots,\; 127.5,\; 130.0]
\label{eq:strikes}
\end{equation}
with a reference forward price of $F = 100$. This grid covers strikes from 30\% out-of-the-money puts to 30\% out-of-the-money calls, which encompasses the liquid portion of most equity option chains.

\subsection{Log-Moneyness Coordinates}

Rather than using raw strike prices, the strike axis is expressed in log-moneyness $m = \log(K/F)$ (Equation~\eqref{eq:logm}), which ranges from $-0.357$ to $+0.262$ on our grid. As discussed in Section~\ref{sec:vol-surface}, log-moneyness centers the surface at the at-the-money level and normalizes across different underlying price levels, making surfaces comparable across assets and time periods.

\subsection{Input Representation}

Each surface is stored as a matrix $\boldsymbol{\Sigma} \in \mathbb{R}^{8 \times 25}$ of implied volatilities. For the reconstruction task, the model receives a two-channel input tensor $\mathbf{X} \in \mathbb{R}^{2 \times 8 \times 25}$:
\begin{itemize}
    \item \textbf{Channel 0}: The masked implied volatility surface, where unobserved points are set to zero.
    \item \textbf{Channel 1}: A binary observation mask $\mathbf{M} \in \{0, 1\}^{8 \times 25}$, where $M_{ij} = 1$ indicates that the implied volatility at tenor $\tau_i$ and strike $K_j$ is observed, and $M_{ij} = 0$ indicates it is missing.
\end{itemize}
The target is the complete surface $\boldsymbol{\Sigma}$, and the training loss is computed only on the missing points (where $M_{ij} = 0$), preventing the model from simply learning the identity function on observed values.

\section{Synthetic Data Generation}
\label{sec:synthetic-data}

The primary training dataset consists of 10,000 volatility surfaces generated from the Heston stochastic volatility model introduced in Section~\ref{sec:heston}. The Heston model was chosen because it is the standard benchmark for stochastic volatility---it captures the essential features of real volatility surfaces (smile, skew, term structure) while remaining analytically tractable via the semi-analytical pricing formula of Equation~\eqref{eq:heston-pricing}.

\subsection{Parameter Sampling}

To generate a diverse dataset that spans realistic market conditions, we sample each surface's parameters uniformly from the ranges shown in Table~\ref{tab:heston-params}.

\begin{table}[H]
    \centering
    \caption{Heston parameter sampling ranges for synthetic data generation.}
    \label{tab:heston-params}
    \begin{tabular}{lccl}
        \toprule
        Parameter & Min & Max & Interpretation \\
        \midrule
        $v_0$ & 0.01 & 0.16 & Spot vol 10\%--40\% \\
        $\kappa$ & 0.5 & 5.0 & Slow to fast mean reversion \\
        $\theta$ & 0.01 & 0.16 & Long-run vol 10\%--40\% \\
        $\xi$ & 0.1 & 0.8 & Low to high smile curvature \\
        $\rho$ & $-0.9$ & $-0.1$ & Moderate to strong equity skew \\
        \bottomrule
    \end{tabular}
\end{table}

These ranges are informed by typical equity calibrations \cite{gatheral2011volsurface}. The correlation $\rho$ is restricted to negative values, reflecting the well-documented leverage effect in equity markets: stock prices and volatility move in opposite directions. We enforce the Feller condition $2\kappa\theta \geq \xi^2$ to ensure that the variance process remains strictly positive, resampling parameter sets that violate it.

\begin{figure}[H]
    \centering
    \includegraphics[width=\textwidth]{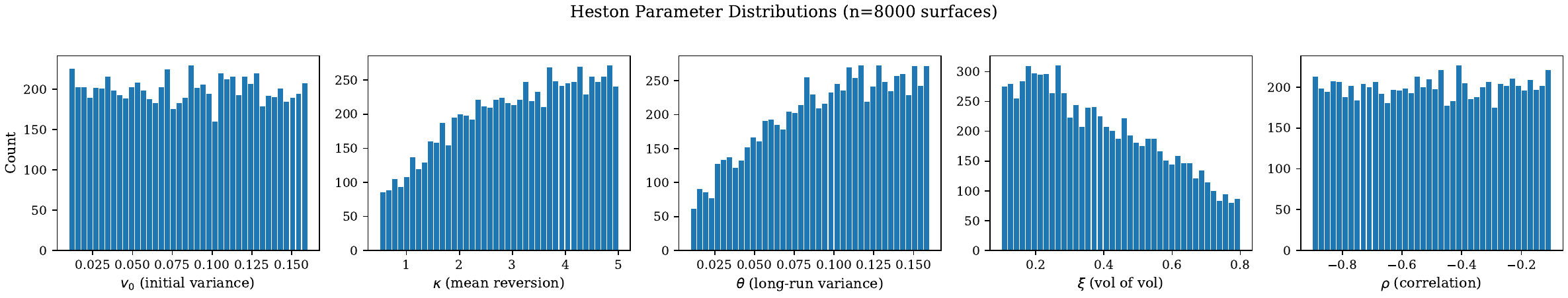}
    \caption{Distribution of Heston parameters across the 10,000 synthetic surfaces. Sampling is uniform within each range, subject to the Feller condition $2\kappa\theta \geq \xi^2$, which slightly shapes the joint distribution of $\kappa$, $\theta$, and $\xi$.}
    \label{fig:heston-params}
\end{figure}

Figure~\ref{fig:heston-params} shows the empirical distribution of the five Heston parameters across the 10,000 generated surfaces.

\subsection{Pricing and Implied Volatility Recovery}

For each parameter set, we compute European call option prices at all 200 grid points $(K_j, \tau_i)$ using the Heston semi-analytical pricing formula (Equation~\eqref{eq:heston-pricing}), with the Albrecher et al.\ \cite{albrecher2007heston} formulation to avoid the ``Little Heston Trap.'' Numerical integration is performed with adaptive quadrature to a tolerance of $10^{-10}$.

The resulting option prices are then converted to implied volatilities using the Newton-Raphson solver described in Section~\ref{sec:implied-vol}, with a convergence tolerance of $10^{-12}$ ensuring machine-precision accuracy in the recovered implied volatilities.

\subsection{Surface Examples}

\begin{figure}[H]
    \centering
    \includegraphics[width=\textwidth]{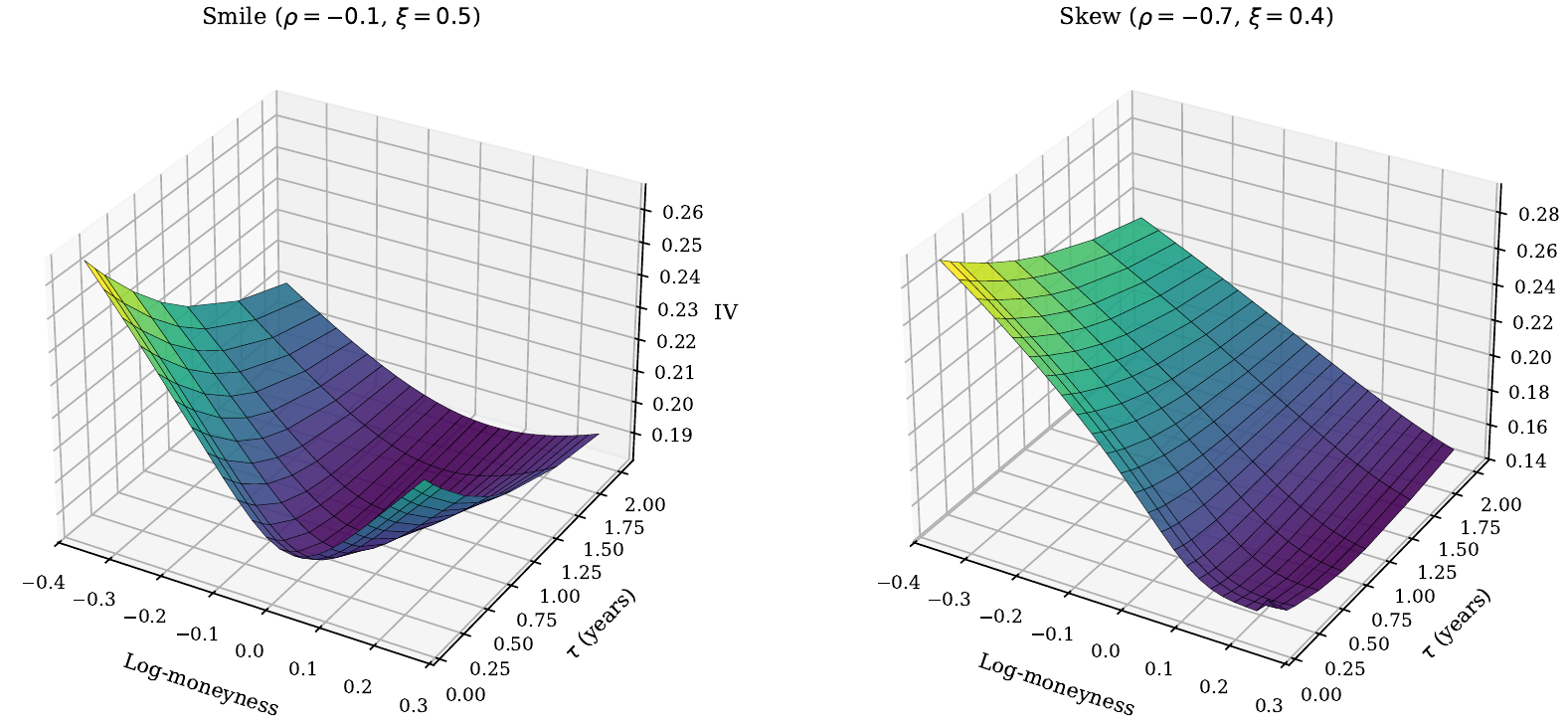}
    \caption{Two Heston volatility surfaces from the synthetic dataset. Left: symmetric smile ($\rho = -0.1$, $\xi = 0.5$). Right: equity skew ($\rho = -0.7$, $\xi = 0.4$). The correlation parameter $\rho$ controls the asymmetry, while the vol-of-vol $\xi$ controls the overall curvature.}
    \label{fig:heston-3d}
\end{figure}

Figure~\ref{fig:heston-3d} shows two representative surfaces from the dataset, illustrating how the Heston parameters control surface shape. With near-zero correlation ($\rho = -0.1$) and high vol-of-vol ($\xi = 0.5$), the surface exhibits a symmetric smile. With strongly negative correlation ($\rho = -0.7$), the characteristic equity skew emerges: implied volatility increases for lower strikes (out-of-the-money puts) and decreases for higher strikes.

\begin{figure}[H]
    \centering
    \includegraphics[width=\textwidth]{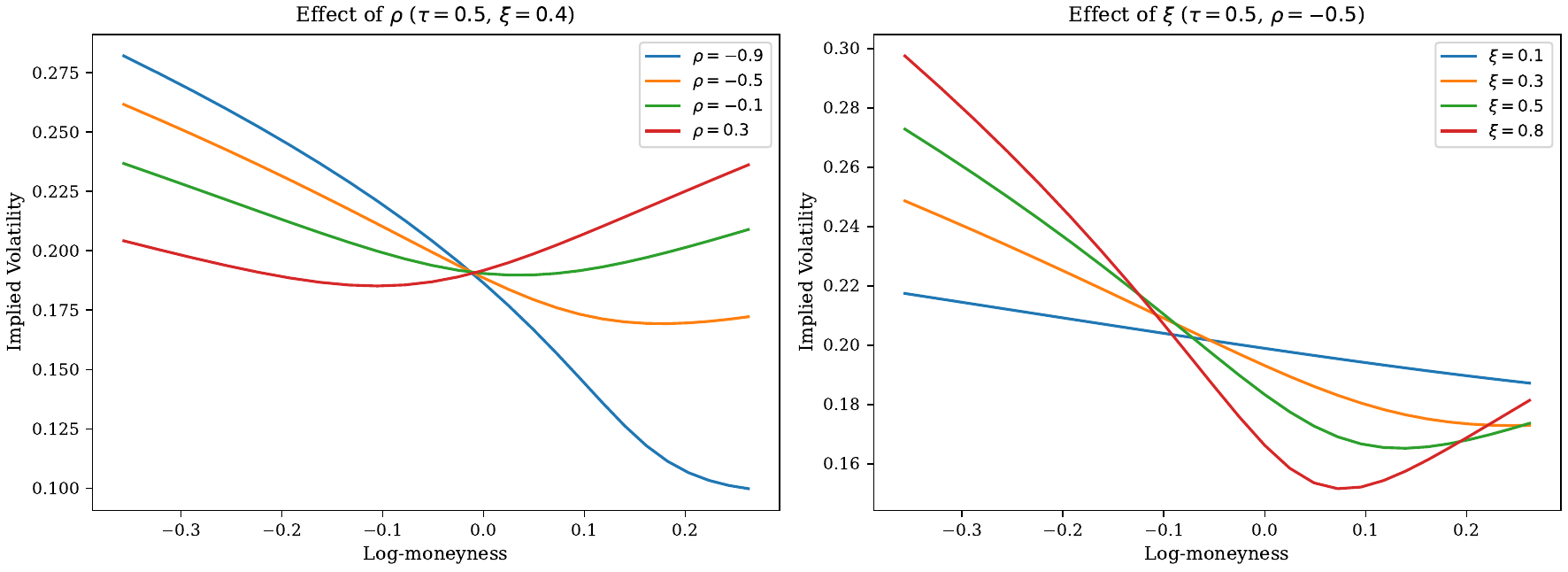}
    \caption{Heston parameter sensitivity at $\tau = 0.5$. Left: sweeping $\rho$ from $-0.9$ to $-0.1$ controls the skew tilt. Right: sweeping $\xi$ from $0.1$ to $0.8$ controls the smile curvature.}
    \label{fig:heston-sensitivity}
\end{figure}

Figure~\ref{fig:heston-sensitivity} isolates the effect of each parameter by sweeping $\rho$ and $\xi$ individually while holding the other parameters fixed.

\subsection{Dataset Scale and Split}

We generate 10,000 surfaces in total, split into training, validation, and test sets:

\begin{table}[H]
    \centering
    \caption{Synthetic dataset split. Each split uses a fixed random seed for reproducibility.}
    \label{tab:synth-split}
    \begin{tabular}{lccc}
        \toprule
        Split & Surfaces & Seed & Purpose \\
        \midrule
        Train & 8,000 & 42 & Model training \\
        Validation & 1,000 & 123 & Early stopping, hyperparameter tuning \\
        Test & 1,000 & 456 & Final evaluation \\
        \bottomrule
    \end{tabular}
\end{table}

Surface generation is parallelized across CPU cores, with each surface requiring approximately 200 characteristic function evaluations (one per grid point). The full dataset of 10,000 surfaces is generated in under one hour on an 8-core machine.

\section{Real Market Data}
\label{sec:real-data}

To validate that models trained on synthetic data can generalize to real markets, and to assess whether real data offers sufficient structure for direct training, we construct a second dataset from historical SPY (S\&P 500 ETF) equity options.

\subsection{Data Source}

We use the historical options dataset compiled by Dubach \cite{dubach2025options}, which covers SPY options from 2008 through 2025. The raw dataset contains approximately 24.7 million option contract records across 18 years, including daily close prices for the underlying.

\subsection{Surface Construction Pipeline}

Constructing a volatility surface from raw option quotes requires several processing steps, since market data arrives as a collection of $(K, \tau, \sigma_{\text{IV}})$ triples at irregular positions rather than on a regular grid.

\paragraph{Filtering.} We apply the following filters to each day's option chain:
\begin{itemize}
    \item \textbf{Moneyness}: $K/S \in [0.70, 1.30]$, matching our grid range.
    \item \textbf{Days to expiry}: $\text{DTE} \in [7, 800]$, excluding very short-dated options (dominated by microstructure effects) and very long-dated options (illiquid).
    \item \textbf{Liquidity}: Bid price $> 0$ and relative bid-ask spread $< 50\%$.
    \item \textbf{IV bounds}: $\sigma_{\text{IV}} \in [0.01, 2.0]$, excluding data errors.
    \item \textbf{OTM selection}: For each strike, we use out-of-the-money options (puts for $K < S$, calls for $K \geq S$), which are more liquid and have tighter spreads than their in-the-money counterparts.
\end{itemize}

\paragraph{Tenor matching.} For each of the 8 standard maturities $\tau_i$, we find the closest available expiry within a 30\% relative tolerance. If two market expiries bracket a standard tenor, we interpolate in total variance space ($\sigma^2 \tau$), which preserves the no-arbitrage constraint that total variance must be non-decreasing in maturity.

\paragraph{Strike interpolation.} Within each matched tenor, we interpolate the market-quoted implied volatilities onto our 25-point log-moneyness grid using cubic spline interpolation. We do not extrapolate beyond the range of available quotes---grid points outside the observed strike range are marked as missing.

\paragraph{Quality filter.} We retain only surfaces with at least 75\% tenor coverage (6 out of 8 tenors present) and at least 70\% average strike coverage across present tenors. This ensures a minimum level of information for each surface.

\subsection{Dataset Statistics}

After filtering, we obtain 3,900 surfaces spanning 2008--2025. The dataset is split temporally (not randomly) to prevent information leakage:

\begin{table}[H]
    \centering
    \caption{Real SPY dataset split. Temporal split ensures no future information leaks into training.}
    \label{tab:real-split}
    \begin{tabular}{lccc}
        \toprule
        Split & Surfaces & Years & Purpose \\
        \midrule
        Train & 2,912 & 2008--2021 & Model training \\
        Validation & 501 & 2022--2023 & Early stopping \\
        Test & 487 & 2024--2025 & Final evaluation \\
        \bottomrule
    \end{tabular}
\end{table}

\begin{figure}[H]
    \centering
    \includegraphics[width=\textwidth]{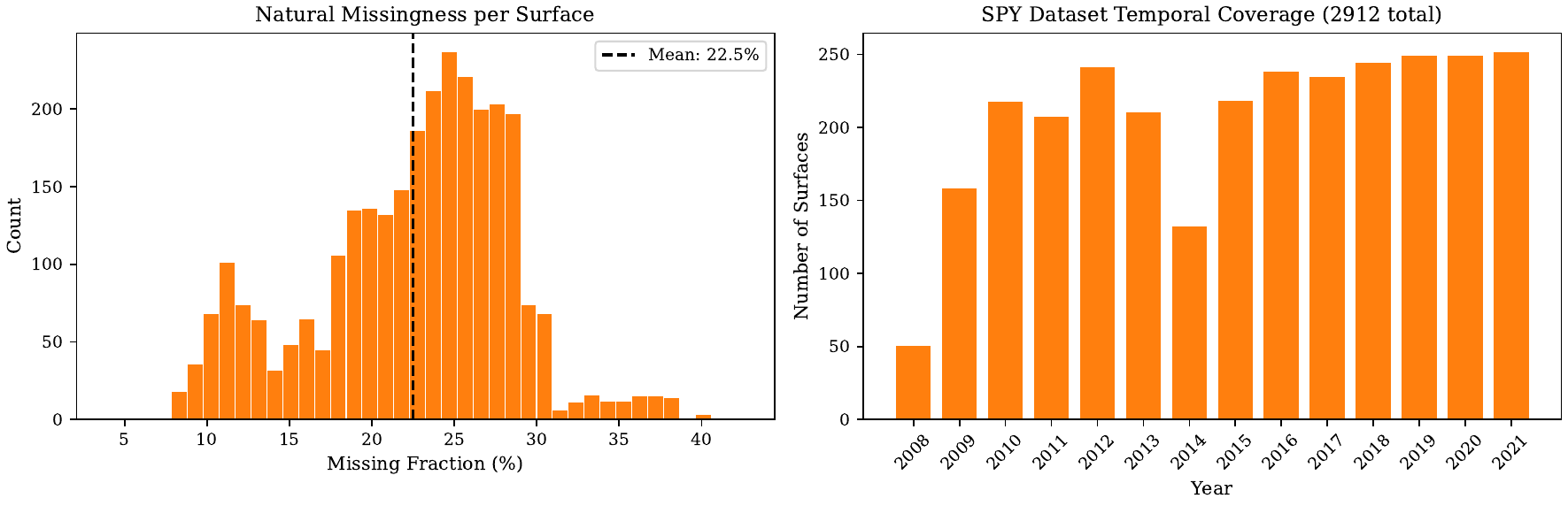}
    \caption{Left: distribution of the fraction of missing grid points across the training split (2,912 SPY surfaces). The mean is approximately 22.5\%. Right: temporal distribution of surfaces by year, reflecting market availability and the growth of listed options over time.}
    \label{fig:real-missingness}
\end{figure}

The real dataset exhibits natural missingness averaging approximately 22.5\% of grid points per surface, concentrated in the wings (deep out-of-the-money strikes) and at short maturities where weekly options may not exist. Figure~\ref{fig:real-missingness} shows the distribution of missing fractions and the temporal coverage.

\section{Synthetic vs Real Surfaces}
\label{sec:synth-vs-real}

While the Heston model captures the main qualitative features of real volatility surfaces---skew, term structure, and smile---there are important structural differences between synthetic and real data.

\begin{figure}[H]
    \centering
    \includegraphics[width=0.9\textwidth]{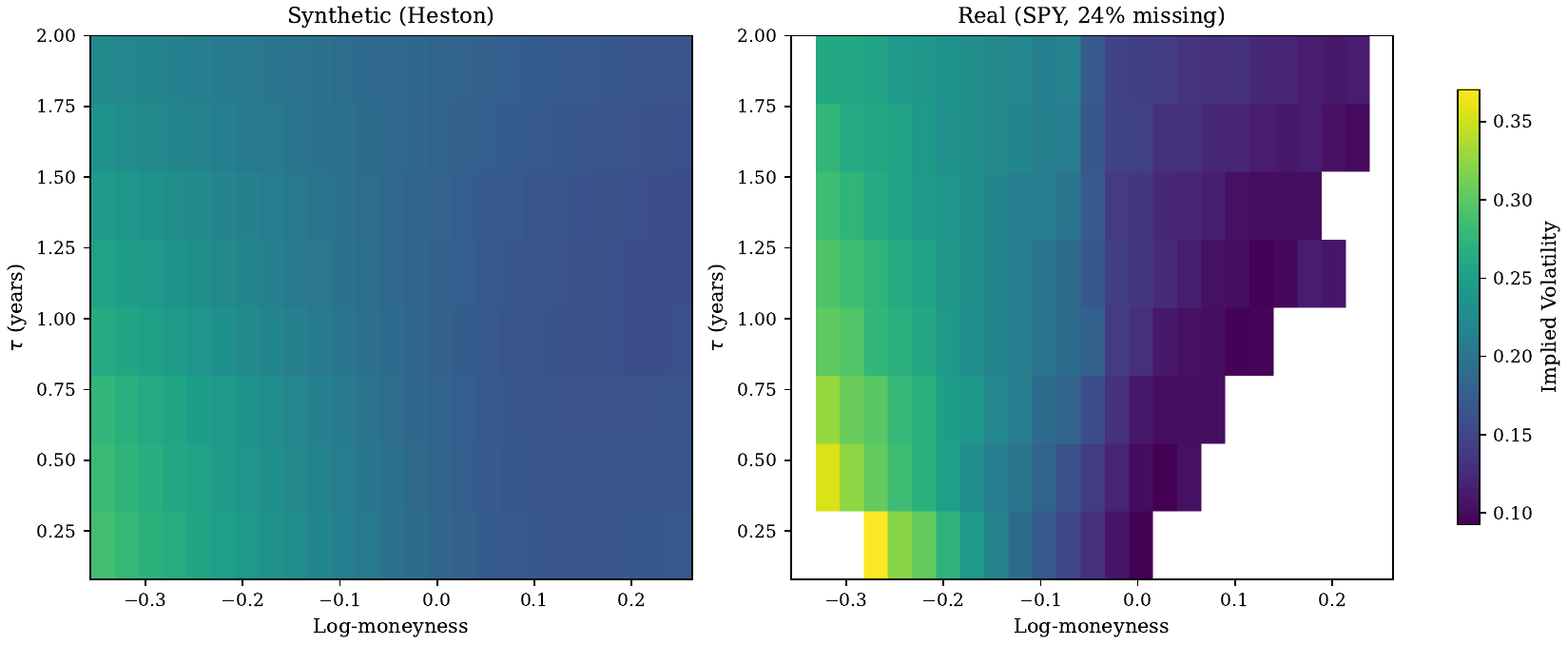}
    \caption{Side-by-side comparison of a synthetic (Heston) and real (SPY) volatility surface on the same grid and colorscale. The real surface exhibits richer local structure and natural missingness.}
    \label{fig:synth-vs-real}
\end{figure}

Figure~\ref{fig:synth-vs-real} shows a representative synthetic surface alongside a representative real SPY surface. The Heston surface is smooth and parametrically determined: given five parameters, the entire surface is fully specified. The real surface, by contrast, exhibits finer local structure---kinks, asymmetries, and maturity-dependent smile shapes---that cannot be captured by any five-parameter model. The real surface also has natural gaps (white regions) where no liquid options were available.

\begin{figure}[H]
    \centering
    \includegraphics[width=0.75\textwidth]{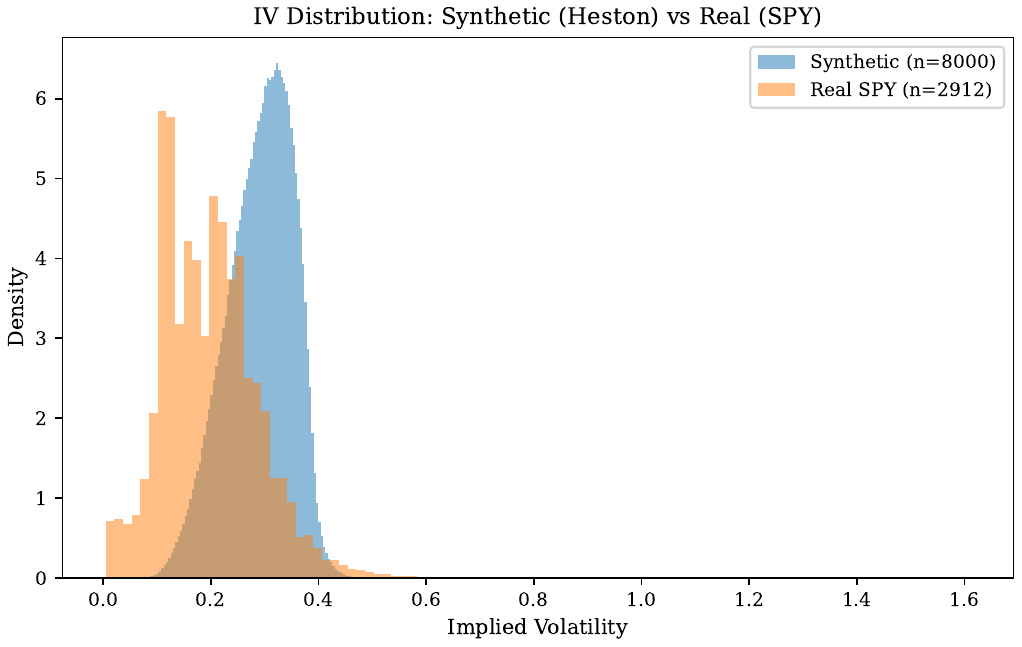}
    \caption{Implied volatility distributions for the synthetic (Heston) and real (SPY) datasets. The synthetic dataset has broader coverage; the real dataset has a concentrated mode and a heavier right tail.}
    \label{fig:iv-distribution}
\end{figure}

Figure~\ref{fig:iv-distribution} compares the implied volatility distributions of the two datasets. The synthetic dataset, generated with $v_0$ and $\theta$ sampled up to 0.16 ($\approx 40\%$ vol), covers a wider IV range than the real SPY data, which is concentrated between 10\% and 50\% with a mode around 15--20\%. The real distribution also exhibits a heavier right tail, reflecting occasional volatility spikes during market stress events (e.g., the 2008 financial crisis, the 2020 COVID crash).

These structural differences have direct implications for transfer learning, which we explore in Chapter~\ref{ch:results}: models pretrained on Heston surfaces must adapt to the different distributional characteristics and finer structure of real market data.

\section{Masking Strategy}
\label{sec:masking}

The core reconstruction task requires models to predict missing implied volatilities given partial observations. We simulate this by applying a binary mask to each complete surface during training.

\subsection{Random Masking}

The default masking strategy drops each grid point independently with probability $p$, where $p = 0.3$ (30\% missing) is the baseline. Formally, the mask $\mathbf{M} \in \{0, 1\}^{8 \times 25}$ is generated as:
\begin{equation}
M_{ij} \sim \text{Bernoulli}(1 - p), \quad \forall\, i \in \{1, \ldots, 8\},\; j \in \{1, \ldots, 25\}
\label{eq:mask}
\end{equation}
where $M_{ij} = 1$ means the point is observed and $M_{ij} = 0$ means it is missing. This produces masks with an expected $p \times 200 = 60$ missing points per surface, though the actual number varies across samples due to randomness.

\subsection{On-the-Fly Masking}

A key design choice is that masks are generated \emph{on the fly} during training---each time a surface is drawn from the dataset, a fresh random mask is generated. This means that the same underlying surface is seen with different observation patterns across epochs, acting as a form of \textbf{data augmentation}. The model cannot memorize which specific points are missing for a given surface; instead, it must learn a general reconstruction strategy that works for any pattern of missing data.

\subsection{Masking as an Evaluation Parameter}

While models are trained with a fixed missing fraction ($p = 0.3$), we evaluate them across a range of sparsity levels from $p = 0.1$ (10\% missing, nearly complete) to $p = 0.9$ (90\% missing, extremely sparse). This tests how gracefully each architecture degrades as observations become scarcer, without retraining. The results of this sweep are presented in Chapter~\ref{ch:results}.

\begin{figure}[H]
    \centering
    \includegraphics[width=\textwidth]{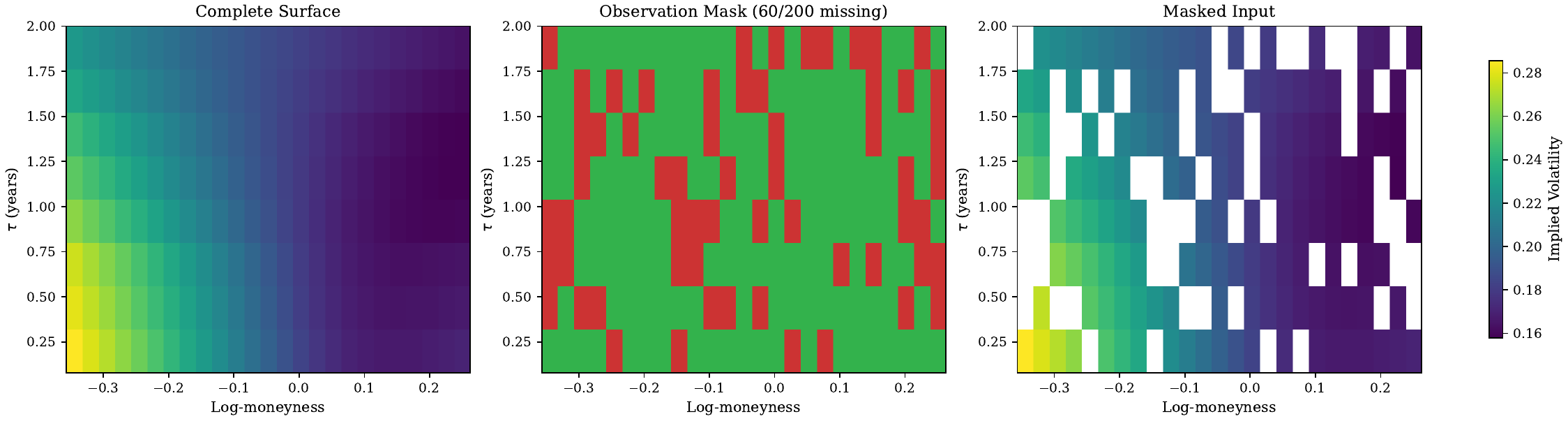}
    \caption{Masking illustration. Left: complete volatility surface. Center: observation mask (green = observed, red = missing, 30\% missing rate). Right: the masked input surface, where missing points are set to zero.}
    \label{fig:masking-illustration}
\end{figure}

Figure~\ref{fig:masking-illustration} illustrates the masking process: a complete Heston surface (left), the observation mask with 30\% of points marked as missing (center), and the resulting masked input that the model receives (right).

\subsection{Real Data Masking}

For real SPY data, the surfaces already have natural missingness from the market. During training, we apply an additional synthetic mask on top of the existing gaps: the observed points are further subsampled at rate $p$. This ensures that the model is trained to reconstruct from even fewer observations than the market naturally provides. The training loss is computed only at points that have valid ground truth (i.e., points that were observed in the original market data but artificially masked during training).

\section{Dataset Pipeline}
\label{sec:pipeline}

The dataset pipeline is designed for reproducibility and efficiency, separating the expensive surface generation step from the fast training loop.

\subsection{Offline Generation}

Surfaces are pre-generated and stored on disk. The \texttt{generate\_and\_save()} function takes a random seed, the number of surfaces, and the grid specification, then:
\begin{enumerate}
    \item Samples Heston parameters (enforcing the Feller condition).
    \item Computes option prices via the characteristic function.
    \item Recovers implied volatilities via Newton-Raphson.
    \item Saves the resulting IV array as a compressed NumPy archive (\texttt{.npz}).
    \item Saves metadata (parameters, grid specification, seed) as JSON.
\end{enumerate}
Using fixed seeds ensures that the exact same dataset is produced on any machine, enabling reproducible experiments.

\subsection{Training-Time Loading}

The \texttt{VolSurfaceDataset} class implements the PyTorch \texttt{Dataset} interface. At initialization, it loads the pre-generated surfaces into memory. Each call to \texttt{\_\_getitem\_\_} returns a four-element tuple:
\begin{enumerate}
    \item \textbf{Input} $\mathbf{X} \in \mathbb{R}^{2 \times 8 \times 25}$: the two-channel masked surface (Section~\ref{sec:surface-representation}).
    \item \textbf{Target} $\boldsymbol{\Sigma} \in \mathbb{R}^{1 \times 8 \times 25}$: the complete IV surface.
    \item \textbf{Mask} $\mathbf{M} \in \{0, 1\}^{8 \times 25}$: the observation mask.
    \item \textbf{Target mask} $\mathbf{T} \in \{0, 1\}^{8 \times 25}$: which points have valid ground truth (all ones for synthetic data; reflects natural missingness for real data).
\end{enumerate}
The mask is generated fresh on each access, providing the data augmentation described in Section~\ref{sec:masking}. For real data, the synthetic mask is combined (via logical AND) with the pre-existing natural missingness mask, ensuring that the model is never asked to reconstruct points for which no ground truth exists.

%% file: chapters/05_methodology.tex

The previous chapter described the data representation, synthetic and real datasets, and masking strategy that frame volatility surface reconstruction as a supervised learning problem. This chapter formalizes the reconstruction task, presents the seven model architectures evaluated in this work, and describes the training protocol, no-arbitrage constraints, and evaluation metrics used throughout the experimental analysis.

\section{Problem Formulation}
\label{sec:problem-formulation}

\subsection{Reconstruction as Supervised Learning}

The volatility surface reconstruction task takes as input a partially observed surface together with an observation mask and produces a complete surface as output. Formally, the model $f_\theta$ maps:
\begin{equation}
f_\theta : \mathbf{X} \in \mathbb{R}^{2 \times 8 \times 25} \;\longrightarrow\; \hat{\boldsymbol{\Sigma}} \in \mathbb{R}^{1 \times 8 \times 25}
\label{eq:reconstruction-map}
\end{equation}
where the two input channels are the masked implied volatility surface and the binary observation mask (as defined in Section~\ref{sec:surface-representation}), and the output is the reconstructed complete surface.

\subsection{Training Objective}

The training loss is computed exclusively on the missing points---those where $M_{ij} = 0$---to prevent the model from simply copying observed values to the output:
\begin{equation}
\mathcal{L}_{\text{MSE}} = \frac{1}{|\{(i,j) : M_{ij} = 0\}|} \sum_{\substack{i,j \\ M_{ij} = 0}} \left( \hat{\Sigma}_{ij} - \Sigma_{ij} \right)^2
\label{eq:mse-loss}
\end{equation}
where $\hat{\Sigma}_{ij}$ is the model's prediction and $\Sigma_{ij}$ is the ground truth implied volatility. For real data, where some grid points lack ground truth entirely, an additional target mask restricts the loss to points with valid ground truth (Section~\ref{sec:pipeline}).

\subsection{Abstract Interface}

All neural network models inherit from a common abstract base class, \texttt{SurfaceReconstructor}, which enforces a standardized interface:
\begin{itemize}
    \item \textbf{Input}: a batch tensor of shape $(B, 2, N_\tau, N_K)$, where $B$ is the batch size.
    \item \textbf{Output}: a batch tensor of shape $(B, 1, N_\tau, N_K)$.
\end{itemize}
This common interface ensures that all models can be trained with the same data pipeline, optimizer, and evaluation code, making comparisons strictly fair.

\subsection{Connection to Image Inpainting}

As discussed in Section~\ref{sec:masked-prediction}, the reconstruction task is structurally analogous to image inpainting and masked autoencoding: given a partially observed 2D grid and a mask, the model must fill in the missing entries. The key differences from generic inpainting are that our task is fully supervised with ground-truth surfaces, and the data has domain-specific structure---smoothness across the strike dimension and monotonicity constraints across the maturity dimension---that can be exploited through the no-arbitrage penalties described in Section~\ref{sec:no-arb-constraints}.

\section{Model Architectures}
\label{sec:architectures}

The central contribution of this thesis is a Transformer encoder-decoder architecture for volatility surface reconstruction, which we propose as a novel application of attention mechanisms to this problem. To rigorously evaluate this proposal, we compare against five neural network baselines and one traditional parametric method, spanning a broad range of inductive biases: from spatially agnostic (MLP) to locally structured (CNN, U-Net) to generative (VAE, Conv VAE), plus the industry-standard SVI parameterization. All neural models are designed to have approximately 288,000 trainable parameters, enabling fair capacity-matched comparisons (Section~\ref{sec:param-normalization}).

This section first presents the baseline architectures (Sections~\ref{sec:mlp}--\ref{sec:unet}), then describes the proposed Transformer model in detail (Section~\ref{sec:transformer}), the generative VAE variants (Section~\ref{sec:vae}), and finally the SVI parametric baseline (Section~\ref{sec:svi-baseline}).

\subsection{Multi-Layer Perceptron (MLP)}
\label{sec:mlp}

The MLP serves as a baseline architecture with no spatial inductive bias. It treats the volatility surface as a flat vector, discarding all 2D spatial structure.

\paragraph{Architecture.} The input tensor $\mathbf{X} \in \mathbb{R}^{2 \times 8 \times 25}$ is flattened to a vector of dimension $2 \times 8 \times 25 = 400$. This vector passes through three hidden layers of 256 units each with ReLU activations, followed by an output layer projecting to 200 dimensions ($8 \times 25$), which is reshaped back to the surface grid:
\begin{equation}
\text{MLP}: \mathbb{R}^{400} \xrightarrow{\text{Linear + ReLU}}_{256} \xrightarrow{\text{Linear + ReLU}}_{256} \xrightarrow{\text{Linear + ReLU}}_{256} \xrightarrow{\text{Linear}} \mathbb{R}^{200}
\label{eq:mlp-arch}
\end{equation}

\paragraph{Key property.} The MLP is \emph{spatially agnostic}: it has no mechanism for exploiting the 2D grid structure or the spatial correlations between neighboring strikes and maturities. Any spatial patterns must be learned entirely from data, making it an important lower bound on what spatial inductive biases contribute.

\paragraph{Parameter count.} Approximately 286k parameters with hidden dimensions $(256, 256, 256)$.

\begin{figure}[H]
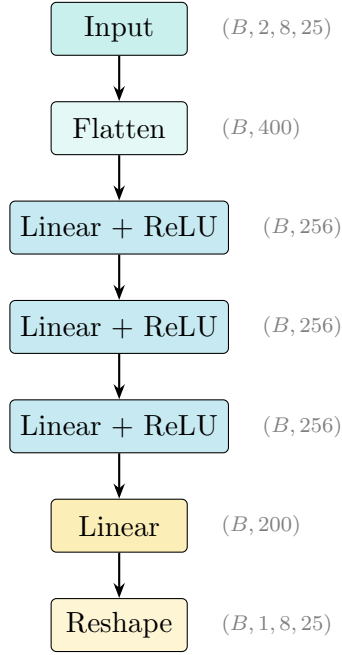

    \centering
    \mlpdiagram
    \caption{MLP architecture. The 2D surface grid is flattened to a vector, processed by three fully connected hidden layers with ReLU activations, and reshaped back to the surface grid. Approximately 286k parameters.}
    \label{fig:mlp-arch}
\end{figure}

\subsection{Convolutional Neural Network (CNN)}
\label{sec:cnn}

The CNN applies local spatial filters to the 2D grid, exploiting the continuity of the volatility smile and term structure through weight sharing and local receptive fields.

\paragraph{Architecture.} The network consists of five convolutional layers, all using $3 \times 3$ kernels with same-padding (padding $= 1$) to preserve the spatial dimensions throughout:
\begin{align}
&\text{Conv2d}(2 \to n_c,\; 3{\times}3,\; \text{pad}=1) + \text{ReLU} \nonumber \\
&\text{Conv2d}(n_c \to n_c,\; 3{\times}3,\; \text{pad}=1) + \text{ReLU} \quad \times\, 3 \nonumber \\
&\text{Conv2d}(n_c \to 1,\; 3{\times}3,\; \text{pad}=1)
\label{eq:cnn-arch}
\end{align}
where $n_c = 104$ is the number of channels in each hidden layer. There is no batch normalization, no pooling, and no activation on the final layer. The spatial dimensions remain $8 \times 25$ throughout the entire forward pass.

\paragraph{Key property.} The CNN has a \emph{local receptive field}: with five $3 \times 3$ layers, each output point aggregates information from a $11 \times 11$ neighborhood (limited by the grid size to the full $8 \times 25$ grid along the maturity axis). This local structure is well-suited to the smooth, continuous nature of volatility surfaces, where nearby points in strike and maturity space have similar implied volatilities.

\paragraph{Parameter count.} Approximately 295k parameters with $n_c = 104$.

\begin{figure}[H]
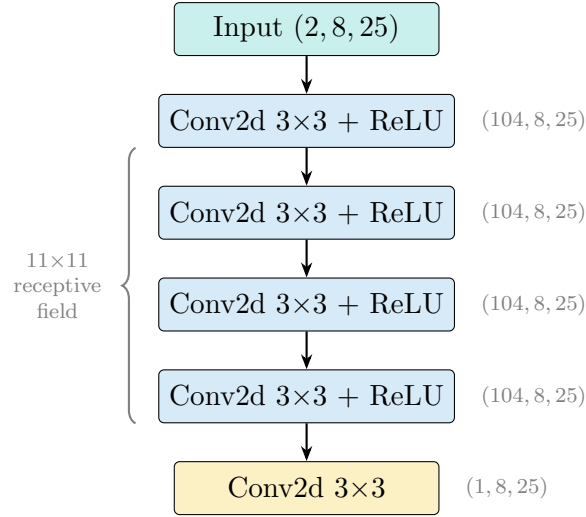

    \centering
    \cnndiagram
    \caption{CNN architecture. Five convolutional layers with $3 \times 3$ kernels and same-padding maintain the $8 \times 25$ spatial dimensions throughout. No pooling or batch normalization is used. The receptive field grows to $11 \times 11$ through successive layers. Approximately 295k parameters with $n_c = 104$.}
    \label{fig:cnn-arch}
\end{figure}

\subsection{U-Net}
\label{sec:unet}

The U-Net \cite{ronneberger2015unet} combines multi-scale feature extraction with skip connections, allowing it to capture both local detail and global context. Originally developed for biomedical image segmentation, the encoder-decoder structure is naturally suited to dense prediction tasks like surface reconstruction.

\paragraph{Architecture.} The network uses two downsampling levels, adapted for the small $8 \times 25$ grid (deeper architectures would reduce spatial dimensions below useful resolution). Each level uses a \emph{ConvBlock} consisting of two consecutive $3 \times 3$ convolutions with ReLU activations:
\begin{equation}
\text{ConvBlock}(c_{\text{in}}, c_{\text{out}}): \quad \text{Conv2d}(c_{\text{in}}, c_{\text{out}}) \to \text{ReLU} \to \text{Conv2d}(c_{\text{out}}, c_{\text{out}}) \to \text{ReLU}
\end{equation}

The full architecture with base channels $b = 24$ is:

\paragraph{Encoder:}
\begin{align}
\mathbf{e}_1 &= \text{ConvBlock}(2, b) & &\text{shape: } (b, 8, 25) \nonumber \\
\mathbf{p}_1 &= \text{MaxPool2d}(2)(\mathbf{e}_1) & &\text{shape: } (b, 4, 12) \nonumber \\
\mathbf{e}_2 &= \text{ConvBlock}(b, 2b) & &\text{shape: } (2b, 4, 12) \nonumber \\
\mathbf{p}_2 &= \text{MaxPool2d}(2)(\mathbf{e}_2) & &\text{shape: } (2b, 2, 6)
\end{align}

\paragraph{Bottleneck:}
\begin{equation}
\mathbf{z} = \text{ConvBlock}(2b, 4b) \qquad \text{shape: } (4b, 2, 6)
\end{equation}

\paragraph{Decoder:}
\begin{align}
\mathbf{u}_1 &= \text{Upsample}(\mathbf{z}) \;\|\; \mathbf{e}_2 & &\text{shape: } (4b + 2b, 4, 12) \nonumber \\
\mathbf{d}_1 &= \text{ConvBlock}(6b, 2b) & &\text{shape: } (2b, 4, 12) \nonumber \\
\mathbf{u}_2 &= \text{Upsample}(\mathbf{d}_1) \;\|\; \mathbf{e}_1 & &\text{shape: } (2b + b, 8, 25) \nonumber \\
\mathbf{d}_2 &= \text{ConvBlock}(3b, b) & &\text{shape: } (b, 8, 25)
\end{align}

\paragraph{Output:}
\begin{equation}
\hat{\boldsymbol{\Sigma}} = \text{Conv2d}(b, 1, \; 1{\times}1)(\mathbf{d}_2) \qquad \text{shape: } (1, 8, 25)
\end{equation}

The $\|$ symbol denotes channel-wise concatenation along the feature dimension. Upsampling uses bilinear interpolation followed by size matching to handle the non-power-of-two spatial dimensions (since $25/2 = 12$ after pooling, recovering the original 25 requires interpolation to the exact target size). Skip connections from the encoder to the decoder preserve fine-grained spatial information that would otherwise be lost through downsampling.

\paragraph{Parameter count.} Approximately 265k parameters with $b = 24$.

\begin{figure}[H]
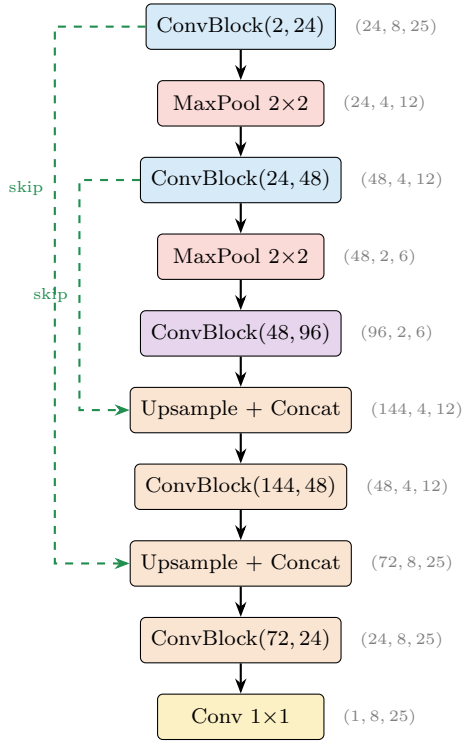

    \centering
    \unetdiagram
    \caption{U-Net architecture adapted for the $8 \times 25$ volatility surface grid. Two downsampling levels with skip connections (dashed green) balance local detail (from encoder features) with global context (from the bottleneck). Approximately 265k parameters with base channels $b = 24$.}
    \label{fig:unet-arch}
\end{figure}

\subsection{Proposed Architecture: Transformer Encoder-Decoder}
\label{sec:transformer}

We propose an encoder-decoder Transformer \cite{vaswani2017attention} for volatility surface reconstruction. The key insight motivating this architecture is that the reconstruction problem has three properties that are naturally addressed by attention:

\begin{enumerate}
    \item \textbf{Variable observation patterns.} The set of observed points changes from surface to surface. Unlike convolutional models, which process a fixed grid regardless of where observations fall, the Transformer encoder processes only the observed tokens through self-attention, adapting its computation to the specific observation pattern.
    
    \item \textbf{Long-range dependencies.} Information at one point on the volatility surface is informative about distant points: the short-tenor smile curvature constrains the long-tenor smile shape, and the ATM term structure constrains the wing behavior. Attention captures these global relationships directly, without requiring information to propagate through multiple convolutional layers.
    
    \item \textbf{Financial coordinate awareness.} Each grid point has a physical meaning---a specific maturity $\tau$ and log-moneyness $m$. By encoding these coordinates as Fourier features, the attention mechanism can learn position-dependent patterns (e.g., that short-tenor points are more informative about smile curvature than long-tenor points).
\end{enumerate}

Related concurrent work by Zhang et al.\ \cite{zhang2025volnp} applies an attention-based neural process to implied volatility surface \emph{fitting} from irregular market quotes, using SABR-generated surfaces \cite{hagan2002sabr} as a structural prior via meta-learning. While the underlying encoder-decoder attention architecture is similar in spirit, the problem formulations are distinct: VolNP fits from a variable number of noisy real quotes at arbitrary positions, whereas our formulation operates on a fixed discretized grid with controlled random masking, enabling systematic evaluation across sparsity levels. Furthermore, we integrate differentiable no-arbitrage penalties directly into the training loss (rather than checking compliance post-hoc), and we conduct a controlled comparison against five alternative architectures at matched parameter counts to isolate the specific contribution of attention to reconstruction quality.

\subsubsection{Token Representation}

The $8 \times 25 = 200$ grid points are flattened into a sequence of 200 tokens. Each token corresponds to a single point $(\tau_i, m_j)$ on the volatility surface and carries two types of information:
\begin{itemize}
    \item \textbf{Value}: the implied volatility $\sigma_{ij}$ at that grid point (set to zero for missing points).
    \item \textbf{Position}: a coordinate encoding of the token's position in $(\tau, m)$ space.
\end{itemize}

\subsubsection{Coordinate Positional Encoding}

Rather than using standard learned or sinusoidal positional encodings indexed by sequence position, we use \emph{coordinate-based} Fourier features inspired by Tancik et al.\ \cite{tancik2020fourier} that encode each token's physical coordinates $(\tau_i, m_j)$ directly. While the original formulation uses random sampled frequencies, we use deterministic log-spaced frequencies, which suffice for our low-dimensional (2D) input space. For a scalar coordinate $x$, the encoding with $L$ frequency bands is:
\begin{equation}
\gamma(x) = \left[ x,\; \sin(\pi x),\; \cos(\pi x),\; \sin(2\pi x),\; \cos(2\pi x),\; \ldots,\; \sin(2^{L-1}\pi x),\; \cos(2^{L-1}\pi x) \right]
\label{eq:fourier-encoding}
\end{equation}
yielding $1 + 2L$ features per coordinate. For the two-dimensional coordinates $(\tau_i, m_j)$, the full encoding concatenates $\gamma(\tau_i)$ and $\gamma(m_j)$:
\begin{equation}
\mathbf{p}_{ij} = \left[ \gamma(\tau_i) \;\|\; \gamma(m_j) \right] \in \mathbb{R}^{d_{\text{coord}}}
\label{eq:coord-encoding}
\end{equation}
where $d_{\text{coord}} = 2(1 + 2L)$. With $L = 8$ frequency bands (frequencies $\pi, 2\pi, \ldots, 128\pi$), this gives $d_{\text{coord}} = 34$ features per token.

This coordinate encoding has two advantages over standard positional encodings. First, it directly encodes the \emph{physical} meaning of each position: the maturity $\tau$ and log-moneyness $m$ are continuous quantities with financial significance, not arbitrary sequence indices. Second, the multi-scale sinusoidal features enable the attention mechanism to distinguish positions at different resolutions, from the coarse distinction between short and long maturities to fine-grained differences between adjacent strikes. The coordinate encoding is computed once from the fixed grid and stored as a non-learnable buffer.

To validate this design choice, we conduct an ablation study (Section~\ref{sec:ablation}) replacing the Fourier encoding with standard learnable positional embeddings of the same dimensionality ($d_{\text{coord}} = 34$). The Fourier encoding reduces reconstruction error by approximately 18\%, confirming that encoding physical coordinates is a meaningful architectural contribution rather than an incidental design choice.

\subsubsection{Encoder}

The encoder processes only the \emph{observed} tokens. Each token's input is formed by concatenating its IV value (a scalar) with its coordinate encoding:
\begin{equation}
\mathbf{x}_{ij}^{\text{enc}} = \left[ \sigma_{ij} \;\|\; \mathbf{p}_{ij} \right] \in \mathbb{R}^{1 + d_{\text{coord}}}
\end{equation}
An input projection maps this to the model dimension:
\begin{equation}
\mathbf{h}_{ij}^{(0)} = \text{Linear}(1 + d_{\text{coord}},\; d_{\text{model}})(\mathbf{x}_{ij}^{\text{enc}})
\end{equation}

The projected tokens pass through $N_{\text{enc}} = 3$ Transformer encoder layers. Each layer consists of multi-head self-attention followed by a position-wise feed-forward network, with pre-layer normalization and residual connections:
\begin{align}
\mathbf{a}^{(\ell)} &= \mathbf{h}^{(\ell-1)} + \text{MHA}\!\left(\text{LN}(\mathbf{h}^{(\ell-1)}),\; \text{LN}(\mathbf{h}^{(\ell-1)}),\; \text{LN}(\mathbf{h}^{(\ell-1)})\right) \\
\mathbf{h}^{(\ell)} &= \mathbf{a}^{(\ell)} + \text{FFN}\!\left(\text{LN}(\mathbf{a}^{(\ell)})\right)
\end{align}
where $\text{MHA}$ denotes multi-head attention with $h = 4$ heads and $\text{FFN}$ is a two-layer network with hidden dimension $d_{\text{ff}} = 256$ and GELU activation. Missing tokens are excluded from the self-attention computation via a \texttt{src\_key\_padding\_mask} that marks positions where $M_{ij} = 0$.

\subsubsection{Decoder}

The decoder takes \emph{all} 200 tokens as queries---both observed and missing---and cross-attends to the encoder's output memory. For missing tokens, the IV value in the query is set to zero (a fixed initialization, not a learned mask token). The query input is:
\begin{equation}
\mathbf{x}_{ij}^{\text{dec}} = \left[ \sigma_{ij} \cdot M_{ij} \;\|\; \mathbf{p}_{ij} \right] \in \mathbb{R}^{1 + d_{\text{coord}}}
\end{equation}
projected through a separate linear layer to $d_{\text{model}}$ dimensions.

The decoder consists of $N_{\text{dec}} = 2$ Transformer decoder layers. Each layer applies self-attention among all query tokens, then cross-attention to the encoder memory:
\begin{align}
\mathbf{a}_{\text{self}}^{(\ell)} &= \mathbf{q}^{(\ell-1)} + \text{MHA}_{\text{self}}\!\left(\text{LN}(\mathbf{q}^{(\ell-1)})\right) \\
\mathbf{a}_{\text{cross}}^{(\ell)} &= \mathbf{a}_{\text{self}}^{(\ell)} + \text{MHA}_{\text{cross}}\!\left(\text{LN}(\mathbf{a}_{\text{self}}^{(\ell)}),\; \text{LN}(\mathbf{m}_{\text{enc}})\right) \\
\mathbf{q}^{(\ell)} &= \mathbf{a}_{\text{cross}}^{(\ell)} + \text{FFN}\!\left(\text{LN}(\mathbf{a}_{\text{cross}}^{(\ell)})\right)
\end{align}
where $\mathbf{m}_{\text{enc}}$ is the encoder output. A \texttt{memory\_key\_padding\_mask} ensures that cross-attention only attends to positions where observations exist. This mechanism is the key to the Transformer's reconstruction ability: missing tokens aggregate information from all observed tokens, weighted by learned attention scores that can capture both local correlations (nearby strikes) and long-range dependencies (across maturities).

\subsubsection{Output Projection}

A final linear layer maps each token's representation from $d_{\text{model}}$ to a scalar predicted implied volatility:
\begin{equation}
\hat{\sigma}_{ij} = \text{Linear}(d_{\text{model}}, 1)(\mathbf{q}_{ij}^{(N_{\text{dec}})})
\end{equation}
The 200 scalar predictions are reshaped back to the $(1, 8, 25)$ surface grid.

\paragraph{Parameter count.} Approximately 288k parameters with $d_{\text{model}} = 64$, $N_{\text{enc}} = 3$, $N_{\text{dec}} = 2$, $h = 4$, and $d_{\text{ff}} = 256$.

\begin{figure}[H]
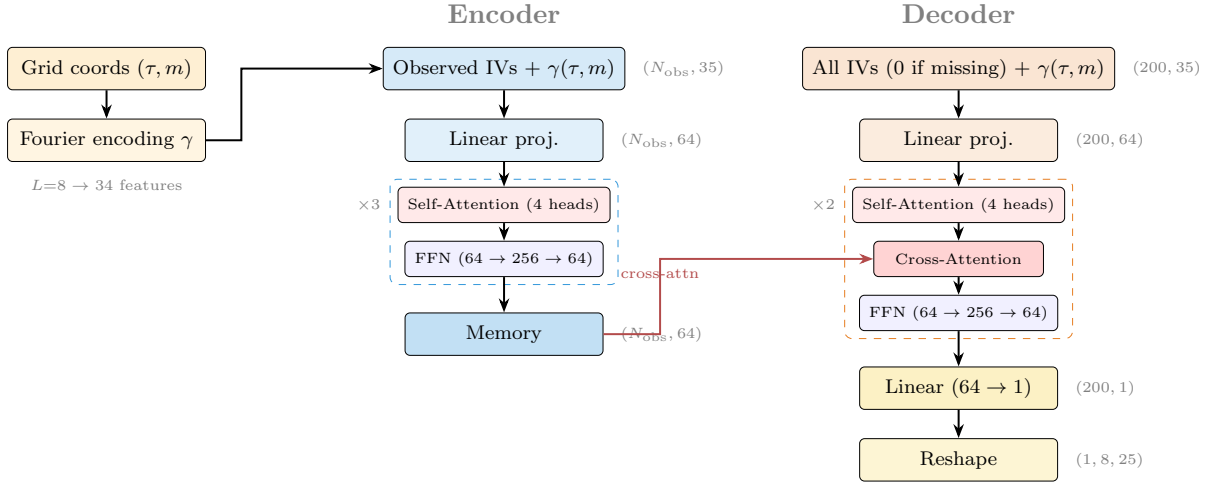

    \centering
    \resizebox{\textwidth}{!}{\transformerdiagram}
    \caption{Transformer architecture for volatility surface reconstruction. Left: coordinate-based Fourier encoding maps grid positions $(\tau, m)$ to 34-dimensional feature vectors. Center: the encoder processes only observed tokens through 3 self-attention layers. Right: the decoder takes all 200 tokens as queries and cross-attends to the encoder memory, reconstructing the full surface. Approximately 288k parameters.}
    \label{fig:transformer-arch}
\end{figure}

\subsection{Variational Autoencoders}
\label{sec:vae}

Variational autoencoders \cite{kingma2014vae} learn a low-dimensional latent representation of the volatility surface manifold. Unlike the discriminative models above, which directly map inputs to outputs, VAEs model the data distribution and reconstruct surfaces by projecting onto the learned manifold. We evaluate two variants: a fully connected (FC) VAE and a convolutional (Conv) VAE.

\subsubsection{FC VAE}

The fully connected VAE flattens the input surface to a vector and processes it through a symmetric encoder-decoder architecture:

\paragraph{Encoder:}
\begin{align}
\mathbf{h}_1 &= \text{ELU}\!\left(\text{Linear}(400, 288)(\text{flatten}(\mathbf{X}))\right) \nonumber \\
\mathbf{h}_2 &= \text{ELU}\!\left(\text{Linear}(288, 144)(\mathbf{h}_1)\right) \nonumber \\
\mathbf{h}_3 &= \text{ELU}\!\left(\text{Linear}(144, 72)(\mathbf{h}_2)\right) \nonumber \\
\boldsymbol{\mu} &= \text{Linear}(72, d_z)(\mathbf{h}_3), \quad \log \boldsymbol{\sigma}^2 = \text{Linear}(72, d_z)(\mathbf{h}_3)
\end{align}

\paragraph{Reparameterization:}
\begin{equation}
\mathbf{z} = \boldsymbol{\mu} + \boldsymbol{\sigma} \odot \boldsymbol{\epsilon}, \quad \boldsymbol{\epsilon} \sim \mathcal{N}(\mathbf{0}, \mathbf{I})
\label{eq:reparam}
\end{equation}
During inference, the stochastic sampling is replaced by the deterministic posterior mean $\mathbf{z} = \boldsymbol{\mu}$.

\paragraph{Decoder:}
\begin{align}
\mathbf{g}_1 &= \text{ELU}\!\left(\text{Linear}(d_z, 72)(\mathbf{z})\right) \nonumber \\
\mathbf{g}_2 &= \text{ELU}\!\left(\text{Linear}(72, 144)(\mathbf{g}_1)\right) \nonumber \\
\mathbf{g}_3 &= \text{ELU}\!\left(\text{Linear}(144, 288)(\mathbf{g}_2)\right) \nonumber \\
\hat{\boldsymbol{\Sigma}} &= \text{reshape}\!\left(\text{Linear}(288, 200)(\mathbf{g}_3)\right)
\end{align}

The encoder progressively compresses the 400-dimensional input through layers of width 288, 144, and 72 to the latent space, while the decoder mirrors this structure in reverse. ELU activations (rather than ReLU) are used throughout, as their non-zero gradient for negative inputs improves gradient flow through the narrow bottleneck layers. The latent dimension is $d_z = 32$, compressing the 200-dimensional surface to a 32-dimensional representation. Approximately 285k parameters.

\subsubsection{Conv VAE}

The convolutional VAE exploits the 2D spatial structure of the surface grid using strided convolutions for downsampling and bilinear interpolation for upsampling:

\paragraph{Encoder:}
\begin{align}
\mathbf{e}_1 &= \text{ReLU}\!\left(\text{Conv2d}(2, b, \; 3{\times}3, \; \text{stride}=1, \; \text{pad}=1)\right) & &(b, 8, 25) \nonumber \\
\mathbf{e}_2 &= \text{ReLU}\!\left(\text{Conv2d}(b, 2b, \; 3{\times}3, \; \text{stride}=2, \; \text{pad}=1)\right) & &(2b, 4, 13) \nonumber \\
\mathbf{e}_3 &= \text{ReLU}\!\left(\text{Conv2d}(2b, 4b, \; 3{\times}3, \; \text{stride}=2, \; \text{pad}=1)\right) & &(4b, 2, 7)
\end{align}

The encoder output is flattened to $4b \times 2 \times 7$ dimensions, from which two linear layers produce $\boldsymbol{\mu}$ and $\log \boldsymbol{\sigma}^2$ of the latent distribution. Reparameterization follows Equation~\eqref{eq:reparam}.

\paragraph{Decoder:}
\begin{align}
\mathbf{g}_0 &= \text{reshape}\!\left(\text{ReLU}\!\left(\text{Linear}(d_z, 4b \times 2 \times 7)\right)\right) & &(4b, 2, 7) \nonumber \\
\mathbf{g}_1 &= \text{ReLU}\!\left(\text{Conv2d}(4b, 2b, \; 3{\times}3, \; \text{pad}=1)\right) \circ \text{Upsample}(4, 13) & &(2b, 4, 13) \nonumber \\
\mathbf{g}_2 &= \text{ReLU}\!\left(\text{Conv2d}(2b, b, \; 3{\times}3, \; \text{pad}=1)\right) \circ \text{Upsample}(8, 25) & &(b, 8, 25) \nonumber \\
\hat{\boldsymbol{\Sigma}} &= \text{Conv2d}(b, 1, \; 3{\times}3, \; \text{pad}=1) & &(1, 8, 25)
\end{align}

Upsampling uses bilinear interpolation to target sizes, avoiding the checkerboard artifacts associated with transposed convolutions. With base channels $b = 32$ and latent dimension $d_z = 16$, the Conv VAE has approximately 273k parameters.

\subsubsection{Training Loss: Evidence Lower Bound}

Both VAE variants are trained by maximizing the evidence lower bound (ELBO), which decomposes into a reconstruction term and a regularization term:
\begin{equation}
\mathcal{L}_{\text{ELBO}} = \mathcal{L}_{\text{MSE}} + \beta \cdot D_{\text{KL}}\!\left(q_\phi(\mathbf{z} | \mathbf{X}) \;\|\; p(\mathbf{z})\right)
\label{eq:elbo}
\end{equation}
where $\mathcal{L}_{\text{MSE}}$ is the standard reconstruction loss (Equation~\eqref{eq:mse-loss}) and the KL divergence for the Gaussian encoder has the closed-form expression:
\begin{equation}
D_{\text{KL}} = -\frac{1}{2} \sum_{k=1}^{d_z} \left(1 + \log \sigma_k^2 - \mu_k^2 - \sigma_k^2\right)
\end{equation}
The weighting factor $\beta = 10^{-4}$ keeps the KL term small relative to the reconstruction loss, prioritizing accurate surface reconstruction over latent space regularity.

\subsubsection{Latent Space Optimization at Inference}

A na\"ive approach to reconstruction with a VAE would encode the partial observation and decode the resulting latent vector. However, encoding a partially observed surface---where missing values are replaced with zeros---distorts the posterior estimate. Instead, we use \emph{latent space optimization}: given a partial observation, we find the latent vector $\mathbf{z}^*$ that minimizes the reconstruction error on observed points:
\begin{equation}
\mathbf{z}^* = \arg\min_{\mathbf{z}} \sum_{\substack{i,j \\ M_{ij} = 1}} \left( \text{decode}(\mathbf{z})_{ij} - \sigma_{ij}^{\text{obs}} \right)^2
\label{eq:latent-opt}
\end{equation}
This is solved by gradient descent: $\mathbf{z}$ is initialized from the encoder's posterior mean and then optimized with Adam for 200 steps at a learning rate of 0.01. The decoder weights are frozen during this optimization. The final reconstruction $\text{decode}(\mathbf{z}^*)$ projects the incomplete observation onto the learned manifold of complete surfaces, naturally filling in missing values with smooth, plausible completions.

\begin{figure}[H]
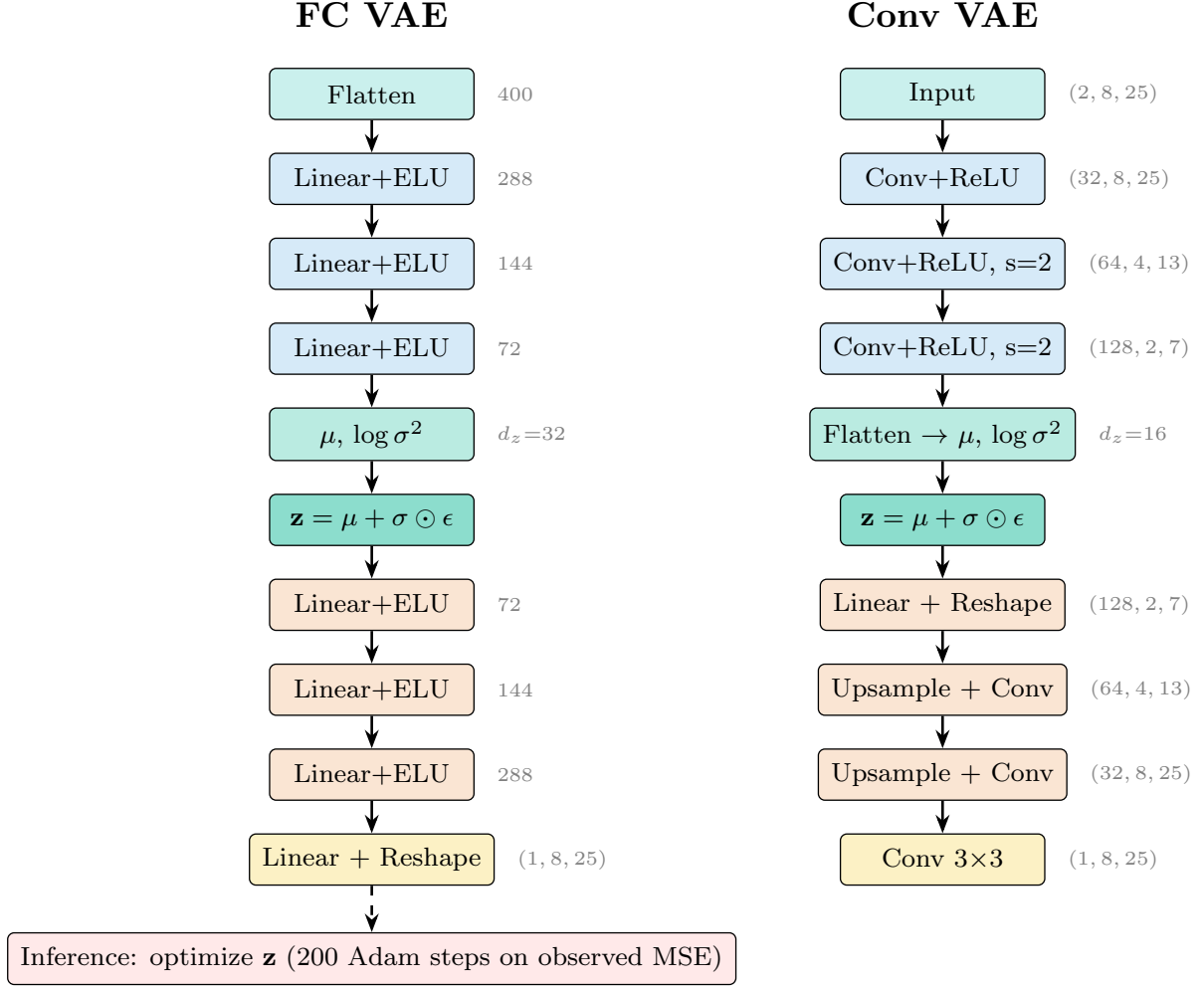

    \centering
    \resizebox{\textwidth}{!}{\vaediagram}
    \caption{VAE architectures. Left: FC VAE with fully connected encoder-decoder and latent dimension $d_z = 32$. Right: Conv VAE with strided convolutions and latent dimension $d_z = 16$. Both use reparameterized sampling during training. At inference, the latent vector $\mathbf{z}$ is optimized via 200 Adam steps to minimize reconstruction error on observed points, projecting partial observations onto the learned surface manifold.}
    \label{fig:vae-arch}
\end{figure}

\subsection{SVI Baseline}
\label{sec:svi-baseline}

The SVI parameterization (Section~\ref{sec:svi-background}) serves as our traditional (non-learning) baseline, using the raw SVI formula (Equation~\eqref{eq:svi-raw}) with five parameters per maturity slice.

\paragraph{Fitting procedure.} For each of the 8 maturity slices independently, the 5 SVI parameters are optimized to minimize the mean squared error between the parameterized and observed implied variances at the observed strikes, using the L-BFGS-B optimizer \cite{byrd1995limited} with bound constraints ($b \geq 0$, $|\rho| \leq 1$, $\sigma > 0$). This gives $5 \times 8 = 40$ parameters per surface. The key limitation is that SVI fits each slice independently with no cross-maturity information sharing, which becomes increasingly severe as sparsity grows (Chapter~\ref{ch:results}).

\section{Parameter Count Normalization}
\label{sec:param-normalization}

A fair comparison between architectures requires controlling for model capacity. A larger model could achieve better performance simply by having more parameters to memorize training patterns, obscuring the contribution of architectural inductive biases. We therefore normalize all neural models to approximately 288,000 trainable parameters by adjusting the width (number of hidden units or channels) of each architecture.

\begin{table}[H]
    \centering
    \caption{Parameter counts for all models, tuned to approximately 288k parameters. The SVI baseline is fundamentally different (per-surface optimization, no trainable neural parameters).}
    \label{tab:param-counts}
    \begin{tabular}{llr}
        \toprule
        Model & Key Width Parameter & Parameters \\
        \midrule
        MLP & hidden\_dims = (256, 256, 256) & $\sim$286k \\
        CNN & n\_channels = 104 & $\sim$295k \\
        U-Net & base\_channels = 24 & $\sim$265k \\
        Transformer & d\_model = 64 & $\sim$288k \\
        FC VAE & hidden\_dims = (288, 144, 72), latent = 32 & $\sim$285k \\
        Conv VAE & base\_channels = 32, latent = 16 & $\sim$273k \\
        \midrule
        SVI & 5 params $\times$ 8 slices & 40/surface \\
        \bottomrule
    \end{tabular}
\end{table}

The SVI baseline cannot be meaningfully compared on parameter count: it has 40 parameters per surface (re-optimized from scratch for each test surface), whereas the neural models have $\sim$288k shared parameters that generalize across all surfaces. This reflects a fundamental difference in approach: SVI is a per-instance optimization method, while the neural models are amortized learners.

\section{Training Protocol}
\label{sec:training-protocol}

All models are implemented in PyTorch \cite{paszke2019pytorch} and trained with a standardized protocol to ensure that performance differences reflect architectural choices rather than training configuration.

\subsection{Optimization}

\begin{table}[H]
    \centering
    \caption{Training hyperparameters. All models use the same configuration except where noted.}
    \label{tab:training-hyperparams}
    \begin{tabular}{ll}
        \toprule
        Hyperparameter & Value \\
        \midrule
        Optimizer & Adam \\
        Learning rate & $10^{-3}$ (Transformer: $10^{-4}$) \\
        Weight decay & 0 \\
        Batch size & 32 \\
        Maximum iterations & 500 \\
        Early stopping patience & 30 iterations \\
        Learning rate schedule & None (constant) \\
        Loss function & MSE (VAEs: ELBO with $\beta = 10^{-4}$) \\
        \bottomrule
    \end{tabular}
\end{table}

The Transformer uses a lower learning rate ($10^{-4}$) because attention-based architectures are sensitive to large gradient updates, particularly in the early stages of training when attention patterns are not yet established. We deliberately avoid learning rate schedulers (such as cosine annealing) to eliminate a confounding variable: different architectures may benefit unequally from scheduled learning rates, making it difficult to attribute performance differences to the architecture itself.

\subsection{Early Stopping}

Training terminates when the validation loss has not improved for 30 consecutive iterations. The model checkpoint with the lowest validation loss is retained for evaluation. This prevents overfitting while allowing sufficient training time for all architectures to converge. In practice, most models converge within 100--200 iterations; the 500-iteration maximum is rarely reached.

\subsection{Fine-Tuning Protocol}

For transfer learning experiments (synthetic $\to$ real data), pretrained models are fine-tuned with reduced learning rates to preserve the representations learned from synthetic data:

\begin{table}[H]
    \centering
    \caption{Fine-tuning hyperparameters for transfer learning from synthetic to real data.}
    \label{tab:ft-hyperparams}
    \begin{tabular}{lcc}
        \toprule
        Model & Learning Rate & Additional Changes \\
        \midrule
        MLP, CNN, U-Net & $10^{-5}$ & None \\
        Transformer & $10^{-4}$ & Dropout $0.0 \to 0.05$ \\
        \bottomrule
    \end{tabular}
\end{table}

The Transformer retains a higher fine-tuning learning rate and adds a small dropout ($0.05$) to prevent overfitting to the smaller real dataset (2,912 training surfaces vs.\ 8,000 synthetic). All other training parameters (batch size, patience, etc.) remain unchanged.

\section{No-Arbitrage Constraints}
\label{sec:no-arb-constraints}

Volatility surfaces are subject to static no-arbitrage conditions that must hold for option prices to be consistent with a single underlying probability measure. Violations of these conditions create riskless profit opportunities and indicate that the surface is not economically meaningful. We incorporate two key constraints as differentiable penalty terms in the training loss.

\subsection{Calendar Spread Constraint}

The \emph{calendar spread} condition requires that total implied variance $w(m, \tau) = \sigma^2(m, \tau) \cdot \tau$ be non-decreasing in maturity $\tau$ for each fixed log-moneyness $m$:
\begin{equation}
w(m, \tau_1) \leq w(m, \tau_2) \quad \text{for all } \tau_1 < \tau_2 \text{ and all } m
\label{eq:calendar-constraint}
\end{equation}
Violation of this condition implies negative forward variance, which would allow a riskless profit by selling a shorter-dated option and buying a longer-dated option at the same strike.

The differentiable penalty is computed using forward differences along the maturity axis:
\begin{equation}
\mathcal{P}_{\text{cal}} = \frac{1}{N} \sum_{i,j} \left[\text{ReLU}\!\left(-\Delta w_{ij}\right)\right]^2, \quad \Delta w_{ij} = w(\tau_{i+1}, m_j) - w(\tau_i, m_j)
\label{eq:calendar-penalty}
\end{equation}
where $N$ is the number of adjacent maturity pairs times the number of strikes. The ReLU ensures that only violations (negative differences) contribute to the penalty, and the squaring makes the penalty smooth and differentiable at the boundary.

\subsection{Butterfly Spread Constraint}

The \emph{butterfly spread} condition requires that total implied variance be convex in log-moneyness $m$ for each fixed maturity $\tau$:
\begin{equation}
\frac{\partial^2 w}{\partial m^2}(m, \tau) \geq 0 \quad \text{for all } m \text{ and } \tau
\label{eq:butterfly-constraint}
\end{equation}
Violation of this condition implies a negative probability density in the risk-neutral distribution, which is economically meaningless and permits arbitrage via butterfly spreads.

For our non-uniformly spaced log-moneyness grid, the second derivative is approximated by finite differences with unequal spacing. Let $h_j^- = m_j - m_{j-1}$ and $h_j^+ = m_{j+1} - m_j$ be the left and right step sizes. The second-order finite difference is:
\begin{equation}
\frac{\partial^2 w}{\partial m^2}\bigg|_{m_j} \approx \frac{2}{h_j^- + h_j^+} \left( \frac{w_{j-1}}{h_j^-} - w_j \left(\frac{1}{h_j^-} + \frac{1}{h_j^+}\right) + \frac{w_{j+1}}{h_j^+} \right)
\label{eq:butterfly-fd}
\end{equation}

The penalty follows the same structure as the calendar penalty:
\begin{equation}
\mathcal{P}_{\text{but}} = \frac{1}{N} \sum_{i,j} \left[\text{ReLU}\!\left(-\frac{\partial^2 w}{\partial m^2}\bigg|_{(\tau_i, m_j)}\right)\right]^2
\label{eq:butterfly-penalty}
\end{equation}

\subsection{Combined Constrained Loss}

The full training loss with no-arbitrage penalties is:
\begin{equation}
\mathcal{L} = \mathcal{L}_{\text{MSE}} + \lambda_{\text{cal}} \cdot \mathcal{P}_{\text{cal}} + \lambda_{\text{but}} \cdot \mathcal{P}_{\text{but}}
\label{eq:combined-loss}
\end{equation}

In practice, we set $\lambda_{\text{cal}} = 0$ because calendar spread violations are negligible across all models (rates below 0.1\%---the smooth maturity axis of the Heston model and the relatively coarse 8-tenor grid make calendar violations rare). The effective constrained loss is therefore:
\begin{equation}
\mathcal{L} = \mathcal{L}_{\text{MSE}} + \lambda \cdot \mathcal{P}_{\text{but}}
\label{eq:effective-loss}
\end{equation}
where $\lambda$ controls the trade-off between reconstruction accuracy and arbitrage-free smoothness. The impact of varying $\lambda$ is analyzed in Chapter~\ref{ch:arbitrage}.

Both penalties are fully differentiable and computed on the \emph{predicted} surface (not the ground truth), providing gradient signal that steers the model toward arbitrage-free outputs during training. Importantly, the penalties operate on the model's output regardless of which points were observed or missing, encouraging global surface coherence.

\section{Evaluation Metrics}
\label{sec:evaluation-metrics}

Model performance is assessed along two axes: \emph{reconstruction accuracy} (how well the model fills in missing values) and \emph{economic consistency} (whether the reconstructed surface respects no-arbitrage conditions).

\subsection{Reconstruction Accuracy}

The primary metric is the root mean squared error on missing points:
\begin{equation}
\text{RMSE}_{\text{miss}} = \sqrt{\frac{1}{|\mathcal{M}|} \sum_{(i,j) \in \mathcal{M}} \left(\hat{\Sigma}_{ij} - \Sigma_{ij}\right)^2}
\label{eq:rmse-miss}
\end{equation}
where $\mathcal{M} = \{(i,j) : M_{ij} = 0\}$ is the set of missing points. This is the most informative metric because it directly measures the model's ability to predict values it has never seen, without contamination from the trivially-accurate observed points.

Additional metrics provide complementary views:
\begin{itemize}
    \item $\text{RMSE}_{\text{obs}}$: RMSE on observed points, measuring how well the model preserves known values.
    \item $\text{MAE}$: Mean absolute error, less sensitive to outliers than RMSE.
    \item $\text{MaxErr}$: Maximum absolute error across all points, capturing worst-case behavior.
\end{itemize}

All metrics are computed per-surface and then averaged across the test set. We also report per-surface RMSE distributions to assess variability---a model with low average RMSE but high variance may produce occasional poor reconstructions that are unacceptable in practice.

\subsection{Regional Analysis}

Reconstruction difficulty varies across the volatility surface: deep out-of-the-money regions have less liquid data and steeper gradients, while near-the-money regions are more densely observed and smoother. To capture this heterogeneity, we partition the grid into regions along both axes.

\begin{table}[H]
    \centering
    \caption{Moneyness and tenor regions for regional error analysis. Log-moneyness boundaries are chosen to separate deep OTM puts, near-the-money, and deep OTM calls.}
    \label{tab:regions}
    \begin{tabular}{lll}
        \toprule
        \multicolumn{3}{l}{\textbf{Moneyness Regions}} \\
        \midrule
        Region & Log-moneyness range & Interpretation \\
        \midrule
        Deep OTM put & $m < -0.2$ & Far left wing \\
        OTM put & $-0.2 \leq m < -0.05$ & Left wing \\
        ATM & $-0.05 \leq m \leq 0.05$ & At-the-money \\
        OTM call & $0.05 < m \leq 0.2$ & Right wing \\
        Deep OTM call & $m > 0.2$ & Far right wing \\
        \midrule
        \multicolumn{3}{l}{\textbf{Tenor Regions}} \\
        \midrule
        Region & Tenor range & Grid points \\
        \midrule
        Short & $\tau \leq 0.25$ yr & $\tau \in \{0.08, 0.17, 0.25\}$ \\
        Medium & $0.25 < \tau \leq 1.0$ yr & $\tau \in \{0.5, 0.75, 1.0\}$ \\
        Long & $\tau > 1.0$ yr & $\tau \in \{1.5, 2.0\}$ \\
        \bottomrule
    \end{tabular}
\end{table}

RMSE is computed separately within each region, producing a $5 \times 3$ error matrix that reveals where each model excels or struggles.

\subsection{Arbitrage Violation Metrics}

We evaluate economic consistency using the same calendar and butterfly conditions described in Section~\ref{sec:no-arb-constraints}, applied to the predicted surfaces at evaluation time. The key metrics are:

\begin{itemize}
    \item \textbf{Violation rate}: The fraction of grid points (or adjacent pairs) where the condition is violated. Reported separately for calendar and butterfly constraints.
    \item \textbf{Mean violation magnitude}: The average size of violations (in total variance units), computed only over points that violate the condition.
    \item \textbf{Expected severity}: The product of violation rate and mean violation magnitude:
    \begin{equation}
    S = r_{\text{viol}} \times \bar{v}
    \label{eq:expected-severity}
    \end{equation}
    This combined metric captures both how frequently and how severely a model violates no-arbitrage conditions. A model with a high violation rate but tiny violations may be preferable to one with rare but large violations; expected severity unifies these dimensions.
\end{itemize}

\paragraph{Ground truth reference.} The ground truth surfaces themselves exhibit a non-zero butterfly violation rate of approximately 8.6\%, arising from the discretization of the continuous Heston model onto a finite grid. The finite-difference approximation of $\partial^2 w / \partial m^2$ on our 25-point log-moneyness grid can register small numerical violations even for the analytically arbitrage-free Heston model. This establishes a floor below which violations are attributable to grid discretization rather than model error, and serves as a reference line in the Pareto analyses of Chapter~\ref{ch:arbitrage}.

%% file: chapters/06_results.tex

The previous chapter described the seven model architectures, training protocol, and evaluation metrics. This chapter presents the experimental results across multiple dimensions: baseline reconstruction accuracy on synthetic data, robustness to increasing sparsity, regional error patterns across the volatility surface, performance on real SPY options, transfer learning dynamics, attention interpretability, and computational efficiency. The no-arbitrage constraint analysis is deferred to Chapter~\ref{ch:arbitrage}.

\section{Synthetic Data Baseline}
\label{sec:synthetic-results}

We begin with the core comparison: all seven models evaluated on the synthetic Heston test set (1,000 surfaces) with 30\% of grid points randomly masked.

\subsection{Overall Performance}

Table~\ref{tab:synthetic-results} summarizes the reconstruction accuracy and arbitrage violation rates for all models.

\begin{table}[H]
    \centering
    \caption{Baseline performance on the synthetic Heston test set (1,000 surfaces, 30\% missing). Models are sorted by RMSE on missing points. All neural models have $\sim$288k parameters; SVI has 40 parameters per surface. Calendar and butterfly columns report violation rates.}
    \label{tab:synthetic-results}
    \input{comparison/table_synthetic}
\end{table}

The Transformer achieves the best reconstruction accuracy with an RMSE of 0.0045 on missing points, followed closely by U-Net (0.0048) and CNN (0.0049). The MLP, lacking spatial inductive bias, trails at 0.0057, 27\% worse than the Transformer. The SVI baseline achieves 0.0065, placing it between the MLP and the VAE variants. Both VAE architectures rank last among the neural models (FC VAE: 0.0070, Conv VAE: 0.0072), suggesting that the information bottleneck imposed by the low-dimensional latent space ($d_z = 32$) limits reconstruction fidelity.

To assess the stability of these rankings, we retrain the top four models (CNN, U-Net, Transformer, MLP) with three different random seeds. Table~\ref{tab:multi-seed} reports the mean and standard deviation of RMSE on missing points.

\begin{table}[H]
    \centering
    \caption{Multi-seed validation (3 seeds). RMSE on missing points reported as mean $\pm$ standard deviation. The Transformer and U-Net are statistically indistinguishable on reconstruction accuracy. U-Net shows the lowest variance across seeds.}
    \label{tab:multi-seed}
    \input{comparison/table_multi_seed}
\end{table}

The multi-seed analysis reveals that the Transformer and U-Net are statistically tied on RMSE ($0.0045 \pm 0.0004$ vs.\ $0.0045 \pm 0.0000$), with overlapping confidence intervals. The ranking of CNN and MLP below them is stable across seeds. Notably, U-Net exhibits remarkably low variance (std $= 0.00004$, an order of magnitude smaller than the other models), suggesting that its architectural inductive biases lead to highly reproducible training dynamics.

The arbitrage columns reveal a striking pattern: all neural models exhibit butterfly violation rates of 43--45\%, while SVI achieves 0.0\% by construction (the SVI parameterization inherently produces convex smiles when properly constrained). Calendar violations are negligible across all models. This accuracy-arbitrage trade-off motivates the constrained training analysis in Chapter~\ref{ch:arbitrage}.

\begin{figure}[H]
    \centering
    \includegraphics[width=0.85\textwidth]{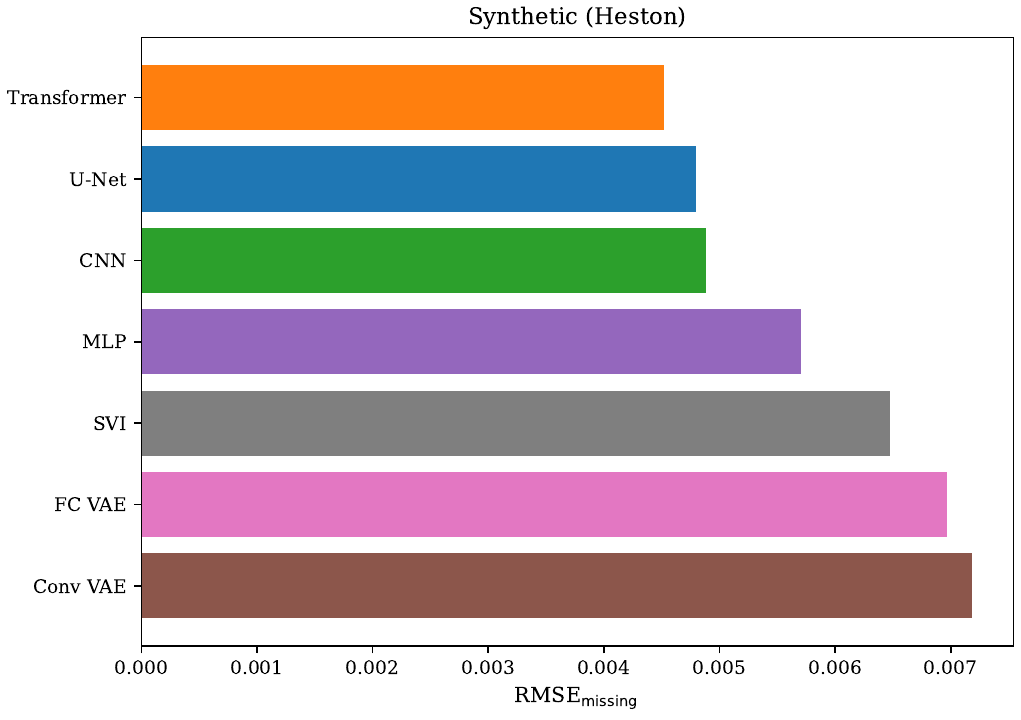}
    \caption{RMSE on missing points for all models on synthetic Heston data (30\% missing). Models are sorted by ascending RMSE. Transformer and U-Net lead, followed closely by CNN.}
    \label{fig:rmse-bar}
\end{figure}

Figure~\ref{fig:rmse-bar} visualizes this RMSE comparison.

\subsection{Visual Comparison}

\begin{figure}[H]
    \centering
    \includegraphics[width=\textwidth]{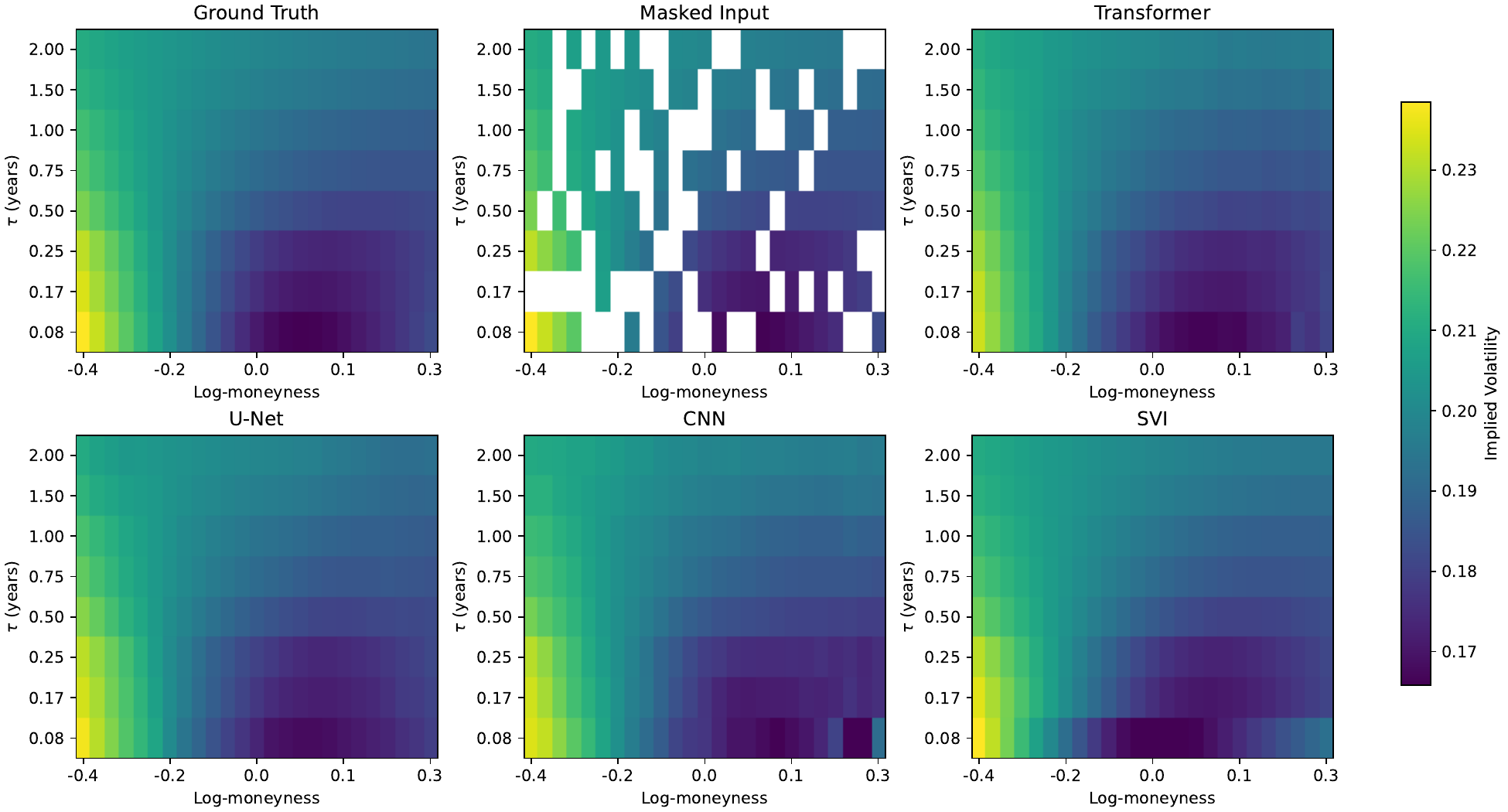}
    \caption{Reconstruction of a representative test surface. Top row: ground truth, masked input (30\% missing, white squares), and Transformer prediction. Bottom row: predictions from U-Net, CNN, and SVI. The vertical axis shows the actual (non-uniformly spaced) tenors. U-Net and Transformer faithfully recover the structure; the CNN shows localized artifacts at short tenors, and SVI overestimates curvature in regions with few observations.}
    \label{fig:sample-reconstruction}
\end{figure}

To provide qualitative intuition beyond aggregate metrics, Figure~\ref{fig:sample-reconstruction} shows a test surface selected for maximum RMSE dispersion among the neural models, reconstructed by each architecture.

Qualitative differences reflect each model's architectural properties. The CNN shows localized artifacts in the short-tenor, high-moneyness region ($\tau \leq 0.25$ years), where the smile curvature is most pronounced and its local $3 \times 3$ receptive field lacks sufficient nearby observations to interpolate accurately. The U-Net, through its cross-scale skip connections, propagates information from data-dense regions to data-sparse ones, producing the visually most faithful reconstruction. The Transformer achieves a comparable result by leveraging its global attention mechanism, which aggregates information from any observed point regardless of spatial distance. SVI, fitting each maturity independently, is especially sensitive to the number of observations per tenor: at maturities with few observed points it overestimates curvature, while at data-rich tenors its fit is competitive.

\begin{figure}[H]
    \centering
    \includegraphics[width=\textwidth]{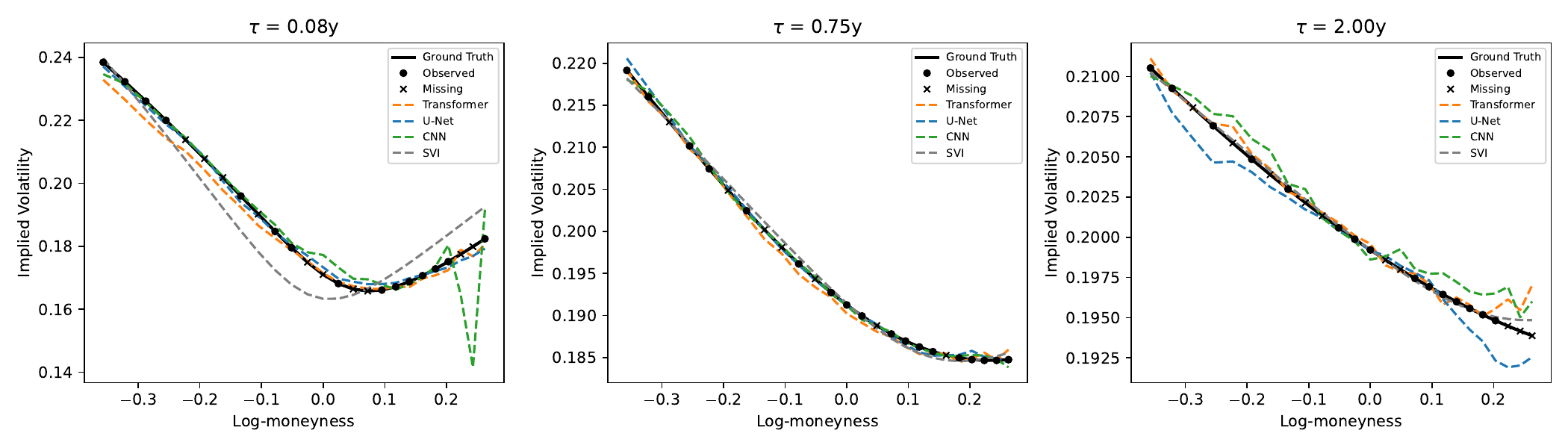}
    \caption{Implied volatility smile slices at three maturities ($\tau = 0.08$, $0.75$, and $2.0$ years). Circles mark observed points; crosses mark missing points. At $\tau = 0.08$ years, SVI overestimates both wings and the CNN oscillates in the right wing, while the Transformer and U-Net track the ground truth closely. At longer tenors, all neural models converge, with the CNN showing the largest deviations in the wings.}
    \label{fig:smile-slices}
\end{figure}

Figure~\ref{fig:smile-slices} shows the reconstructed implied volatility smiles at three representative maturities, overlaid with the ground truth.

\subsection{Per-Surface RMSE Distribution}

\begin{figure}[H]
    \centering
    \includegraphics[width=\textwidth]{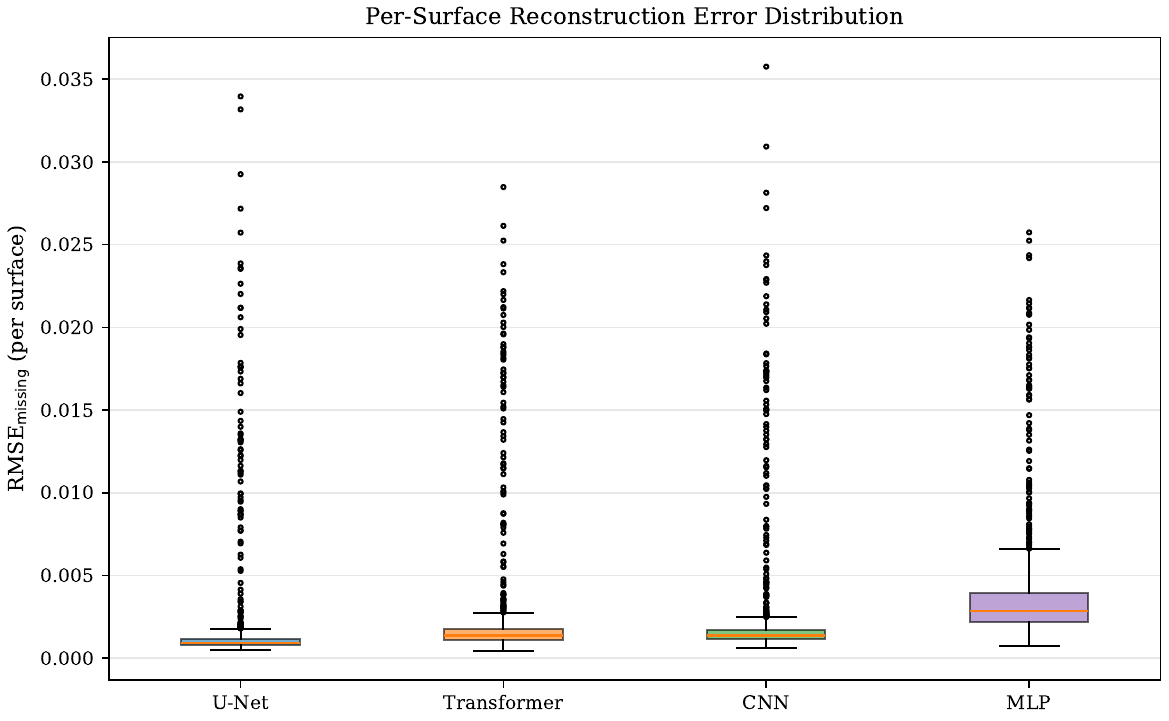}
    \caption{Distribution of per-surface RMSE across the 1,000 test surfaces for the four main models. The U-Net has both the lowest median and the narrowest interquartile range. The MLP shows the greatest dispersion, indicating occasional poor reconstructions.}
    \label{fig:rmse-boxplots}
\end{figure}

Aggregate metrics can mask variability across individual surfaces. Figure~\ref{fig:rmse-boxplots} shows the distribution of per-surface RMSE values for each model.

The U-Net achieves not only the lowest median but also the most consistent performance: its interquartile range ($3.8 \times 10^{-4}$) is the narrowest among all four models, followed by CNN ($5.1 \times 10^{-4}$) and Transformer ($6.2 \times 10^{-4}$). The MLP shows an interquartile range of $1.8 \times 10^{-3}$, nearly five times wider than the U-Net's, together with longer upper tails indicating occasional surfaces where reconstruction quality degrades substantially. Note that while the Transformer achieves the best \emph{average} RMSE (Table~\ref{tab:synthetic-results}), the U-Net exhibits lower variability across surfaces, consistent with the low cross-seed variance observed in Table~\ref{tab:multi-seed}.

\subsection{Discussion}

Three findings emerge from the synthetic baseline:

\begin{enumerate}
    \item \textbf{Spatial inductive bias matters.} The CNN and U-Net outperform the MLP by 14--17\% in RMSE, confirming that exploiting the 2D grid structure through local convolutions provides a meaningful advantage for volatility surface reconstruction.
    
    \item \textbf{Global attention matches the best convolutional models.} The proposed Transformer achieves RMSE statistically tied with U-Net across multiple seeds ($0.0045 \pm 0.0004$ vs.\ $0.0045 \pm 0.0000$). While the single-run results suggest a 6\% advantage, the multi-seed analysis shows that this difference is within the Transformer's training variance. The Transformer's advantage over convolutional models emerges more clearly under structured missingness and high sparsity (Sections~\ref{sec:sparsity} and~\ref{sec:structured-masking}).
    
    \item \textbf{Generative bottleneck limits fidelity.} The VAE variants rank last among neural models. The 32-dimensional latent space, while sufficient to capture the broad structure of Heston surfaces, loses the fine-grained detail needed for precise reconstruction. This is the cost of the generative modeling objective: the KL regularization and low-dimensional bottleneck smooth away high-frequency features.
\end{enumerate}
\section{Robustness to Sparsity}
\label{sec:sparsity}

A critical practical question is how gracefully each model degrades as observations become scarcer. We evaluate all models across missing fractions from 10\% to 90\%, \emph{without retraining}---the same models trained at 30\% missing are tested at all sparsity levels.

\begin{figure}[H]
    \centering
    \includegraphics[width=\textwidth]{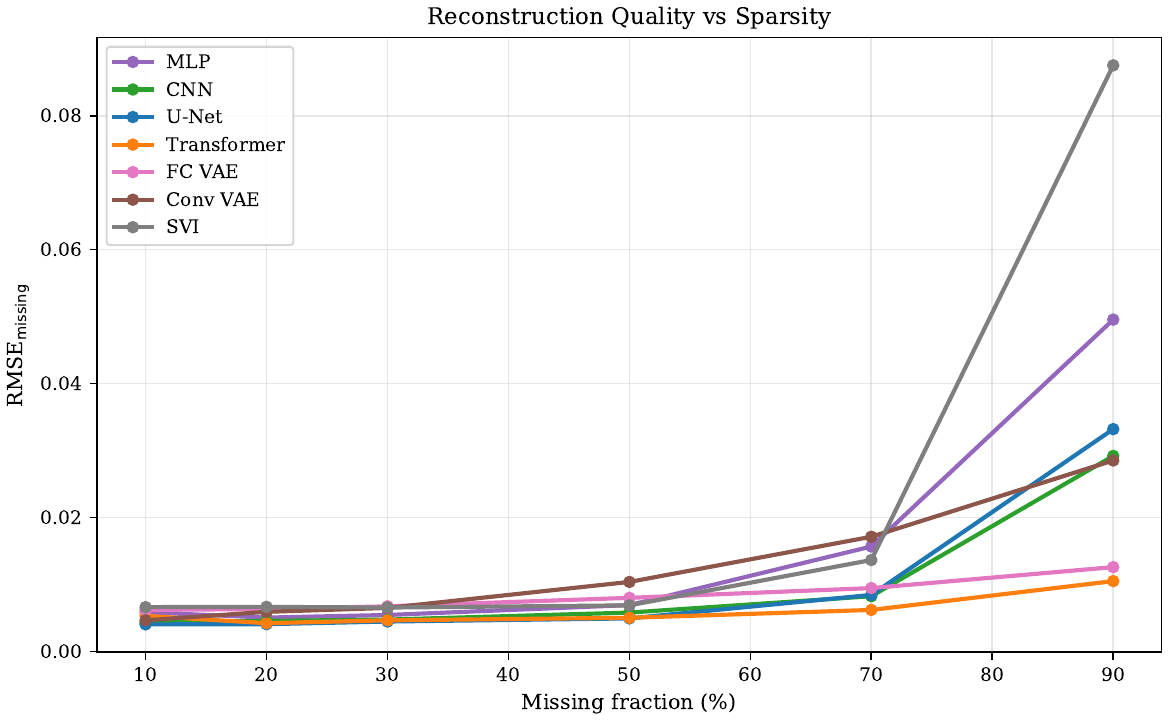}
    \caption{RMSE degradation as a function of missing fraction (10\%--90\%). All models are trained at 30\% missing and evaluated without retraining. The Transformer degrades most gracefully, achieving 9$\times$ lower RMSE than SVI at 90\% missing. SVI degrades catastrophically beyond 70\%.}
    \label{fig:masking-degradation}
\end{figure}

Figure~\ref{fig:masking-degradation} reveals sharply different degradation profiles:

\paragraph{Transformer.} The Transformer degrades most gracefully of all models, maintaining usable RMSE even at extreme sparsity. At 90\% missing (only 20 observed points out of 200), it achieves approximately 9$\times$ lower RMSE than SVI. This robustness stems from the attention mechanism's ability to aggregate information from any number of observed tokens: whether the encoder receives 180 or 20 tokens, the same cross-attention mechanism allows the decoder to reconstruct the surface.

\paragraph{CNN and U-Net.} The convolutional models degrade smoothly but faster than the Transformer. Their local receptive fields become a liability at extreme sparsity: when observed points are sparse and scattered, the $3 \times 3$ convolutional filters may not have enough nearby observations within their receptive field to produce accurate predictions. The U-Net's multi-scale architecture partially mitigates this through its bottleneck, which compresses the surface to a global representation.

\paragraph{SVI.} SVI degrades catastrophically beyond approximately 70\% missing. This is a direct consequence of its per-slice architecture: with only $\sim$6 observations per maturity slice (at 70\% missing), fitting 5 SVI parameters becomes severely underdetermined. At 90\% missing, some slices may have only 2--3 observations, making the fit meaningless.

\paragraph{VAEs.} The VAE variants show an interesting crossover: they perform relatively poorly at low sparsity (where the latent bottleneck limits fidelity) but degrade more slowly than convolutional models at high sparsity. The learned manifold of complete surfaces acts as a strong prior---the latent optimization projects the sparse observations onto the nearest plausible complete surface, even when very few observations are available.

This experiment represents the strongest validation of the proposed Transformer architecture. The attention mechanism provides not just the best absolute accuracy, but the most graceful degradation under data scarcity, precisely the regime most relevant to real options markets, where deep out-of-the-money strikes and off-cycle maturities often lack quotes. The $9\times$ advantage over SVI at extreme sparsity directly addresses the core motivation of this thesis: that attention-based models can leverage global surface structure to reconstruct volatility surfaces from severely incomplete observations, far beyond the capability of per-slice parametric methods.

\subsection{Structured Masking}
\label{sec:structured-masking}

The random masking protocol used above drops points uniformly across the grid. In practice, missingness is structured: deep out-of-the-money options are the least liquid and most likely to be missing, while at-the-money options are almost always available. To evaluate robustness under realistic missingness patterns, we repeat the sparsity sweep using a \emph{combined} masking strategy that always removes the two deepest OTM strikes ($|\log m| > 0.3$) and then applies random masking at the specified fraction on the remaining points.

\begin{figure}[H]
    \centering
    \includegraphics[width=\textwidth]{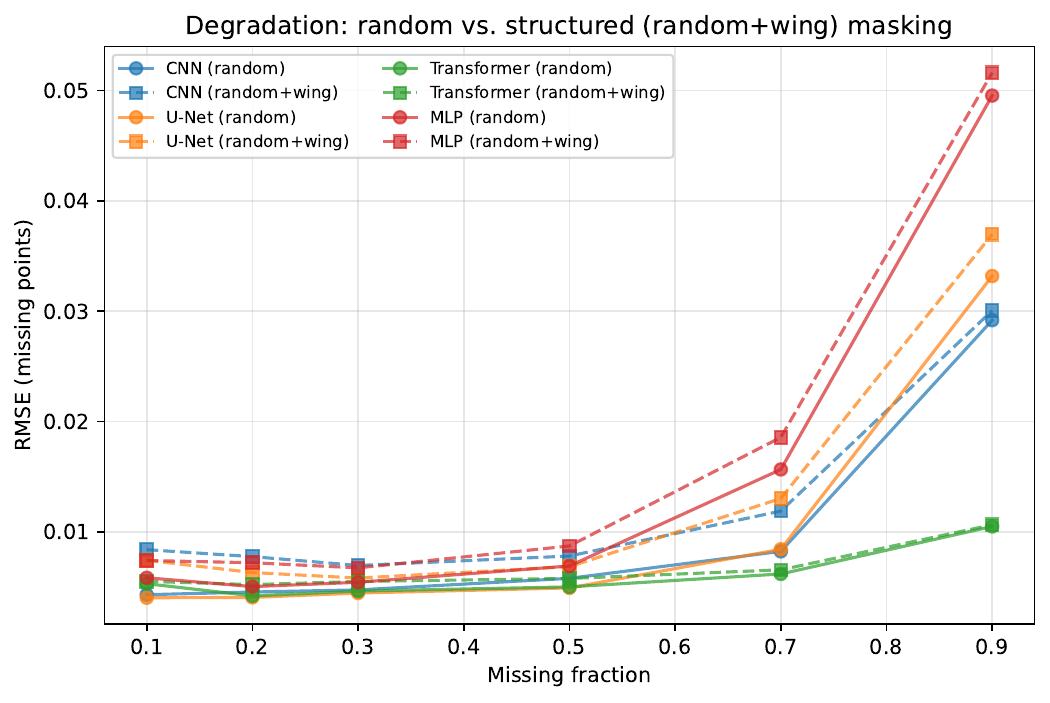}
    \caption{RMSE degradation under random masking (solid) vs.\ structured random+wing masking (dashed). The Transformer is nearly unaffected by structured missingness, while convolutional models degrade significantly when wing information is absent.}
    \label{fig:masking-comparison}
\end{figure}

Figure~\ref{fig:masking-comparison} reveals a striking difference in robustness. At 30\% missing, the Transformer's RMSE increases by only 0.0009 (from 0.0046 to 0.0055) when wing points are structurally absent, a degradation of less than 20\%. In contrast, the CNN degrades by 47\% (0.0047 to 0.0070) and the U-Net by 30\% (0.0045 to 0.0058). At 90\% missing, the gap narrows as random masking dominates, but the Transformer still leads by a wide margin (0.0107 vs.\ 0.0301 for CNN, 0.0370 for U-Net).

This result highlights a fundamental difference between the architectures. The convolutional models rely on spatial locality: they reconstruct missing points from their grid neighbors, so removing entire columns (the wings) eliminates the local context they depend on. The Transformer's cross-attention mechanism, combined with coordinate-based positional encoding, can interpolate into unseen regions by attending to distant observed points. Where the multi-seed analysis shows the Transformer and U-Net tied on accuracy under standard conditions, structured masking reveals that the Transformer is substantially more robust to the patterns of missingness encountered in real options markets.

\section{Regional Error Analysis}
\label{sec:regional}

Reconstruction difficulty is not uniform across the volatility surface. To understand where each model excels and struggles, we decompose the error by moneyness and tenor region (region definitions in Table~\ref{tab:regions}).

\begin{table}[H]
    \centering
    \caption{RMSE by moneyness and tenor region for the top four models (synthetic data, 30\% missing). Upper block: moneyness regions. Lower block: tenor regions.}
    \label{tab:regional-results}
    \input{comparison/table_regional}
\end{table}

\begin{figure}[H]
    \centering
    \includegraphics[width=\textwidth]{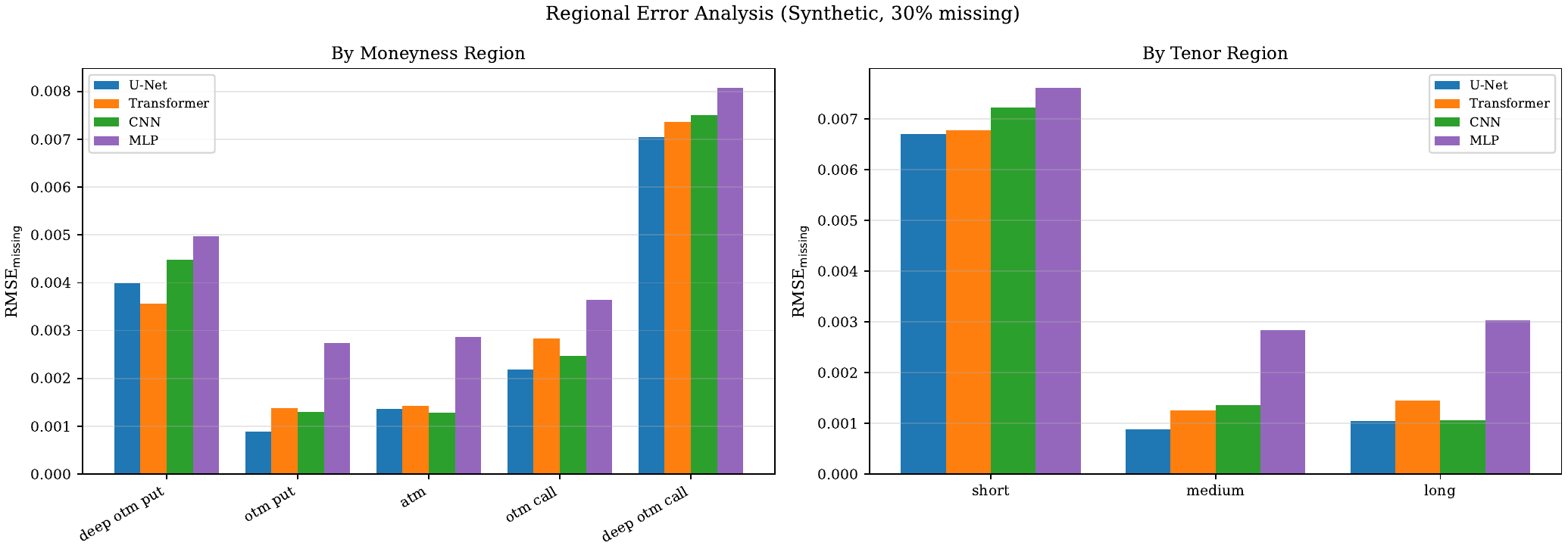}
    \caption{Regional RMSE comparison. Left: by moneyness region (deep OTM put through deep OTM call). Right: by tenor region (short, medium, long). Short tenor and deep OTM wings are universally the most challenging regions.}
    \label{fig:regional-bar}
\end{figure}

\subsection{Moneyness Patterns}

The error distribution across moneyness reveals a clear U-shape: the at-the-money and near-OTM regions are easiest (RMSE 0.0009--0.0015 for the best models), while the deep OTM wings are hardest (RMSE 0.0042--0.0081). This pattern reflects the shape of the volatility smile: the steep gradients in the wings create larger interpolation errors when observations are missing.

The Transformer achieves the best performance in the deep OTM put region (0.0042 vs.\ U-Net's 0.0049), where global attention can leverage information from the observed ATM region and other maturities to constrain the wing shape. The U-Net excels in the near-ATM regions (OTM put: 0.0009, ATM: 0.0013), where its local convolutional filters are well-suited to the smooth, slowly varying surface shape. The MLP shows the largest performance gap in the wings (0.0081 in deep OTM call vs.\ 0.0067--0.0069 for the convolutional models), confirming that spatial structure is most valuable in the regions with the steepest gradients.

\subsection{Tenor Patterns}

The tenor decomposition reveals that short maturities ($\tau \leq 0.25$ years) are universally the most challenging region, with RMSE values 5--8$\times$ higher than medium or long maturities. This is a well-known phenomenon in options markets: short-dated options have steeper, more curved smiles that are harder to interpolate. The Heston model amplifies this effect at short maturities, where the smile shape is particularly sensitive to the vol-of-vol parameter $\xi$.

Medium and long tenors are nearly perfectly reconstructed by all models (RMSE $\leq$ 0.0015 for U-Net, Transformer, and CNN), reflecting the smoother, flatter smiles at longer maturities.

\subsection{Spatial Error Maps}

\begin{figure}[H]
    \centering
    \includegraphics[width=\textwidth]{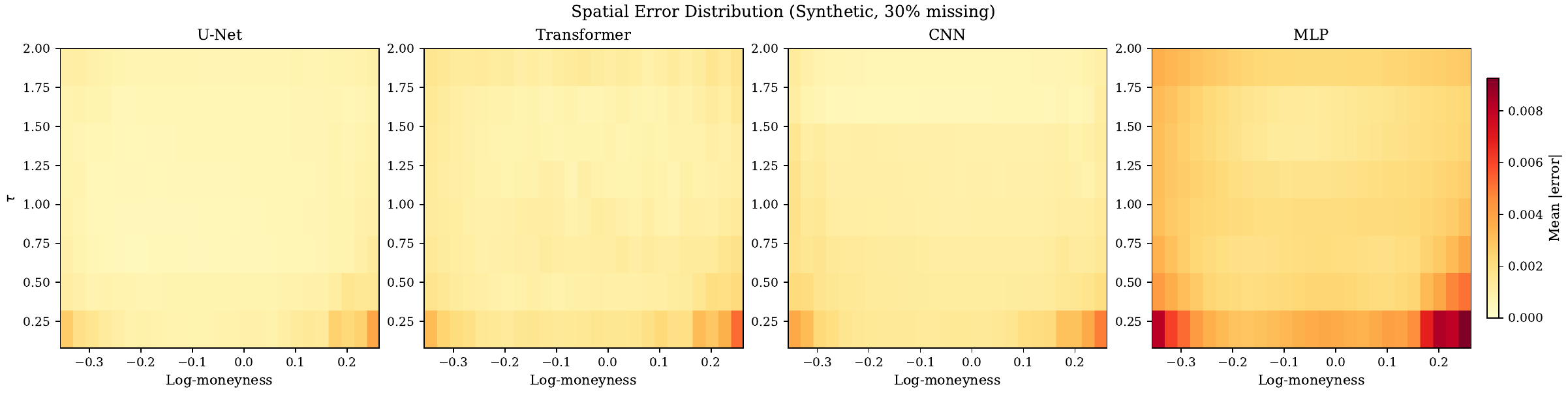}
    \caption{Mean absolute error at each grid point (log-moneyness $\times$ tenor) averaged over the test set. The Transformer and U-Net show concentrated errors in the short-tenor wings. The MLP shows elevated errors across the entire surface, with the worst performance in the deep OTM wings at short maturity.}
    \label{fig:error-heatmaps}
\end{figure}

Figure~\ref{fig:error-heatmaps} provides a detailed spatial view of reconstruction errors across the full grid for each model.

The heatmaps confirm that the short-tenor, deep-OTM corners of the grid are the universal bottleneck. For the Transformer and U-Net, errors are tightly concentrated in these corners, with the rest of the surface nearly perfectly reconstructed. The MLP, by contrast, shows elevated errors diffused across the entire surface, including the ATM region where other models achieve near-zero error. This visualization makes vivid the benefit of spatial inductive biases: the convolutional and attention-based models focus their limited error budget on the intrinsically hardest regions, while the MLP spreads errors uniformly because it treats all grid positions identically.

\section{Real Data Results}
\label{sec:real-results}

We evaluate the three best-performing architectures (CNN, U-Net, Transformer) on real SPY equity options data (Section~\ref{sec:real-data}), comparing three training strategies: from-scratch training on real data only, fine-tuning from synthetic pretraining, and the SVI baseline. The MLP and both VAE variants are excluded from the real data evaluation: the MLP consistently underperformed the convolutional models on synthetic data, while the VAEs' latent-space optimization at inference (200 Adam steps per surface) makes them computationally expensive and their synthetic RMSE was 40--60\% worse than the top models.

\begin{table}[H]
    \centering
    \caption{Performance on the real SPY test set (487 surfaces). Top block: models trained from scratch on real data. Middle block: models pretrained on synthetic data and fine-tuned on real data. Bottom: SVI baseline. Calendar and butterfly columns report violation rates.}
    \label{tab:real-results}
    \input{comparison/table_real}
\end{table}

\begin{figure}[H]
    \centering
    \includegraphics[width=0.85\textwidth]{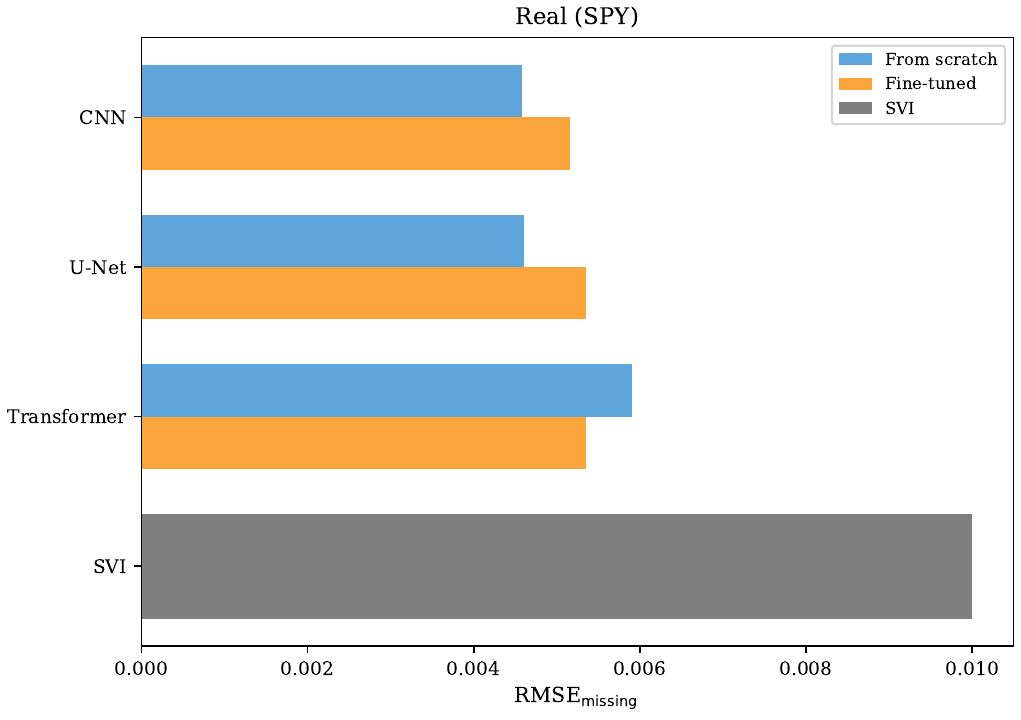}
    \caption{RMSE on missing points for real SPY data. For each neural model, both from-scratch and fine-tuned results are shown. All neural models substantially outperform SVI ($\sim$2$\times$ lower RMSE).}
    \label{fig:rmse-bar-real}
\end{figure}

Figure~\ref{fig:rmse-bar-real} shows the RMSE comparison across the different training strategies.

\subsection{From-Scratch vs.\ Fine-Tuned}

A notable finding is that \textbf{from-scratch training outperforms fine-tuning for CNN and U-Net} on RMSE (0.0046 vs.\ 0.0052--0.0054). While pretraining on a larger dataset is generally expected to help, the distributional gap between synthetic Heston surfaces and real SPY surfaces (documented in Section~\ref{sec:synth-vs-real}) is large enough that the pretrained representations need substantial adaptation.

The proposed Transformer is the exception: fine-tuning improves its RMSE from 0.0059 to 0.0054, making it the only architecture that unambiguously benefits from synthetic pretraining. We attribute this to two factors. First, the Transformer's from-scratch performance on the smaller real dataset (2,912 training surfaces) is weaker than the convolutional models, leaving more room for improvement. Second, the attention mechanism's flexible parameterization benefits from the structural priors learned during synthetic pretraining---such as the general shape of the volatility smile and term structure---which provide a better initialization than random weights. This suggests that pretraining is most valuable for architectures with higher representational capacity that struggle with limited data.

\subsection{Neural Models vs.\ SVI}

All neural models outperform SVI by a factor of approximately 2$\times$ on RMSE (best: 0.0046 vs.\ SVI: 0.0100). This gap is larger than on synthetic data (where SVI achieved 0.0065), reflecting two factors: real surfaces have more complex local structure that the five-parameter SVI model cannot capture, and real data has natural missingness patterns that are more structured (concentrated in wings and short maturities) than random masking.

\subsection{Butterfly Violation Rates}

Fine-tuned models exhibit notably lower butterfly violation rates than their from-scratch counterparts: CNN drops from 43.9\% to 33.4\%, and U-Net from 42.0\% to 35.3\%. This suggests that pretraining on arbitrage-free synthetic surfaces teaches the models to produce smoother, more convex smiles---a structural property that persists through fine-tuning even though the distributional properties change. SVI achieves the lowest butterfly rate (2.1\%), with the residual violations attributable to grid discretization effects.

\section{Transfer Learning Analysis}
\label{sec:transfer}

To understand the dynamics of knowledge transfer from synthetic to real data, we examine the three stages of transfer learning: from-scratch performance, zero-shot transfer (applying the synthetic-trained model directly to real data), and fine-tuned performance.

\begin{figure}[H]
    \centering
    \includegraphics[width=\textwidth]{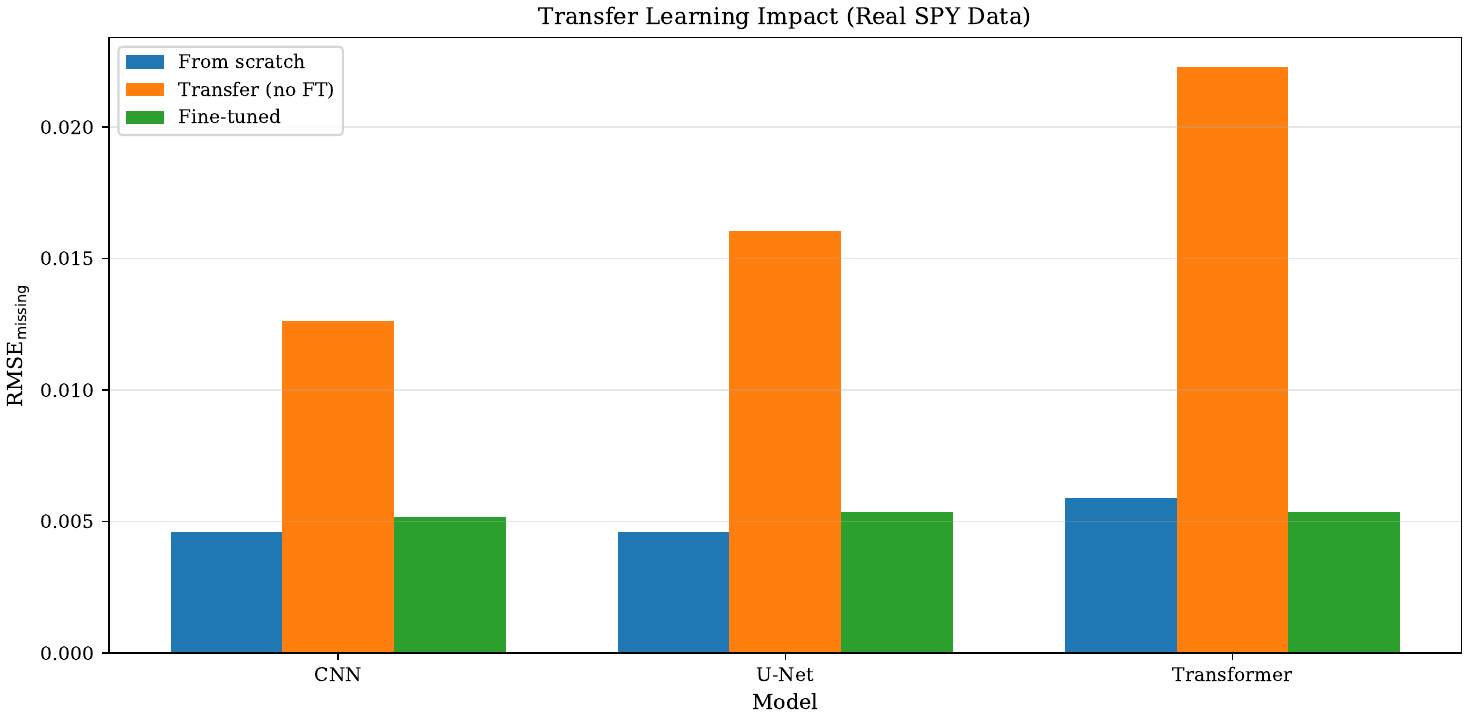}
    \caption{Transfer learning waterfall for three architectures. Stage 1: trained from scratch on real data. Stage 2: zero-shot transfer (synthetic model applied to real data without fine-tuning). Stage 3: fine-tuned on real data. Zero-shot transfer is consistently negative; fine-tuning recovers to from-scratch levels.}
    \label{fig:transfer-waterfall}
\end{figure}

\subsection{Negative Zero-Shot Transfer}

Zero-shot transfer---applying a model trained on synthetic Heston data directly to real SPY data---produces RMSE values \emph{worse} than from-scratch training on real data alone. This negative transfer reflects the distributional mismatch between the two domains: the Heston model, while capturing the qualitative features of real surfaces (skew, term structure), produces quantitatively different volatility levels, smile shapes, and local structures. The synthetic data distribution (IV values uniformly spanning 10--40\%) differs substantially from the real SPY distribution (concentrated around 15--20\% with heavy right tail), causing the pretrained models to systematically mispredict volatility levels.

\subsection{Fine-Tuning Recovery}

Fine-tuning recovers the zero-shot degradation, bringing all models back to approximately from-scratch performance levels. For the Transformer, fine-tuning slightly surpasses from-scratch (0.0054 vs.\ 0.0059), making it the only architecture that unambiguously benefits from synthetic pretraining.

\subsection{Structural vs.\ Distributional Transfer}

Although RMSE does not improve through transfer learning for CNN and U-Net, the butterfly violation rates tell a different story. Fine-tuned models consistently produce smoother surfaces with fewer arbitrage violations than their from-scratch counterparts (Table~\ref{tab:real-results}). This suggests that pretraining transfers \emph{structural} knowledge---the general convexity and smoothness properties of volatility surfaces---even when it fails to transfer \emph{distributional} knowledge about specific volatility levels and smile shapes. The pretrained models have learned that volatility surfaces should be smooth and convex, and this prior persists through fine-tuning as a regularization effect.

\subsection{Synthesis of Real Data Results}

The real data experiments on SPY reveal a more nuanced picture than the synthetic results, which reinforces the external validity of the conclusions. Four findings deserve highlighting. First, from-scratch CNN and U-Net achieve the best RMSE (0.0046), demonstrating that 2,912 real surfaces are sufficient for convolutional architectures to learn effective representations without pretraining. Second, the Transformer does not dominate on accuracy over real data as it does under extreme sparsity on synthetic data; its advantage manifests in transfer learning, where it is the only architecture that improves with pretraining. Third, synthetic pretraining transfers \emph{structural} properties (smoothness, convexity) but not \emph{distributional} ones (IV levels, smile shapes), which explains why fine-tuned models have lower arbitrage rates even when they do not always have better RMSE. Fourth, SVI loses by a factor of $2\times$ on RMSE against all neural models, but retains the advantage of near-zero arbitrage violations, a trade-off that is analyzed in depth in Chapter~\ref{ch:arbitrage}.

\section{Transformer Attention Analysis}
\label{sec:attention}

The Transformer's attention mechanism provides a unique opportunity for interpretability: by examining the cross-attention weights in the decoder, we can see which observed points each missing point ``looks at'' during reconstruction.

\begin{figure}[H]
    \centering
    \includegraphics[width=\textwidth]{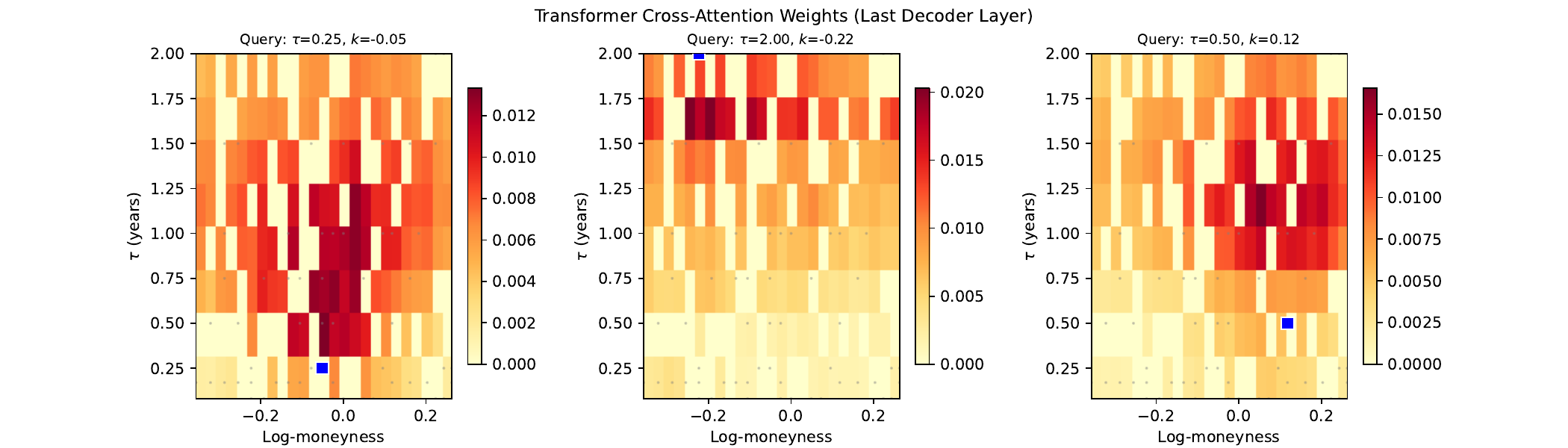}
    \caption{Cross-attention weights for three representative missing tokens (blue squares): an ATM short-tenor point, an OTM put at long tenor, and an OTM call at medium tenor. Brighter colors indicate higher attention weight. Gray dots mark other missing positions. The attention patterns align with financial intuition: each missing point attends primarily to nearby strikes at the same maturity and to the same strike at adjacent maturities.}
    \label{fig:attention-heatmap}
\end{figure}

Figure~\ref{fig:attention-heatmap} reveals attention patterns that align with financial intuition:

\begin{itemize}
    \item \textbf{Local smile structure.} Missing points attend most strongly to nearby observed strikes at the same maturity, reflecting the smooth continuity of the volatility smile. This is analogous to local interpolation, but learned rather than prescribed.
    
    \item \textbf{Term structure coupling.} Significant attention weight is also directed to the same strike at adjacent maturities, reflecting the term structure relationship. This cross-maturity attention is precisely the mechanism that SVI lacks and that gives neural models their advantage under sparsity.
    
    \item \textbf{Adaptive attention span.} The attention patterns are not fixed templates---they adapt to the specific observation pattern and surface shape. Points in data-sparse regions cast wider attention spans, aggregating information from more distant observations, while points in data-dense regions attend more locally.
\end{itemize}

These patterns are consistent with the hypothesis that the Transformer has learned financially meaningful relationships, although attention weights are descriptive and do not constitute causal evidence of such learning. The local smile attention reflects interpolation along the strike axis, the cross-maturity attention captures term structure coupling, and the adaptive attention span acts as a data-dependent receptive field---three mechanisms that together explain why the architecture excels under sparsity. This interpretability is a distinctive advantage of the attention-based approach: unlike convolutional or fully connected models, the Transformer allows inspecting \emph{which} observations contribute to the reconstruction of each missing point and with \emph{what} relative weight, offering a degree of interpretability that convolutional and fully connected models do not provide.

\section{Computational Efficiency}
\label{sec:computational}

Practical deployment requires not only accuracy but also efficient inference. Table~\ref{tab:computational} reports single-surface inference latency, batch throughput, and GPU memory usage for all models.

\begin{table}[H]
    \centering
    \caption{Computational performance. Latency and throughput measured on an NVIDIA RTX 4080 SUPER GPU (batch size 1 for latency, batch size 256 for throughput). SVI runs on CPU (single-threaded). GPU memory is the incremental allocation for the model and a single inference pass.}
    \label{tab:computational}
    \input{comparison/table_computational}
\end{table}

\subsection{Neural Model Performance}

All neural models achieve sub-millisecond inference latency, making them suitable for real-time applications. The MLP is the fastest at 0.1\,ms per surface (17,567 surfaces/second), followed by the CNN at 0.1\,ms (11,468 surfaces/second). The Transformer is the slowest neural model at 0.9\,ms (1,169 surfaces/second), approximately 9$\times$ slower than the CNN, reflecting the quadratic cost of attention over 200 tokens. However, at over 1,000 surfaces per second, the Transformer is still fast enough for any practical application---a trading desk processing thousands of options chains per day would require only seconds of compute time.

GPU memory usage is uniformly low across all models (10--12\,MB), well within the capacity of even entry-level GPUs. This confirms that the $\sim$288k parameter scale used in this work is far from the memory-limited regime.

\subsection{SVI Performance}

SVI is dramatically slower than all neural models: 13.1\,ms per surface (76 surfaces/second), which is 172$\times$ slower than the MLP and 15$\times$ slower than the Transformer. This overhead comes from the iterative L-BFGS-B optimization required for each individual surface---there is no amortization across surfaces. SVI's computational cost also scales with the number of observed points per slice, making it even slower for densely observed surfaces.

\subsection{VAE Inference Caveat}

The timings in Table~\ref{tab:computational} reflect a single forward pass through each model. For the VAE variants, full inference includes an additional latent optimization step (200 Adam iterations on the latent vector $\mathbf{z}$, as described in Section~\ref{sec:vae}), which increases the effective inference time by approximately two orders of magnitude. This makes the VAE inference pipeline comparable in cost to SVI, partially negating the amortization advantage of neural models.

\subsection{Accuracy--Efficiency Trade-Off}

Combining the accuracy results from Section~\ref{sec:synthetic-results} with the computational analysis:
\begin{itemize}
    \item \textbf{Best accuracy}: Transformer (RMSE 0.0045, 0.9\,ms).
    \item \textbf{Best efficiency}: MLP (RMSE 0.0057, 0.1\,ms) or CNN (RMSE 0.0049, 0.1\,ms).
    \item \textbf{Best trade-off}: CNN---within 9\% of the Transformer's accuracy at 9$\times$ the speed.
\end{itemize}
For latency-critical applications, the CNN offers the most compelling balance of accuracy and speed. For applications where accuracy is paramount, particularly under high sparsity, where the Transformer's advantage over convolutional models widens from 6\% to over $2\times$, the proposed Transformer is the clear choice, with inference latency that remains well within real-time requirements.

\section{Transformer Ablation Study}
\label{sec:ablation}

To quantify the contribution of the Fourier coordinate encoding (Section~\ref{sec:transformer}), we train an otherwise identical Transformer replacing the fixed sinusoidal encoding with learnable positional embeddings of the same dimensionality ($d_{\text{coord}} = 34$). All other hyperparameters (architecture, learning rate, training protocol) remain unchanged.

\begin{table}[H]
    \centering
    \caption{Ablation: Fourier coordinate encoding vs.\ learnable positional embeddings. Both models have $\sim$288k parameters and identical architecture except for the positional encoding.}
    \label{tab:ablation}
    \begin{tabular}{lcc}
        \toprule
        Positional Encoding & RMSE$_\text{miss}$ & Parameters \\
        \midrule
        Fourier (proposed) & \textbf{0.0045} & 288,129 \\
        Learnable & 0.0055 & 288,639 \\
        \bottomrule
    \end{tabular}
\end{table}

The Fourier encoding reduces RMSE by 18\% compared to learnable embeddings. This confirms that encoding the physical coordinates $(\tau, \log m)$ through multi-scale sinusoidal features provides meaningful inductive bias for this task: the model benefits from knowing the continuous financial meaning of each grid position rather than learning position representations from scratch. The result aligns with the findings of Tancik et al.\ \cite{tancik2020fourier}, who showed that Fourier features help MLPs learn high-frequency functions in low-dimensional domains.

%% file: comparison/table_synthetic.tex
\begin{tabular}{lrrrrrr}
\toprule
Model & Params & RMSE$_\text{miss}$ & MAE & Max Error & Calendar & Butterfly \\
\midrule
Transformer & 288k & 0.0045 & 0.0012 & 0.2005 & 0.0\% & 45.0\% \\
U-Net & 265k & 0.0048 & 0.0008 & 0.2641 & 0.0\% & 44.9\% \\
CNN & 295k & 0.0049 & 0.0011 & 0.2464 & 0.0\% & 45.1\% \\
MLP & 286k & 0.0057 & 0.0025 & 0.2380 & 0.0\% & 44.3\% \\
SVI & 40/surf & 0.0065 & 0.0008 & 0.3251 & 0.0\% & 0.0\% \\
FC VAE & 285k & 0.0070 & 0.0030 & 0.2568 & 0.0\% & 45.0\% \\
Conv VAE & 273k & 0.0072 & 0.0024 & 0.3316 & 0.0\% & 43.2\% \\
\bottomrule
\end{tabular}

%% file: comparison/table_multi_seed.tex
\begin{tabular}{lccc}
\toprule
Model & RMSE$_\text{miss}$ (mean $\pm$ std) & Original & $n$ \\
\midrule
  CNN & $0.0049 \pm 0.0002$ & 0.0049 & 3 \\
  U-Net & $0.0045 \pm 0.0000$ & 0.0048 & 3 \\
  Transformer & $0.0045 \pm 0.0004$ & 0.0045 & 3 \\
  MLP & $0.0054 \pm 0.0001$ & 0.0057 & 3 \\
\bottomrule
\end{tabular}

%% file: comparison/table_regional.tex
\begin{tabular}{lrrrr}
\toprule
Region & U-Net & Transformer & CNN & MLP \\
\midrule
deep otm put & 0.0066 & 0.0053 & 0.0069 & 0.0059 \\
otm put & 0.0009 & 0.0014 & 0.0013 & 0.0027 \\
atm & 0.0009 & 0.0014 & 0.0013 & 0.0028 \\
otm call & 0.0023 & 0.0028 & 0.0028 & 0.0036 \\
deep otm call & 0.0060 & 0.0071 & 0.0070 & 0.0076 \\
\midrule
short & 0.0075 & 0.0074 & 0.0083 & 0.0078 \\
medium & 0.0009 & 0.0013 & 0.0013 & 0.0028 \\
long & 0.0010 & 0.0015 & 0.0010 & 0.0031 \\
\bottomrule
\end{tabular}

%% file: comparison/table_real.tex
\begin{tabular}{lrrrrr}
\toprule
Model & RMSE$_\text{miss}$ & Test MSE & MAE & Calendar & Butterfly \\
\midrule
CNN & 0.0046 & $8.31 \times 10^{-6}$ & 0.0019 & 1.4\% & 43.9\% \\
Transformer & 0.0059 & $1.35 \times 10^{-5}$ & 0.0023 & 0.5\% & 44.9\% \\
U-Net & 0.0046 & $7.94 \times 10^{-6}$ & 0.0018 & 3.1\% & 42.0\% \\
\midrule
CNN (FT) & 0.0052 & $1.56 \times 10^{-5}$ & 0.0029 & 0.8\% & 33.4\% \\
Transformer (FT) & 0.0054 & $1.27 \times 10^{-5}$ & 0.0025 & 0.2\% & 42.3\% \\
U-Net (FT) & 0.0054 & $1.99 \times 10^{-5}$ & 0.0033 & 1.0\% & 35.3\% \\
\midrule
SVI & 0.0100 & $7.13 \times 10^{-5}$ & 0.0059 & 2.2\% & 2.1\% \\
\bottomrule
\end{tabular}

%% file: comparison/table_computational.tex
\begin{tabular}{lrrrr}
\toprule
Model & Params & Latency (ms) & Throughput (surf/s) & GPU Memory (MB) \\
\midrule
Transformer & 288k & 0.9 & 1169 & 11.8 \\
CNN & 295k & 0.1 & 11468 & 10.9 \\
U-Net & 265k & 0.2 & 4486 & 11.0 \\
MLP & 285k & 0.1 & 17567 & 10.2 \\
FC VAE & 284k & 0.1 & 6850 & 10.2 \\
Conv VAE & 273k & 0.2 & 5387 & 10.5 \\
SVI & 40/surf & 13.1 & 76 & -- \\
\bottomrule
\end{tabular}

%% file: chapters/07_arbitrage.tex

Chapter~\ref{ch:results} demonstrated that neural network models achieve substantially better reconstruction accuracy than the SVI baseline. However, the results also revealed a significant shortcoming: all neural models exhibit butterfly spread violation rates of 43--45\%, while SVI achieves near-zero violations by construction. This chapter investigates whether differentiable no-arbitrage penalties can close this gap, and whether the accuracy--arbitrage trade-off is truly a trade-off at all.

\section{The Arbitrage Problem in Neural Reconstructions}
\label{sec:arb-problem}

The butterfly violation rates reported in Table~\ref{tab:synthetic-results} present a practical challenge. A reconstructed volatility surface with 45\% of its grid points violating the convexity condition implies negative probability density in the risk-neutral measure at those points---an economically meaningless result. While these violations are typically small in magnitude (mean violation $\sim 10^{-2}$ in total variance units), they are pervasive, affecting nearly half of all interior grid points.

SVI avoids this problem entirely: the parametric form of Equation~\eqref{eq:svi-raw} naturally produces convex smiles when $b > 0$ and $\sigma > 0$. But SVI pays for this guarantee with 33--44\% worse RMSE compared to the best neural models (CNN, U-Net, and Transformer). The central question of this chapter is: \emph{can we retain the accuracy of neural models while approaching the arbitrage-free quality of SVI?}

\subsection{Calendar vs.\ Butterfly Violations}

As noted in Section~\ref{sec:no-arb-constraints}, two static no-arbitrage conditions apply to volatility surfaces: calendar spread (total variance non-decreasing in maturity) and butterfly spread (total variance convex in log-moneyness). In our experiments, calendar violations are negligible---below 0.1\% for all models---reflecting the relatively coarse 8-tenor grid and the inherent monotonicity of total variance in the Heston model. The remainder of this chapter focuses exclusively on butterfly violations, which are the dominant source of arbitrage opportunities in the reconstructed surfaces.

\section{Constraint Methodology}
\label{sec:constraint-method}

The no-arbitrage penalty formulation was described in detail in Section~\ref{sec:no-arb-constraints}. We briefly recap the key elements and the experimental design for this chapter.

\subsection{Augmented Loss Function}

Since calendar violations are negligible, the effective training loss reduces to:
\begin{equation}
\mathcal{L} = \mathcal{L}_{\text{MSE}} + \lambda \cdot \mathcal{P}_{\text{but}}
\end{equation}
where $\mathcal{P}_{\text{but}}$ is the butterfly convexity penalty (Equation~\eqref{eq:butterfly-penalty}) and $\lambda \geq 0$ controls the strength of the constraint. At $\lambda = 0$, training is unconstrained (the baseline from Chapter~\ref{ch:results}). As $\lambda$ increases, the model is increasingly penalized for producing non-convex smiles, steering predictions toward arbitrage-free surfaces at the potential cost of reconstruction accuracy.

The penalty is a \emph{soft} constraint: it adds a cost for violations but does not enforce strict satisfaction. This approach has two advantages over hard projection methods (which would project the output onto the feasible set after each forward pass). First, the penalty is fully differentiable, providing gradient signal that trains the model to internalize the constraint rather than relying on post-hoc correction. Second, it allows the model to trade off small violations against reconstruction accuracy, rather than rigidly enforcing a condition that may conflict with the training data (recall that the ground truth itself has $\sim$8.6\% butterfly violations from grid discretization).

\subsection{Expected Severity and Ground Truth Reference}

We use expected severity $S = r_{\text{viol}} \times \bar{v}$ (defined in Section~\ref{sec:evaluation-metrics}) as the primary arbitrage metric, as it captures both the frequency and magnitude of violations. The ground truth surfaces exhibit approximately 8.6\% butterfly violations due to grid discretization (Section~\ref{sec:evaluation-metrics}), establishing a floor below which further reduction is not meaningful for this grid resolution.

\section{Constraint Impact at Fixed Lambda}
\label{sec:fixed-lambda}

We first examine the effect of a moderate constraint strength ($\lambda = 0.1$) applied uniformly to all six neural architectures. Each model is retrained from scratch with the augmented loss function; all other hyperparameters remain identical to the unconstrained baseline (Table~\ref{tab:training-hyperparams}), with early stopping patience of 30 iterations and a maximum of 500 iterations. The results in this section correspond to a single random seed per configuration; Section~\ref{sec:lambda-sweep} validates these findings with multiple seeds.

\begin{table}[H]
    \centering
    \caption{Constraint impact at $\lambda = 0.1$ for all neural models. RMSE, butterfly violation rate, and expected severity are shown for the unconstrained ($\lambda = 0$) and constrained ($\lambda = 0.1$) settings. SVI is shown as a reference. Models are sorted by unconstrained RMSE.}
    \label{tab:constraint-impact}
    \input{comparison/table_arbitrage}
\end{table}

\begin{figure}[H]
    \centering
    \includegraphics[width=\textwidth]{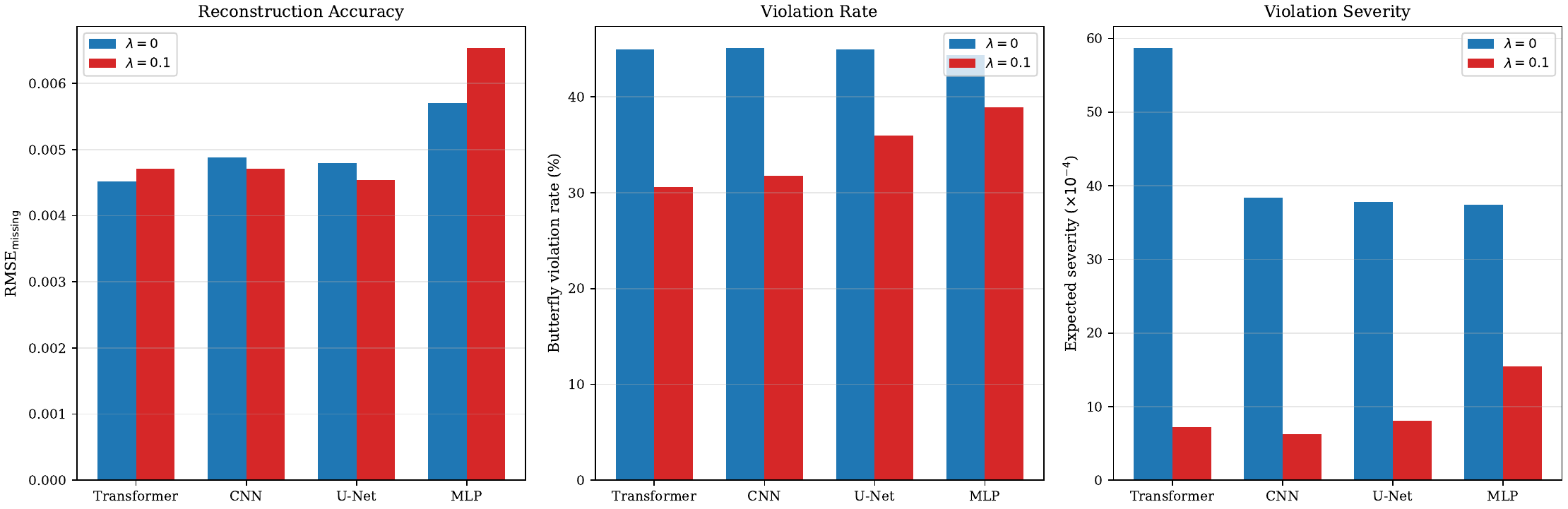}
    \caption{Constraint impact at $\lambda = 0.1$ across all six neural architectures. Left: RMSE on missing points. Center: butterfly violation rate. Right: expected severity. The CNN, U-Net, and Conv VAE \emph{improve} RMSE with constraints (green arrows), while reducing severity by 5--8$\times$. The MLP degrades substantially.}
    \label{fig:constraint-impact}
\end{figure}

\subsection{The Regularization Effect}

The most striking finding is that three architectures---CNN, U-Net, and Conv VAE---\emph{improve} their RMSE when trained with the butterfly penalty:

\begin{itemize}
    \item \textbf{U-Net}: RMSE decreases from 0.0048 to 0.0045 ($-6\%$), while expected severity drops by $4.7\times$ (from $3.78 \times 10^{-3}$ to $8.07 \times 10^{-4}$).
    \item \textbf{CNN}: RMSE decreases from 0.0049 to 0.0047 ($-4\%$), with $6.1\times$ severity reduction.
    \item \textbf{Conv VAE}: RMSE decreases from 0.0072 to 0.0060 ($-17\%$), with $7.7\times$ severity reduction. While the percentage improvement is the largest, the Conv VAE's baseline RMSE is substantially worse than the other architectures.
\end{itemize}

This is not a trade-off---for these architectures, the constraint is \emph{free}: it simultaneously improves accuracy and reduces arbitrage violations. The butterfly penalty acts as a \emph{regularizer}, enforcing smooth, convex smile shapes that happen to generalize better. The mechanism is intuitive: without the penalty, convolutional models can fit high-frequency noise in the training data that manifests as non-convex wiggles in the reconstructed smiles. The convexity penalty smooths these wiggles, and since the underlying Heston surfaces are indeed convex, the smoothed predictions are closer to the ground truth.

\subsection{Architecture-Dependent Response}

Not all architectures respond equally well to constraints:

\begin{itemize}
    \item \textbf{Transformer}: RMSE increases slightly from 0.0045 to 0.0047 ($+4\%$), but severity decreases by $8.1\times$---the largest severity reduction of any model. The Transformer's global attention mechanism already produces relatively smooth surfaces, so the marginal benefit of the convexity prior is smaller, and the constraint slightly limits the model's ability to fit sharp features at short tenors.
    
    \item \textbf{FC VAE}: RMSE remains flat (0.0070 to 0.0069), with $3.3\times$ severity reduction. The latent bottleneck already regularizes the output, so the additional convexity constraint has limited incremental effect on accuracy.
    
    \item \textbf{MLP}: RMSE increases substantially from 0.0057 to 0.0065 ($+14\%$), the worst response of any model. The MLP lacks spatial structure to efficiently enforce a spatially defined constraint like convexity. With no convolutional filters to propagate the gradient signal spatially, the MLP must learn pointwise adjustments that satisfy the joint constraint across neighboring strikes---a difficult optimization problem that harms overall accuracy.
\end{itemize}

\subsection{Severity as the Key Metric}

The expected severity panel in Figure~\ref{fig:constraint-impact} provides the clearest picture: at $\lambda = 0.1$, all models reduce their expected severity by at least $2.4\times$ (MLP) and up to $8.1\times$ (Transformer). Even the MLP, despite its RMSE degradation, substantially reduces the economic impact of its violations. This motivates the use of expected severity rather than violation rate as the primary arbitrage metric---it captures the economically relevant combination of frequency and magnitude.

\section{Lambda Sweep and Pareto Frontiers}
\label{sec:lambda-sweep}

The previous section showed varied responses to a fixed constraint strength: from RMSE improvements (CNN, U-Net) to substantial degradation (MLP). To identify the optimal operating point per architecture, we train the three best architectures (Transformer, U-Net, CNN) at six constraint strengths: $\lambda \in \{0, 0.01, 0.05, 0.1, 0.5, 1.0\}$. Each configuration is trained three times with different random seeds, and the reported metrics are averaged across the three runs to reduce noise from training variance.\footnote{The lambda sweep seeds are independent of those used in the multi-seed validation of Table~\ref{tab:multi-seed}. Minor differences in the baseline ($\lambda = 0$) values between the two experiments are within the expected cross-seed variance.}

\begin{figure}[H]
    \centering
    \includegraphics[width=\textwidth]{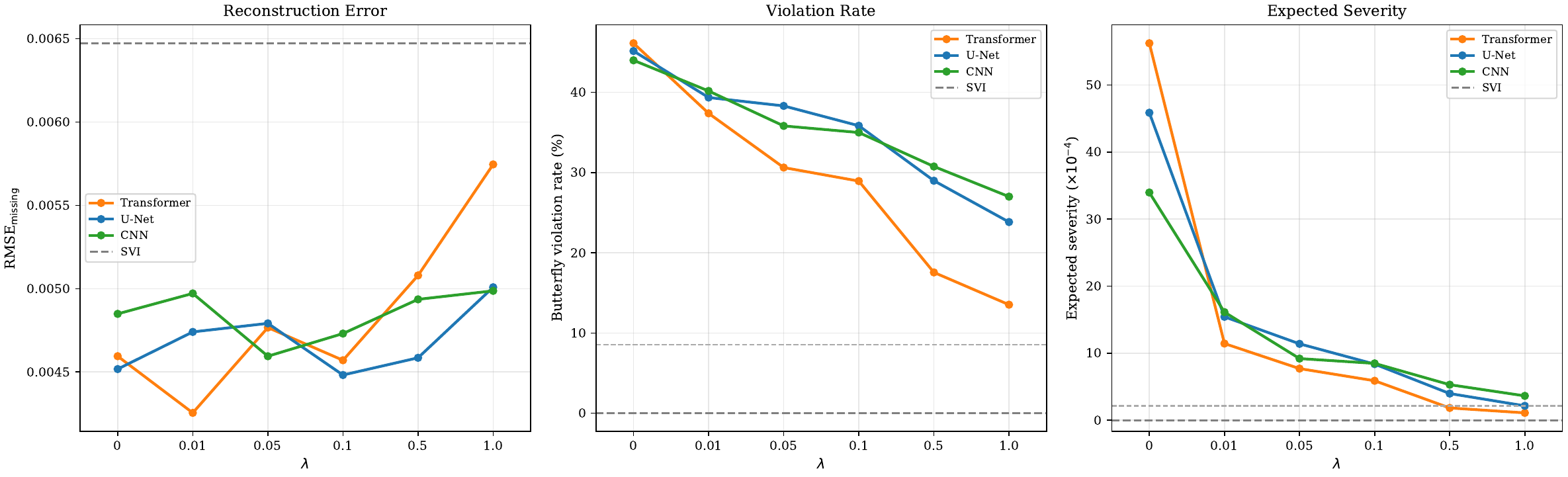}
    \caption{Effect of constraint strength $\lambda$ on reconstruction error (left), butterfly violation rate (center), and expected severity (right). Each point is the average of three random seeds. The dashed lines mark the SVI reference and ground truth butterfly rate ($\sim$8.6\%). All models reduce violations monotonically with $\lambda$, while the RMSE response is architecture-dependent.}
    \label{fig:pareto-lambda}
\end{figure}

\subsection{Violation Rate and Expected Severity}

The center panel of Figure~\ref{fig:pareto-lambda} shows that all three models reduce butterfly violation rates monotonically as $\lambda$ increases, confirming that the penalty provides fine-grained control over arbitrage quality. The Transformer achieves the steepest reduction, dropping from 46\% at $\lambda = 0$ to 14\% at $\lambda = 1.0$. The CNN and U-Net follow similar trajectories, reaching 27\% and 24\% respectively at $\lambda = 1.0$.

The right panel shows expected severity, which combines violation frequency and magnitude. The severity reduction is dramatic for all models: the Transformer drops by $51\times$ from $\lambda = 0$ to $\lambda = 1.0$, while the U-Net and CNN achieve $22\times$ and $9\times$ reductions respectively. At $\lambda \geq 0.5$, all models approach expected severities within an order of magnitude of SVI's near-zero level.

\subsection{RMSE Response: The Accuracy Cost}

The left panel reveals the architecture-dependent accuracy cost of constraints. The U-Net and CNN exhibit remarkable stability: their RMSE remains essentially flat across the entire $\lambda$ range (0.0044--0.0050), confirming that the convexity penalty acts as a regularizer rather than a competing objective. The CNN even \emph{improves} its RMSE at $\lambda = 0.05$ (0.0047 vs.\ 0.0050 unconstrained).

The Transformer shows a more complex behavior. At $\lambda = 0.01$, the regularization effect produces its best RMSE (0.0042), but beyond $\lambda = 0.1$ the RMSE increases steadily, reaching 0.0056 at $\lambda = 1.0$, a 30\% increase from the unconstrained baseline. It is worth noting that at $\lambda \geq 0.5$, the Transformer reaches the maximum of 500 training iterations without fully converging (the best checkpoint falls in the last iterations), suggesting that part of the degradation at high $\lambda$ may be due to more difficult optimization rather than solely an intrinsic trade-off between accuracy and convexity.

\subsection{Per-Model Analysis}

\paragraph{U-Net.} The U-Net exhibits the most favorable accuracy--arbitrage profile. Its RMSE remains flat at 0.0044 from $\lambda = 0$ through $\lambda = 0.1$, while expected severity drops by $5.5\times$ over the same range. Even at $\lambda = 1.0$, RMSE increases only to 0.0047, a modest 7\% cost for a $22\times$ severity reduction. This confirms that the convexity penalty is essentially free for the U-Net across a wide range of constraint strengths.

\paragraph{Transformer.} The Transformer achieves the lowest unconstrained RMSE (0.0043) and benefits from mild regularization at $\lambda = 0.01$ (RMSE 0.0042). However, it is the most sensitive to high $\lambda$ values, with RMSE rising to 0.0056 at $\lambda = 1.0$. As noted above, at $\lambda \geq 0.5$ the training does not converge within the 500-iteration budget, so these RMSE values represent an upper bound. The optimal operating point for the Transformer is $\lambda = 0.1$, where training converges normally, RMSE remains at 0.0047 (still better than the unconstrained CNN) while severity drops by $9.5\times$.

\paragraph{CNN.} The CNN closely parallels the U-Net but with slightly higher baseline RMSE. It benefits from moderate constraints: at $\lambda = 0.05$, RMSE improves from 0.0050 to 0.0047 while severity drops by $3.7\times$. At high $\lambda$, the CNN maintains stable RMSE (0.0048--0.0050), showing that convolutional architectures efficiently internalize the convexity constraint without sacrificing reconstruction quality.

\section{Constrained Models on Real Data}
\label{sec:constrained-real}

The regularization effect observed on synthetic data motivates applying constraints during fine-tuning on real SPY data. The real data experiments focus on the three best-performing architectures from the synthetic evaluation---CNN, U-Net, and Transformer---as these consistently outperformed the MLP and both VAE variants. The VAEs were excluded because their latent-space optimization inference procedure is computationally expensive and their synthetic RMSE was 40--60\% worse than the convolutional models. The MLP was excluded due to its poor response to constraints (Section~\ref{sec:fixed-lambda}).

Table~\ref{tab:real-results} already showed that fine-tuned models have lower butterfly rates than from-scratch models (33--35\% vs.\ 42--44\%), suggesting that synthetic pretraining implicitly teaches some smoothness. This is consistent with the Heston training surfaces being analytically convex: the fine-tuned models inherit a prior toward convex smiles from the synthetic pretraining phase.

Adding an explicit butterfly penalty ($\lambda = 0.1$) during fine-tuning further reduces violations, as shown in Table~\ref{tab:real-constrained}.

\begin{table}[H]
    \centering
    \caption{Effect of butterfly penalty ($\lambda = 0.1$) on fine-tuned models evaluated on real SPY data. Unlike synthetic data, constraints increase RMSE for all architectures, but substantially reduce butterfly violation rates.}
    \label{tab:real-constrained}
    \input{comparison/table_real_constrained}
\end{table}

Unlike the synthetic setting, where convolutional models improved RMSE with constraints, the real-data results show increased RMSE for all three architectures when the butterfly penalty is applied. The Transformer shows the best trade-off: a 24\% RMSE increase in exchange for a 49\% reduction in butterfly violations (from 42.3\% to 21.6\%). The CNN and U-Net show larger RMSE costs (27\% and 37\% respectively) for more modest violation reductions.

This difference from the synthetic results is likely explained by the nature of the data. Heston-generated surfaces are analytically smooth and convex, so a convexity penalty aligns with the ground truth and acts as a regularizer. Real SPY surfaces are noisier, contain genuinely non-convex features such as sharp short-tenor skew and earnings-related effects, and the training set is smaller (2,912 surfaces vs.\ 8,000). The penalty still reduces violations substantially, but forcing convexity in regions where the data is genuinely non-convex comes at a cost to reconstruction accuracy that is absent in the synthetic setting.

\section{Discussion}
\label{sec:arb-discussion}

\subsection{Constraints as Free Regularization}

The central finding of this chapter is that, for convolutional architectures, \emph{no-arbitrage constraints do not necessarily imply a trade-off}. The lambda sweep confirms that the CNN and U-Net maintain stable RMSE from $\lambda = 0$ through $\lambda = 1.0$ while achieving $9$--$22\times$ severity reductions. This changes the practical recommendation from ``how much accuracy are you willing to sacrifice for arbitrage-free surfaces?'' to ``always use a constraint penalty for convolutional models, as it is strictly beneficial.''

The Transformer exhibits a mild accuracy--arbitrage trade-off at high $\lambda$, but benefits from mild regularization at $\lambda = 0.01$ (its best RMSE). At the recommended $\lambda = 0.1$, the Transformer's RMSE of 0.0047 remains better than the unconstrained CNN, while severity drops by $9.5\times$.

\subsection{Connection to Physics-Informed Neural Networks}

The regularization effect of no-arbitrage penalties connects to the broader literature on physics-informed neural networks (PINNs) \cite{raissi2019pinn}, where physical laws are incorporated as soft penalty terms in the training loss. In PINNs, physical constraints often improve generalization by restricting the hypothesis space to physically plausible solutions. Our finding is analogous: the butterfly convexity condition is a ``financial physics'' constraint that restricts the hypothesis space to economically meaningful surfaces, and this restriction improves generalization because the true data-generating process respects the constraint.

\subsection{Practical Recommendations}

Based on the results of this chapter:

\begin{enumerate}
    \item \textbf{Always use constraints for CNN and U-Net.} The convexity penalty at $\lambda = 0.05$ to $\lambda = 0.1$ is free---it maintains RMSE while reducing expected severity by $4$--$6\times$.
    
    \item \textbf{Use moderate constraints for the Transformer.} A penalty of $\lambda = 0.1$ provides the best accuracy--severity trade-off, reducing severity by $9.5\times$ while keeping RMSE at 0.0047, still better than the unconstrained CNN.
    
    \item \textbf{Monitor expected severity, not just violation rate.} Violation rate alone is misleading: a model with 30\% violations of magnitude $10^{-4}$ is far less problematic than one with 5\% violations of magnitude $10^{-2}$. Expected severity captures this distinction.
\end{enumerate}

Overall, constrained neural models---particularly the U-Net at $\lambda = 0.05$--$0.1$ and the Transformer at $\lambda = 0.1$---combine the accuracy of neural networks with expected severity reduced by one to two orders of magnitude toward the SVI level, resolving the apparent tension between accuracy and economic consistency that motivated this chapter.

%% file: comparison/table_arbitrage.tex
\begin{tabular}{lrrrrrr}
\toprule
 & \multicolumn{2}{c}{RMSE$_\text{miss}$} & \multicolumn{2}{c}{Butterfly rate} & \multicolumn{2}{c}{Expected severity} \\
\cmidrule(lr){2-3} \cmidrule(lr){4-5} \cmidrule(lr){6-7}
Model & $\lambda=0$ & $\lambda=0.1$ & $\lambda=0$ & $\lambda=0.1$ & $\lambda=0$ & $\lambda=0.1$ \\
\midrule
Transformer & 0.0045 & 0.0047 & 45.0\% & 30.6\% & $5.87 \times 10^{-3}$ & $7.26 \times 10^{-4}$ \\
CNN & 0.0049 & 0.0047 & 45.1\% & 31.8\% & $3.84 \times 10^{-3}$ & $6.27 \times 10^{-4}$ \\
U-Net & 0.0048 & 0.0045 & 44.9\% & 36.0\% & $3.78 \times 10^{-3}$ & $8.07 \times 10^{-4}$ \\
MLP & 0.0057 & 0.0065 & 44.3\% & 38.9\% & $3.74 \times 10^{-3}$ & $1.54 \times 10^{-3}$ \\
FC VAE & 0.0070 & 0.0069 & 45.0\% & 37.1\% & $4.43 \times 10^{-3}$ & $1.33 \times 10^{-3}$ \\
Conv VAE & 0.0072 & 0.0060 & 43.2\% & 30.2\% & $3.20 \times 10^{-3}$ & $4.18 \times 10^{-4}$ \\
\midrule
SVI & 0.0065 & -- & 0.0\% & -- & $3.15 \times 10^{-11}$ & -- \\
\bottomrule
\end{tabular}

%% file: comparison/table_real_constrained.tex
\begin{tabular}{llrrrr}
\toprule
Model & Setting & RMSE$_\text{miss}$ & Butterfly & $\Delta$RMSE & $\Delta$Butterfly \\
\midrule
CNN & FT & 0.0052 & 33.4\% & -- & -- \\
CNN & FT + $\lambda{=}0.1$ & 0.0066 & 24.2\% & +27\% & $-$28\% \\
\midrule
U-Net & FT & 0.0054 & 35.3\% & -- & -- \\
U-Net & FT + $\lambda{=}0.1$ & 0.0074 & 28.5\% & +37\% & $-$19\% \\
\midrule
Transformer & FT & 0.0054 & 42.3\% & -- & -- \\
Transformer & FT + $\lambda{=}0.1$ & 0.0067 & 21.6\% & +24\% & $-$49\% \\
\bottomrule
\end{tabular}

%% file: chapters/08_conclusion.tex
This chapter summarizes the contributions and key findings of this thesis, discusses its limitations, and outlines directions for future work.

\section{Summary of Contributions}

This thesis addressed the problem of reconstructing complete implied volatility surfaces from incomplete market observations, framed as a two-dimensional signal reconstruction task. The main contributions are summarized below.

\begin{enumerate}
    \item We proposed an encoder-decoder Transformer with coordinate-based Fourier positional encoding for volatility surface reconstruction. The encoder processes only the observed grid points, while the decoder reconstructs all positions through cross-attention to the encoded observations. On synthetic data, this architecture achieved reconstruction error statistically tied with the best convolutional model (U-Net) across multiple seeds, and degraded most gracefully under increasing sparsity, achieving $9\times$ lower error than SVI at 90\% missing data. An ablation study confirmed that the Fourier positional encoding accounts for an 18\% RMSE improvement over learnable embeddings.

    \item We compared six deep learning architectures (MLP, CNN, U-Net, Transformer, FC-VAE, Conv-VAE) and the SVI parametric baseline, all matched at approximately 288,000 parameters. Multi-seed experiments validated the ranking stability: the Transformer and U-Net are statistically tied at the top (RMSE $0.0045 \pm 0.0004$ and $0.0045 \pm 0.0000$), with CNN and MLP trailing. This controlled setup isolated the effect of architectural inductive bias: spatial structure (CNN, U-Net) provides 14--17\% gains over the spatially agnostic MLP, while the generative bottleneck of VAEs limits their reconstruction fidelity.

    \item We formulated calendar spread and butterfly convexity conditions as differentiable penalty terms and integrated them into the training loss. A systematic sweep over penalty strengths revealed that for convolutional architectures these constraints act as regularizers: the CNN improved its RMSE by 6\% at $\lambda = 0.05$, while the U-Net maintained its accuracy virtually unchanged. Both architectures simultaneously reduced expected arbitrage severity by $4$--$6\times$. The Transformer traded a 9\% RMSE increase for a $9.5\times$ severity reduction at $\lambda = 0.1$. These results demonstrate that accuracy and arbitrage compliance are not necessarily in opposition.

    \item All models were trained at a fixed 30\% missing rate and evaluated across the range of 10\% to 90\% without retraining. Beyond random masking, we evaluated models under structured missingness (combined random and wing masking, simulating the systematic absence of deep OTM quotes). The Transformer proved nearly unaffected by structured masking (less than 20\% degradation), while convolutional models suffered 30--47\% RMSE increases when wing information was absent. This robustness advantage, rather than raw accuracy, is the Transformer's strongest differentiator.

    \item We validated all models on 3,900 SPY options surfaces spanning 2008 to 2025. All neural models achieved roughly $2\times$ lower error than SVI. A transfer learning analysis revealed that from-scratch training on real data outperformed fine-tuning from synthetic pretraining for CNN and U-Net, while only the Transformer benefited from pretraining on reconstruction accuracy. However, fine-tuned models consistently exhibited lower butterfly arbitrage rates, indicating that synthetic pretraining imparts useful structural priors (smoothness, convexity) even when it does not improve accuracy.
\end{enumerate}
\section{Key Findings and Discussion}

Beyond the individual contributions, several cross-cutting findings emerge from this work.

\textbf{The Transformer matches the best convolutional models on accuracy and excels on robustness.} Multi-seed experiments show that the Transformer and U-Net are statistically tied on reconstruction RMSE under standard conditions ($0.0045 \pm 0.0004$ vs.\ $0.0045 \pm 0.0000$). The Transformer's advantage emerges under challenging conditions: it degrades most gracefully under increasing sparsity ($9\times$ lower error than SVI at 90\% missing), and is nearly unaffected by structured wing masking (less than 20\% degradation vs.\ 30--47\% for convolutional models). It is also the only architecture to benefit from synthetic-to-real transfer learning on reconstruction accuracy. These properties make the Transformer the preferred architecture when robustness to varying and structured missingness is important.

\textbf{Fourier positional encoding is a meaningful architectural contribution.} The ablation study replacing Fourier coordinate features with learnable positional embeddings increased RMSE by 18\%. The multi-scale sinusoidal encoding of physical coordinates $(\tau, \log m)$ provides inductive bias that helps the attention mechanism distinguish positions at different resolutions, confirming the value of encoding domain-specific structure into the architecture.

\textbf{Graceful degradation under sparsity.} A distinctive strength of the Transformer is its behavior as the fraction of missing data increases. While all models were trained at a fixed 30\% missing rate, the Transformer's attention mechanism adapts naturally to any number of observed tokens, maintaining the lowest error across the full 10\%--90\% range. At 90\% missing data, it achieves $9\times$ lower error than SVI, which becomes underdetermined when individual maturity slices have too few observations. This graceful degradation is arguably the single most practically relevant finding, as real-world surfaces frequently exhibit high levels of missingness.

\textbf{Accuracy versus arbitrage is not a binary trade-off.} The conventional expectation is that imposing constraints on a model's output must come at the cost of reconstruction accuracy. On synthetic data, our results show that this is architecture-dependent: convolutional models (CNN, U-Net) maintain or improve their accuracy when trained with mild no-arbitrage penalties, as the convexity constraint aligns with the analytically smooth Heston ground truth and acts as a regularizer. On real data, the same constraint strength reduces violation rates substantially but increases RMSE for all architectures. This difference reflects the nature of the data: real SPY surfaces contain genuinely non-convex features (sharp short-tenor skew, earnings effects) that conflict with the convexity penalty. The regularization effect is thus data-dependent, working best when the ground truth respects the imposed constraint.

\textbf{Inductive bias is the key differentiator.} The most important factor in model performance is not the number of parameters but the type of structural assumption each architecture makes. Models without spatial bias (MLP) perform worst and respond poorly to spatial constraints. Models with local spatial bias (CNN, U-Net) provide strong baselines and benefit most from constraint regularization. Global attention (Transformer) excels under sparsity and structured missingness by capturing cross-maturity dependencies. Generative models (VAEs) produce smooth surfaces but sacrifice fine-grained fidelity due to their information bottleneck.

\textbf{Architecture choice depends on deployment context.} No single model dominates across all criteria. All neural models achieve sub-millisecond inference on GPU, but with meaningful differences: for latency-critical applications, the CNN offers inference times of 0.1 ms with competitive accuracy, while the Transformer provides the best accuracy at 0.9 ms per surface. For the best balance between accuracy and arbitrage compliance, the U-Net with a mild constraint penalty ($\lambda = 0.05$--$0.1$) offers an attractive operating point. All neural models are at least $15\times$ faster than SVI, which requires iterative optimization per surface.

\textbf{Transfer learning is nuanced.} Synthetic-to-real transfer does not uniformly help. The Heston model generates surfaces with different distributional characteristics than real SPY surfaces, and zero-shot transfer (applying a synthetically trained model directly to real data) degrades performance for all architectures. Fine-tuning recovers and sometimes exceeds from-scratch performance, but only for the Transformer. For convolutional models, the real dataset is sufficient on its own, and the inductive bias already provides the structural priors that pretraining would otherwise supply. The one consistent benefit of pretraining is lower arbitrage rates, suggesting that structural properties (smoothness, convexity) transfer across domains even when distributional properties do not.

\textbf{Connection to broader themes.} This work sits at the intersection of several active research areas: domain-constrained learning (as in physics-informed neural networks), masked prediction (as in masked autoencoders), and sim-to-real transfer (as in robotics). The finding that domain constraints can serve as regularizers rather than trade-offs may be relevant beyond finance, to any reconstruction task where the output must satisfy known structural properties.

\subsection{Practical Recommendations}

Based on the full set of experiments, we offer the following concrete guidance for practitioners:

\begin{enumerate}
    \item \textbf{Default choice: U-Net with mild butterfly penalty ($\lambda = 0.05$--$0.1$).} This combination achieves top-tier reconstruction accuracy while substantially reducing arbitrage violations. The constraint acts as a free regularizer for convolutional architectures on smooth underlying data.
    
    \item \textbf{High-sparsity or structured missingness: Transformer.} When significant portions of the surface are missing---particularly in the wings, as is common in real markets---the Transformer's global attention mechanism provides substantially better robustness than convolutional alternatives (less than 20\% degradation under wing masking vs.\ 30--47\% for CNN/U-Net).
    
    \item \textbf{Latency-critical applications: CNN.} At 0.1\,ms per surface (9$\times$ faster than the Transformer) with RMSE within 9\% of the best model, the CNN offers the most compelling speed--accuracy trade-off.
    
    \item \textbf{Always prefer neural models over SVI for reconstruction.} All neural models outperform SVI by at least 2$\times$ on RMSE, with the gap widening dramatically under sparsity. SVI remains valuable only when hard arbitrage-free guarantees are non-negotiable.
\end{enumerate}

\section{Limitations}

Several limitations of this work should be acknowledged.

\textbf{Single underlying asset.} All real-data experiments use SPY (S\&P 500 ETF) options, which exhibit a characteristic skew driven by equity index dynamics. It is unclear whether the findings generalize to other asset classes with different surface characteristics, such as foreign exchange options (which tend to exhibit symmetric smiles), commodity options (which reflect backwardation and contango dynamics), or single-stock options (which are subject to discrete events such as earnings announcements).

\textbf{Fixed grid resolution.} The $8 \times 25$ grid used throughout this work is relatively coarse. Higher-resolution grids would better approximate continuous surfaces but would also change the computational trade-offs: the Transformer's quadratic attention cost would grow more rapidly than the CNN's linear cost, potentially altering the performance ranking. Additionally, the approximately 8.6\% ground-truth butterfly violation rate observed in the synthetic data is a direct artifact of finite-difference approximations on this coarse grid.

\textbf{No temporal modeling.} Each volatility surface is treated as an independent sample. In practice, surfaces evolve smoothly over time, and today's surface is highly informative about tomorrow's. Exploiting this temporal structure could improve both accuracy and smoothness.

\textbf{Limited multi-seed validation.} We validated ranking stability with three random seeds for the top four models. While the results are consistent (the Transformer and U-Net are statistically tied, with CNN and MLP clearly behind), a larger number of seeds would provide tighter confidence intervals and enable formal statistical significance testing.

\textbf{Limited training masking.} All models are trained exclusively with uniform random masking at 30\%. While we evaluate under both random and structured (random+wing) patterns, training does not include structured masking or multiple rates simultaneously. Training with mixed masking strategies might improve robustness and alter the relative rankings.

\textbf{Soft constraints only.} We explore only soft penalty-based constraint enforcement. Alternative approaches such as architecturally guaranteed constraints (e.g., input-convex neural networks for butterfly, monotone layers for calendar) were not tested and could provide hard guarantees without accuracy trade-offs.

\textbf{Limited hyperparameter tuning.} To ensure a fair comparison, we deliberately kept training hyperparameters fixed across architectures (with the exception of learning rate for the Transformer). Per-architecture tuning might narrow the performance gaps between models.

\textbf{Synthetic data generator.} All synthetic experiments use the Heston model, which produces smooth, parametrically determined surfaces. A different generator, such as rough volatility models, local stochastic volatility, or empirical resampling, could produce surfaces with different characteristics (sharper features, non-convex regions) that favor different architectures. Similarly, combining a more realistic generator with training masking strategies that mimic market liquidity could jointly reduce the gap between synthetic and real data performance.

\section{Future Work}

Several directions for future research emerge from this work.

\textbf{Temporal models.} Incorporating sequences of surfaces through recurrent architectures or temporal Transformers could exploit the strong autocorrelation of volatility surfaces over time. Yesterday's surface could serve as an informative prior for today's reconstruction, potentially improving both accuracy and smoothness.

\textbf{Higher-resolution and irregular grids.} Testing on finer grids (e.g., $20 \times 50$) would assess scalability and might change the CNN-Transformer trade-off. Extending to irregular grids, where observations are not aligned to a fixed lattice, would better reflect real market data and could further favor the Transformer's ability to process unstructured token sets.

\textbf{Hard constraint enforcement.} Architecturally guaranteeing no-arbitrage properties, for example through input-convex neural networks for butterfly convexity or monotone output layers for calendar spread conditions, would provide hard guarantees that soft penalties cannot. Comparing such approaches with the penalty-based method studied here is a natural next step.

\textbf{Realistic missingness patterns.} Training with structured masking that mimics real market liquidity patterns (more missing in the wings and short tenors, less near at-the-money) might close the gap between synthetic and real data performance.

\textbf{Multi-asset generalization.} Evaluating on foreign exchange, commodity, and single-stock options would test whether the architectural conclusions drawn from equity index options hold across asset classes with fundamentally different surface characteristics.

\textbf{Ensemble and hybrid methods.} Combining the strengths of different architectures, for example using a CNN for fast initial estimation and a Transformer for refinement in sparse regions, could offer better accuracy-latency trade-offs than any single model.

\textbf{Generative improvements.} Replacing the VAE decoder with more expressive generative models such as diffusion models could address the information bottleneck that limits VAE reconstruction fidelity, while retaining the benefits of latent-space optimization at inference time.

\textbf{Production deployment.} Packaging the best-performing models as real-time services for trading desks and benchmarking against commercial volatility surface fitting tools would provide a practical validation of the approach beyond academic metrics.

\bigskip

In summary, this thesis demonstrates that reconstructing volatility surfaces from incomplete observations is a problem where architecture choice and the incorporation of domain knowledge matter more than model scale. The attention mechanism does not universally dominate, but it offers the greatest robustness under the conditions that characterize real markets: sparse data, structured missingness, and shifting distributions. No-arbitrage constraints, far from being an inevitable cost, can act as regularizers that simultaneously improve accuracy and economic consistency. These findings suggest that the paradigm of combining flexible architectures with domain-specific constraints may be useful beyond the financial domain, in any reconstruction problem where the output must respect known structural properties.